\documentclass[11pt]{article}

\usepackage[english]{babel}
\usepackage[utf8]{inputenc}
\usepackage{natbib}
\usepackage{amsmath}
\usepackage{graphicx}
\usepackage{color}
\usepackage{xcolor}
\usepackage{svg}
\usepackage{natbib}
\usepackage{url}
\usepackage{float}
\usepackage{amsfonts}
\usepackage{booktabs}
\usepackage{bbm}
\usepackage{listings}
\usepackage{lineno}
\usepackage[toc,page]{appendix}
\usepackage{enumitem}
\usepackage{amssymb}
\usepackage{pifont}
\usepackage{subcaption}
\usepackage{tikz}
\usepackage{authblk} 
\usepackage{appendix}
\usepackage[format=plain, labelfont={bf,it},
            textfont=it]{caption}
\newtheorem{theorem}{Theorem}

\newtheorem{definition}{Definition}
\newtheorem{algorithm}{Algorithm}
\newtheorem{remark}{Remark}
\newtheorem{example}{Example}

\usepackage[a4paper,margin=2.54cm]{geometry}

\usepackage{hyperref}
\hypersetup{
    colorlinks=true,
    linkcolor=blue,
    filecolor=red,
    citecolor=cyan,
    urlcolor=cyan,
}

\makeatletter
\def\Let@{\def\\{\notag\math@cr}}
\makeatother

\newcommand{\ssum}[3][i]{\sum_{#1=#2}^{#3}}

\newcommand{\sgn}{\mathrm{sign}}
\newcommand{\sign}[1]{\mathrm{sign}(#1)}

\newcommand{\diag}[1]{\mathrm{diag}(#1)}

\newcommand{\bsy}[1]{\boldsymbol{#1}}
\newcommand{\mbf}[1]{\mathbf{#1}}
\newcommand{\mbb}[1]{\mathbb{#1}}
\newcommand{\mca}[1]{\mathcal{#1}}
\newcommand{\trm}[1]{\textrm{#1}}

\newcommand*{\medcap}{\mathbin{\scalebox{1.4}{\ensuremath{\cap}}}}%
\title{Detection of Change Points in Piecewise Polynomial Signals Using Trend Filtering}

\author[1]{Reza Valiollahi Mehrizi}
\author[1]{Shojaeddin Chenouri}
\affil[1]{Department of Statistics and Actuarial Science, University of Waterloo}

\date{}

\begin{document}

\maketitle

\thispagestyle{empty}
\begin{abstract}
  While many approaches have been proposed for discovering abrupt changes in piecewise constant signals, few methods are available to capture these changes in piecewise polynomial signals. In this paper, we propose a change point detection method, PRUTF, based on trend filtering. By providing a comprehensive dual solution path for trend filtering, PRUTF allows us to discover change points of the underlying signal for either a given value of the regularization parameter or a specific number of steps of the algorithm. We demonstrate that the dual solution path constitutes a Gaussian bridge process that enables us to derive an exact and efficient stopping rule for terminating the search algorithm. We also prove that the estimates produced by this algorithm are asymptotically consistent in pattern recovery. This result holds even in the case of staircases (consecutive change points of the same sign) in the signal. Finally, we investigate the performance of our proposed method for various signals and then compare its performance against some state-of-the-art methods in the context of change point detection. We apply our method to three real-world datasets including the UK House Price Index (HPI), the GISS surface Temperature Analysis (GISTEMP) and the Coronavirus disease (COVID-19) pandemic.   

\end{abstract}



\section{Introduction}
\label{sec:introduction.proj1}
The problem of change point detection is more than sixty years old. It was first studied by \cite{page1954continuous,page1955test},  and since then, has been of interest to many scientists including statisticians. Many of the earlier developments concerned the existence of at most one change point; however, considerable attention in recent years has been given to multiple change point analysis, which has found applications in many fields such as finance and econometrics, \cite{bai2003computation, hansen2001new}, bioinformatics and genomics, \cite{futschik2014multiscale, lavielle2005using}, climatology, \cite{liu2010impacts, pezzatti2013fire}, and technology, \cite{siris2004application, oudre2011segmentation, lung2012distributed, ranganathan2012pliss, galceran2017multipolicy}. Consequently, there is a vast and rich literature on the subject. In the following, we only review a body of literature on a retrospective change point framework closely related to our work and refer the interested readers to \cite{eckley2011analysis, lee2010change, horvath2014extensions, truong2018review} for more comprehensive reviews.

We consider the univariate signal plus noise model
\begin{align}\label{fmodel.proj1}
    y_{_i}=f_{_i}+\varepsilon_{_i}, \qquad\qquad i=1,\,\ldots,\,n,
\end{align}
where $f_{_i}=f(i/n)$ is a deterministic and unknown signal with equally spaced input points over the interval $[0,\,1]$. The error terms $\varepsilon_{_1},\,\ldots,\,\varepsilon_{_n}$ are assumed to be independently and identically distributed Gaussian random variables with mean zero and finite variance $\sigma^2$.
We assume that $f(\cdot)$ undergoes $J_{_0}$ unknown and distinct changes at point fractions $0=\omega_{_0}<\omega_{_1}< \ldots< \omega_{_{J_0}}< \omega_{_{J_0+1}}=1$, where the number of change point fractions, $J_{_0}$ can grow with the sample size $n$. Additionally, we assume that $f(\cdot)$ is a piecewise polynomial function of order $r \in \mbb N$. These assumptions imply that, associated with $\omega_{_0}, \ldots, \omega_{_{J_0+1}}$, there are change points locations $0=\tau_{_0}<\tau_{_1}< \ldots< \tau_{_{J_0}}< \tau_{_{J_0+1}}=n$, which partition the entire signal $\mbf f=(f_{_1}, \ldots, f_{_n})$ into $J_0+1$ segments.  More specifically, any subsignal of $\mbf f$ within segments created by the change points follows an $r$-degree polynomial structure with or without a continuity constraint at the change points. For more detail, see Figure \ref{fig:coor-removal}. Change in the level of a piecewise constant signal, known as the canonical multiple change point problem, and change in the slope of a piecewise linear signal are examples of the problem under consideration in this paper.  In change point analysis, the objective is to estimate the number of change points, $J_{_0}$, as well as their locations $\bsy \tau=\{\tau_{_1},\,\ldots,\,\tau_{_{_{J_0}}}\}$ based on the observations $\mbf y=(y_{_1},\,\ldots,\,y_{_n})$. 


The canonical multiple change point problem, where the signal $\mbf f$ is modelled as a piecewise constant function, has been extensively studied in the literature. In this framework, there are many approaches and we only attempt to list a selection of them here. The majority of these techniques seek to identify all change points at once by solving an optimization problem consisting of a loss function, often the negative log-likelihood, and a penalty criterion.   \cite{yao1988estimating, yao1989least} used the square error loss along with the Schwarz Information Criterion (SIC) as a penalty function to consistently estimate the bounded number of change points and their locations for the data drawn from a Gaussian distribution. Within the same setting,
incorporation of various penalty functions including Modified Information Criterion (MIC) \cite{pan2006application}, modified Bayesian Information Criterion (mBIC) \cite{zhang2007modified}, Simultaneous Information Theoretic Criterion (SITC) \cite{wu2008simultaneous} and modified SIC \cite{ciuperca2011general, ciuperca2014model}, have been studied. Specific algorithms such as Optimal Partitioning \cite{auger1989algorithms}, Segment Neighbourhood \cite{jackson2005algorithm}, and pruning approaches such as PELT \cite{killick2012optimal} and PDPa \cite{rigaill2015pruned} are developed to solve such optimization problems. 

Apart from penalty-based techniques, another frequently used class of change point detection approaches encompasses the greedy search procedures in which they search sequentially for one single change point at a time. The most popular methods in this class are Binary Segmentation \cite{vostrikova1981detecting} and its variants such as Circular Binary Segmentation (CBS) \cite{olshen2004circular}, and Wild Binary Segmentation (WBS) \cite{fryzlewicz2014wild}. In recent years, researchers have attempted to improve Binary Segmentation's performance from statistical and computational viewpoints. \cite{fryzlewicz2018tail} suggested a backward (bottom-up) mechanism, called Tail Greedy Unbalanced Haar (TGUH), which is computationally fast and statistically consistent in estimating both the number and the locations of change points.
Also, \cite{fryzlewicz2018detecting} introduced Wild Binary Segmentation 2 (WBS2) to deal with the shortcomings of WBS in datasets with frequent changes. It has been shown that the method is fast in run time and accurate in detection.

Beyond the canonical change point problem, signals in which $f$ is modelled as a piecewise polynomial of order $r\geq 1$ have attracted less attention in the literature despite many applications. For instance, piecewise linear signals are applied in monitoring patient health (\cite{aminikhanghahi2017survey}, \cite{stasinopoulos1992detecting}), climate change (\cite{robbins2011changepoints}), and finance (\cite{bianchi1999comparison}). In such a framework,  \cite{bai1997estimating} introduced a method based on Wald-type sequential tests,  and \cite{maidstone2017optimal} devised a dynamic programming applied to an $\ell_{_0}$-penalized least square procedure. In continuous piecewise linear models, \cite{kim2009ell_1} developed a methodology called $\ell_{_1}$-trend filtering. Furthermore, \cite{baranowski2019narrowest} put forward the method of Narrowest Over Threshold (NOT), and \cite{anastasiou2019detecting} developed an approach called Isolate-Detect (ID) which both provide asymptotically consistent estimators of the number and locations of change points. 

Our goal in this paper is to introduce a unifying method covering the canonical change point problem and beyond. More precisely, the method is cable of detecting change points in piecewise polynomial signals of order $r$ ($r=0,\,1,\,2,\,\ldots$) with and without continuity constraint at the locations of change points.

The detection of change points in a sequence of data can be formulated as a penalized regression fitting problem. According to our notation, the quantity $f_{\tau}-f_{\tau+1}$ is nonzero if the signal $ f$ undergoes a change at point $\tau$, and is zero otherwise. Moreover, if we assume that change points are sparse, that is, the number of locations where $f$ changes, $J_{_0}$, is much smaller than the number of observations $n$, change points can be estimated using the one-dimensional fused lasso problem
\begin{align*}
    \min_{\mbf f\in \mathbb{R}^n}
    \frac{1}{2}\, \big\|\,\mbf y- \mbf f \, \big\|_2^2 \,+\, \lambda\, \ssum{1}{n-1} \big| f_{_{i+1}}-f_{_i} \big|\,,
\end{align*}
where $\mbf f=(f_{_1},\,\ldots,\,f_{_n})$.

This formulation of the canonical change point problem was first considered in \cite{huang2005detection} and was applied to analyze a DNA copy number dataset. \cite{harchaoui2010multiple} considered the same formulation and proved the consistency of the respective change point estimates when the number of change points is bounded. Employing sparse fused lasso which is composed of both the $\ell_{_1}$-norm and the total variation seminorm penalties, \cite{rinaldo2009properties} proposed a sparse piecewise constant fit and established the consistency of the corresponding estimates when the variance of the noise terms vanishes, and the minimum magnitude of jumps is bounded from below. However, \cite{rojas2014change} argued that the consistency results achieved by \cite{rinaldo2009properties} are incorrect when a frequently viewed pattern, called {\it staircase}, exists in the signal. The staircase phenomenon occurs in a piecewise constant model when there are either two consecutive downward jumps or upward jumps in its mean structure. The staircase pattern will be discussed in more detail in Section \ref{sec:pattern.recovery.proj1}. Additionally, \cite{qian2016stepwise} showed that the lasso problem of \cite{tibshirani1996regression} when derived by transforming fused lasso does not satisfy the Irrepresentable Condition (\cite{zhao2006model}) that is necessary and sufficient for exact pattern recovery. In particular, \cite{qian2016stepwise} proposed an approach called preconditioned fused lasso based on the puffer transformation of \cite{jia2015preconditioning} and established that it can recover the exact pattern with probability approaching one. 


A similar approach to that of the piecewise constant signals can be considered for estimating change points in piecewise polynomial signals. 
In particular, a positive integer $\tau$ is a change location in an $r$-th degree piecewise polynomial signal $f$ if $\tau$-th element of the vector $\mbf D^{(r+1)}\, \mbf f$ is non-zero, denoted by $[\,\mbf D^{(r+1)}\, \mbf f\,]_{\tau}\neq 0$. Here $\mbf D^{(r+1)}$ is a penalty matrix that will be defined in Section \ref{sec:dual.tf.proj1}. Hence, change points can be estimated from nonzero elements of $\mbf D^{(r+1)}\, \widehat{\mbf f}$, where $\widehat{\mbf f}$ is the solution of 
\begin{align}\label{tf.obj.proj1}
    \min_{\mbf f\in \mathbb{R}^{^n}}
    \frac{1}{2}\|\,\mathbf{y}-\mbf f \,\|_{_2}^2+\lambda \,\|\mathbf{D}^{(r+1)}\mbf f \|_{_1}\,.
\end{align}

The aforementioned problem was first studied by \cite{steidl2006splines} in the context of image processing and was called {\it higher order total variation regularization}. It was later rediscovered by \cite{kim2009ell_1} and termed {\it trend filtering} in the nonparametric regression setting. \cite{kim2009ell_1} specifically explored linear trend filtering ($r=1$) which fits piecewise linear models. \cite{tibshirani2014adaptive} extensively studied trend filtering and compared its performance with smoothing splines \cite{green1993nonparametric} and locally adaptive regression splines \cite{mammen1997locally} in the context of nonparametric regression. \cite{tibshirani2014adaptive} also established that trend filtering enjoys desirable and strong theoretical properties of locally adaptive regression splines while being computationally less intensive due to its banded penalty matrix. Moreover, trend filtering has an adaptive knot selection property, which makes it well suited for change point analysis. 

From a computational and algorithmic standpoint, \cite{kim2009ell_1} described Primal-Dual Interior Point (PDIP) method for deriving the estimates of the linear trend filtering problem at a fixed value of $\lambda$. This can be easily carried over to the trend filtering problem of any order. 
 \cite{wang2014falling} suggested an algorithm based on a falling factorial basis while \cite{ramdas2016fast} derived an algorithm based on the Alternating Direction Method of Multipliers (ADMM) discussed in \cite{boyd2011distributed}.
The computational complexity of all these algorithms is of order $O(n)$.

In this paper, we develop a new methodology called  {\it Pattern Recovery Using Trend Filtering} (PRUTF) for identifying unknown change points in piecewise polynomial signals with no continuity restriction at change point locations. Therefore, a change point is defined as a sudden jump in the signal and its all derivatives up to order $r$. Figure \ref{fig:coor-removal} displays such change points for various $r$. In this paper, we make the following contributions.

\begin{itemize}
    \item We propose a generic dual solution path algorithm along with the regularization parameter for trend filtering. This solution path, whose basic idea is borrowed from \cite{tibshirani2011solution} enables us to determine change points at each level of the regularization parameter. Our algorithm, PRUTF, is different from that of \cite{tibshirani2011solution} as we remove $(r+1)$ coordinates of dual variables after identifying each change point. This adjustment to the algorithm allows us to have independent dual variables between each pair of neighbouring change points. Besides, the elimination of $(r+1)$ coordinates at each step leads to faster implementation of the algorithm.
    
    \item We establish a stopping criterion that plays an essential rule in the PRUTF algorithm used to find change points. Notably, we show that the dual variables of trend filtering between consecutive change points constitute a Gaussian bridge process. This finding allows us to introduce a threshold for terminating our proposed algorithm.
    
    \item If the signal contains a staircase pattern, we prove that the method is statistically inconsistent, which makes it unfavourable. Explaining the reason for this end, we modify the PRUTF algorithm to produce estimates consistent in terms of both the number and location of change points.
\end{itemize}


This paper is organized as follows: we first describe how to characterize the dual optimization problem of trend filtering. In Section \ref{sec:solution-path-algorithm}, we develop our main algorithm, PRUTF, to use in constructing the dual solution path of trend filtering and, in turn, identifying the locations of change points. Section \ref{sec:property.solution.path.proj1} discusses the properties of this dual solution path. We establish that the dual variables derived from the solution path form a Gaussian bridge process that makes them favourable for statistical inference. Applying these properties, we develop a stopping rule for the change point search algorithm in Section \ref{sec:stop.rule.proj1}. The quality of the PRUTF algorithm is validated in terms of pattern recovery of the true signal in Section \ref{sec:pattern.recovery.proj1}. It is established that the proposed technique in its naive form fails to consistently identify the true signal when a special pattern, called staircase, is present in the signal. Section \ref{sec:modified.trend.filtering.proj1} elaborates on how to modify the algorithm in order to estimate the true pattern consistently. Simulation results, and real-world applications are presented in Section \ref{sec:simulation.proj1}. We explore the performance of our proposed method for signals with frequent change points as well as models with dependent error terms in Section \ref{sec:model_misspecification.proj1}.  We conclude the paper with a discussion in Section \ref{sec:discussion.proj1}.

\section{Notations and Fundamental Concepts}\label{sec:notations.concepts.proj1}

\subsection{Notations}\label{sec:notations.proj1}
We begin this section with setting up notations that will be used throughout this article. For an $m\times n$ matrix $\mbf A$, we denote its rows by $\mbf A_{_1},\ldots,\mbf A_{_m}$ and express the matrix as $\mbf A=(\mbf A_{_1}^{^T},\ldots,\mbf A_{_m}^{^T})^T$. Now for the set of indices $\mca I=\{i_{_1},\,\ldots,\,i_{_k}\}\subseteq\{1,\,\ldots,\,m\}$, the notation $\mbf A_{_{\mca I}}=(\mbf A_{i_{_1}}^{^T},\,\ldots,\,\mbf A_{i_{_k}}^{^T})^T$ represents the submatrix of $\mbf A$ whose row labels are in the set $\mca I$. In a similar manner, for a vector $\bsy a$ of length $m$, we let $\bsy a_{_{\mca I}}=( a_{_{i_{_1}}},\ldots, a_{_{i_{_k}}})^{^T}$ denote a subvector of $\bsy a$ whose coordinate labels are in $\mca I$. We write $\mbf A_{_{-\mca I}}$ and $\bsy a_{_{-\mca I}}$ to denote $\mbf A_{_{\{1,\,\ldots,\,m\} \backslash \mca I}}$ and $\bsy a_{_{\{1,\,\ldots,\,m\} \backslash \mca I}}$\,, respectively, where
$\mca J\backslash\mca I$ is the set of indices in $\mca J$ but not in $\mca I$.
Furthermore, for selecting $i$-th row of $\mbf A$, the notation $[\mbf A]_i$ and for its  $(i,j)$-th element the notation $[\mbf A]_{ij}$ are used. Also, $[\bsy a]_i$ extracts the $i$-th elements of the vector $\bsy a$. We write $\diag{\mbf A}$ to denote the vector of the main diagonal entries of the matrix $\mbf A$. Moreover, for a real number $x$, $\lfloor x\rfloor$ denotes the greatest integer less than or equal $x$, and $\lceil x \rceil$ denotes the least integer greater or equal $x$. For a set $A$, the indicator function is denoted by $\mathbbm{1}(A).$

\subsection{The Dual Problem of Trend Filtering}
\label{sec:dual.tf.proj1}

\begin{figure}[!ht]
\begin{subfigure}{.32\textwidth}
  \centering
  \includegraphics[width=1\linewidth]{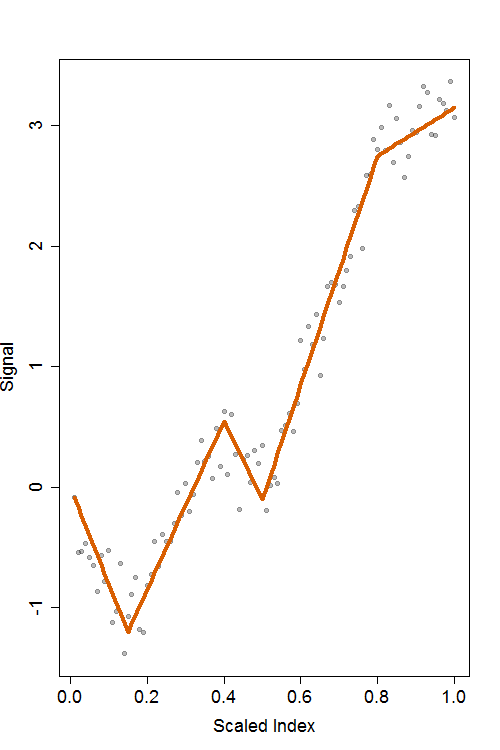}  
  \caption{Linear}
\end{subfigure}
\begin{subfigure}{.32\textwidth}
  \centering
  \includegraphics[width=1\linewidth]{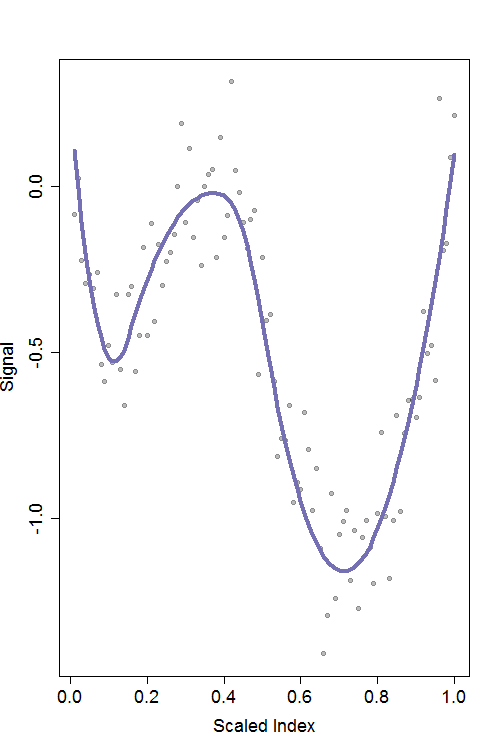}  
  \caption{Quadratic}
\end{subfigure}
\begin{subfigure}{.32\textwidth}
  \centering
  \includegraphics[width=1\linewidth]{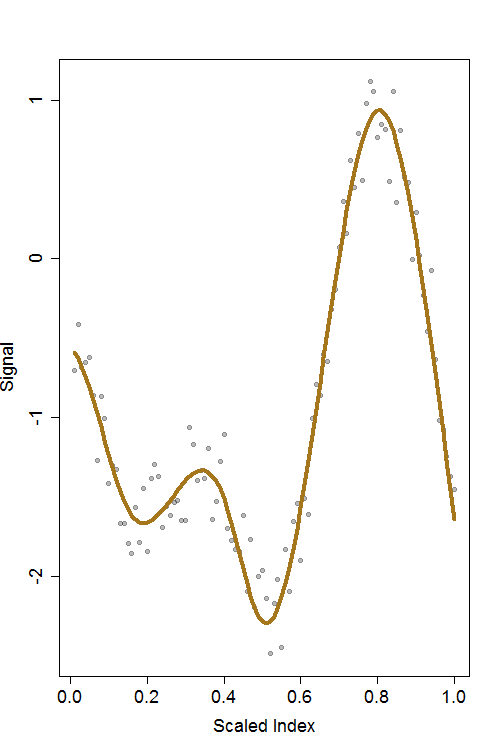}  
  \caption{Cubic}
\end{subfigure}
\caption[]{Trend filtering solutions for $r=1,\,2,\,3$ producing (a) piecewise linear, (b) piecewise quadratic and (c) piecewise cubic fits, respectively.}
\label{fig:tf-splines}
\end{figure}

Recall the trend filtering problem  
\begin{align}\label{tf2.obj.proj1}
    \min_{\mbf f\in \mathbb{R}^{^n}}
    \frac{1}{2}\|\mathbf{y}-\mbf f \|_{_2}^2+\lambda \|\mathbf{D}^{(r+1)}\mbf f \|_{_1},
\end{align}
where $\lambda\geq 0$ is the regularization parameter for controlling the effect of smoothing, and the $(n-r-1)\times n$ penalty matrix $\mathbf{D}^{(r+1)}$ is the difference operator of order $(r+1)$. For $r=0$, the first order difference matrix $\mathbf{D}^{(1)}$ is defined as 
\begin{align*}
    \mathbf{D}^{(1)}=\begin{pmatrix}
    -1 & 1 & 0 & \ldots & 0 & 0 \\
    0 & -1 & 1 & \ldots & 0 & 0 \\
    \vdots & & & & & \vdots \\
    0 & 0 & 0 & \ldots & -1 & 1 \\
    \end{pmatrix},
\end{align*}
and for $r\geq 1$, the difference operator of order $r+1$ can be recursively computed by $\mathbf{D}^{(r+1)}=\mathbf{D}^{(1)}\times \mathbf{D}^{(r)}$. Notice that, in this matrix multiplication, we  consider only the submatrix consisting of the first $n-r-2$ rows and $n-r-1$ columns of the matrix $\mathbf{D}^{(1)}$. Figure \ref{fig:tf-splines} displays the trend filtering fits for $r=0,1,2$ for simulated data.

Although the objective function in the $r$-th order trend filtering \eqref{tf2.obj.proj1} is strictly convex and thus the minimization has a guaranteed unique solution, the penalty term is not differentiable in $\mbf f $, so solving the optimization in its current form is difficult.  
To overcome this difficulty, we follow the argument in \cite{tibshirani2011solution} and convert this optimization problem into its dual form. Since the objective function in the primal problem is strictly convex with no constraint, the strong duality holds, meaning that the primal and the dual solutions coincide.

The trend filtering problem \eqref{tf2.obj.proj1} can be rewritten as 
\begin{equation*}
 \min_{\mbf f \in \mathbb{R}^{^n}}
    \frac{1}{2}\|\mathbf{y}-\mbf f \|_{_2}^2+\lambda \|\mathbf{z}\|_{_1}, \quad \text{subject to } \mathbf{z}=\mathbf{D}\mbf f \,,
\end{equation*}
where, for ease in the notation, we use $\mathbf{D}=\mathbf{D}^{(r+1)}$. For any given $\lambda>0$, the Lagrangian is
\begin{align*}
 \mathcal{L}(\mbf f ,\,\mathbf{z},\,\mathbf{u})&=\frac{1}{2}\|\mathbf{y}-\mbf f \,\|_{_2}^2+\lambda \|\mathbf{z}\|_{_1}+\mathbf{u}^T(\mathbf{D}\mbf f -\mathbf{z})
\end{align*}
and, thus the dual function is given by
\begin{equation*}
g(\mathbf{u})= \inf_{\mbf f \in \mathbb{R}^{^n},\,\mathbf{z}\in\mathbb{R}^{^m}}\mathcal{L}(\mbf f ,\,\mathbf{z},\,\mathbf{u}),
\end{equation*}
which is a concave function defined on $\mathbb{R}^{^m}$, where $m=n-r-1$ and takes values in the extended real line $\mathbb{R}\cup\lbrace -\infty,\,\infty\rbrace$. The vectors $\mbf f $ and $\mathbf{u}$ are called the primal and dual variables, respectively. 
Taking the derivative of the Lagrangian $\mathcal{L}(\mbf f ,\,\mathbf{z},\,\mathbf{u})$ with respect to $\mbf f $ and setting it to be equal to zero, we obtain
\begin{align}\label{primal.dual.exact.proj1}
    \mbf f =\mathbf{y-D}^T\mathbf{u}.
\end{align}
Now substituting this back into the Lagrangian $\mathcal{L}(\mbf f ,\,\mathbf{z},\,\mathbf{u})$, and performing certain algebraic manipulations, we obtain
\begin{align*}
\mathcal{L}^\ast(\mathbf{z},\,\mathbf{u})&=\inf_{\mbf f \in \mathbb{R}^{^n}}\mathcal{L}(\mbf f ,\,\mathbf{z},\,\mathbf{u})\\
&=-\frac{1}{2}\|\mathbf{y}-\mathbf{D}^T\mathbf{u}\|_{_2}^2+\frac{1}{2}\|\mathbf{y}\|^2+\lambda \|\mathbf{z}\|_{_1}-\mathbf{u}^T\mathbf{z}\,.
\end{align*}  
Minimizing $\mathcal{L}^\ast(\mathbf{z},\,\mathbf{u})$, or equivalently maximizing $\mathbf{u}^T\mathbf{z}-\lambda \|\mathbf{z}\|_{_1}$, with respect to $\mathbf{z}\in\mathbb{R}^{^m}$ leads us to the dual function $g(\mathbf{u})$. Notice that $\sup\limits_{\mathbf{z}}\lbrace\mathbf{u}^T\mathbf{z}-\lambda \|\mathbf{z}\|_{_1}\rbrace$ is the conjugate of the function $\lambda \|\mathbf{z}\|_{_1}$ in the context of conjugate convex functions. See \cite{brezis2010functional} and \cite{boyd2004convex}. This conjugate function is given by
\begin{align*}
    \sup\limits_{\mathbf{z}}\lbrace\mathbf{u}^T\mathbf{z}-\lambda \|\mathbf{z}\|_{_1}\rbrace=\begin{cases}
    0 & \text{ if }\|\mathbf{u}\|_{_\infty}\le \lambda\\
    \infty & \text{ otherwise\,.}
    \end{cases}
\end{align*}
From all these, the dual function is given as
\begin{align*}
g(\mathbf{u})=  -\frac{1}{2}\|\mathbf{y}-\mathbf{D}^T\mathbf{u}\|_{_2}^2+\frac{1}{2}\|\mathbf{y}\|^2 \quad \textrm{ for }\quad \|\mathbf{u}\|_{_\infty}\leq\lambda\,,
\end{align*}
and, thus the dual problem is to find the maximum of the dual function $g(\mathbf{u})$. This is equivalent to 
\begin{equation}\label{tf.dual.obj.proj1}
\min\limits_{\mathbf{u}\in \mathbb{R}^{^m}} \frac{1}{2}\|\mathbf{y}-\mathbf{D}^T\mathbf{u}\|_{_2}^2   \quad \textrm{subject to }\quad \|\mathbf{u}\|_{_\infty}\leq\lambda\,.
\end{equation}  
The constraint in \eqref{tf.dual.obj.proj1} is an $\ell_{_\infty}$-ball or a hypercube centered at the origin with the boundaries given by the set $\lbrace -\lambda,\,+\lambda\rbrace^{m}$. Since the matrix $\mbf D$ is full row rank, the problem \eqref{tf.dual.obj.proj1} is strictly convex and has a unique solution, see \cite{ali2019generalized}. In addition, notice that the dimension of the dual vector $\mbf u$ is $m$, which is smaller than that of the primal vector $\mbf f $ and may lead to relatively faster computations. The connection between the primal and the dual solutions is given by the equations 
\begin{align}\label{primal.to.dual}
    \hspace{-1.45cm}\widehat{\mbf u}_{_\lambda}=\lambda \,\widehat{\bsy\gamma},
\end{align}
\vspace{-1.2cm}
\begin{align}\label{dual.to.primal}
    \widehat{\mbf f }_{_\lambda}=\mathbf{y}-\mathbf{D}^T\widehat{\mathbf{u}}_{_\lambda}\,,
\end{align}
where $\widehat{\bsy\gamma} \in \mathbb{R}^{^m}$ is a subgradient of $\| \mbf x\|_{_1}$ computed at $\mbf x=\mbf D\widehat{\mbf f }_{_\lambda}$. This subgradient is given by
\begin{align}\label{gamma.subgrad}
    \widehat{\gamma}_{_i}\in \left\{
    \begin{array}{lcl}
        \{+1\} & \textrm{if} & [\mathbf{D\widehat{\mbf f }_{_\lambda}}]_i>0  \\
        \{-1\} & \textrm{if} & [\mathbf{D\widehat{\mbf f }_{_\lambda}}]_i<0  \\
         \,[-1,+1] & \textrm{if} &  [\mathbf{D\widehat{\mbf f }_{_\lambda}}]_i=0\,. 
    \end{array}
    \right.
\end{align}
The statements in Equations \eqref{primal.to.dual}-\eqref{gamma.subgrad}  are equivalent to the KKT optimality conditions of the primal problem \eqref{tf2.obj.proj1}. 
The dual problem \eqref{tf.dual.obj.proj1} demonstrates that  $\mbf D^T\widehat{\mbf u}_{_\lambda}$ is the projection, $P_{_{\mbb C}}(\mbf y)$, of $\mathbf{y}$ onto the convex polyhedron (or hypercube here) $\mathbb{C}=\{\mbf x\in \mathbb{R}^{^m}:\, \|\mbf x\|_{_\infty}\leq \lambda\}\,$.
From this, the primal solution \eqref{dual.to.primal} can be rewritten in the form of $(\mbf I- P_{_{\mathbb{C}}})\,(\mbf y)$, representing the residual projection map of $\mbf y$ onto the polyhedron $\mathbb{C}$. 

Our idea of applying trend filtering to discover change points in piecewise polynomial signals is inspired by \cite{rinaldo2009properties} and its correction \cite{rinaldo2014corrections}, in which change point detection is studied using fused lasso. Besides extending to piecewise polynomial signals, the novelty of our work is in providing an exact stopping criterion, which is based on the Gaussian bridge property of the trend filtering dual variables. In addition, we propose an algorithm which, unlike that proposed in \cite{rinaldo2009properties}, always produces consistent change points even in the presence of staircase patterns.

\section{Solution Path of Trend Filtering and PRUTF Algorithm}\label{sec:solution-path-algorithm}

In this section, we construct and study the solution path of dual variables $\widehat{\mbf u}_{\lambda}$ as the regularization parameter decreases from $\lambda=\infty$ to $\lambda=0$. In the following, the PRUTF algorithm is given to compute the entire dual solution path. This dual solution path identifies the corresponding primal solution using \eqref{dual.to.primal}. For any given $\lambda$, we call any coordinate of $\widehat{\mbf u}_{\lambda}$ a boundary coordinate if it is a vertex of the polyhedron $\mathbb{C}= \big\{ \mbf x\in \mathbb{R}^{^m}:\, \|\mbf x\|_{_\infty}\leq \lambda \big\}\,$, meaning that its absolute value becomes $\lambda$. In the process of constructing the solution path, for any $\lambda$, we trace several sets, introduced below.
\begin{itemize}
\item The set $\mca B=\mca B(\lambda)$, called the boundary set,  contains the boundary coordinates identified by $\widehat{\mbf u}_{\lambda}$.
\item The vector $\mbf s_{\mca B}=\mbf s_{\mca B}(\lambda)$, called the sign vector, represents collectively the signs of the boundary points in $\mca B(\lambda)$. 
\item The set $\mca A=\mca A(\lambda)$, called the augmented boundary set,  contains the boundary coordinates in $\mca B(\lambda)$  as well as the first $r_{_a}=\lfloor (r+1)/2\rfloor$ coordinates immediately after. 
\item  The vector $\mbf s_{\mca A}=\mbf s_{\mca A}(\lambda)$ represents collectively the signs of the augmented boundary points in $\mca A(\lambda)$. 
\end{itemize}

\begin{figure}[!t]
\begin{center}
\begin{subfigure}{.43\textwidth}
  \centering
  \includegraphics[width=1\linewidth]{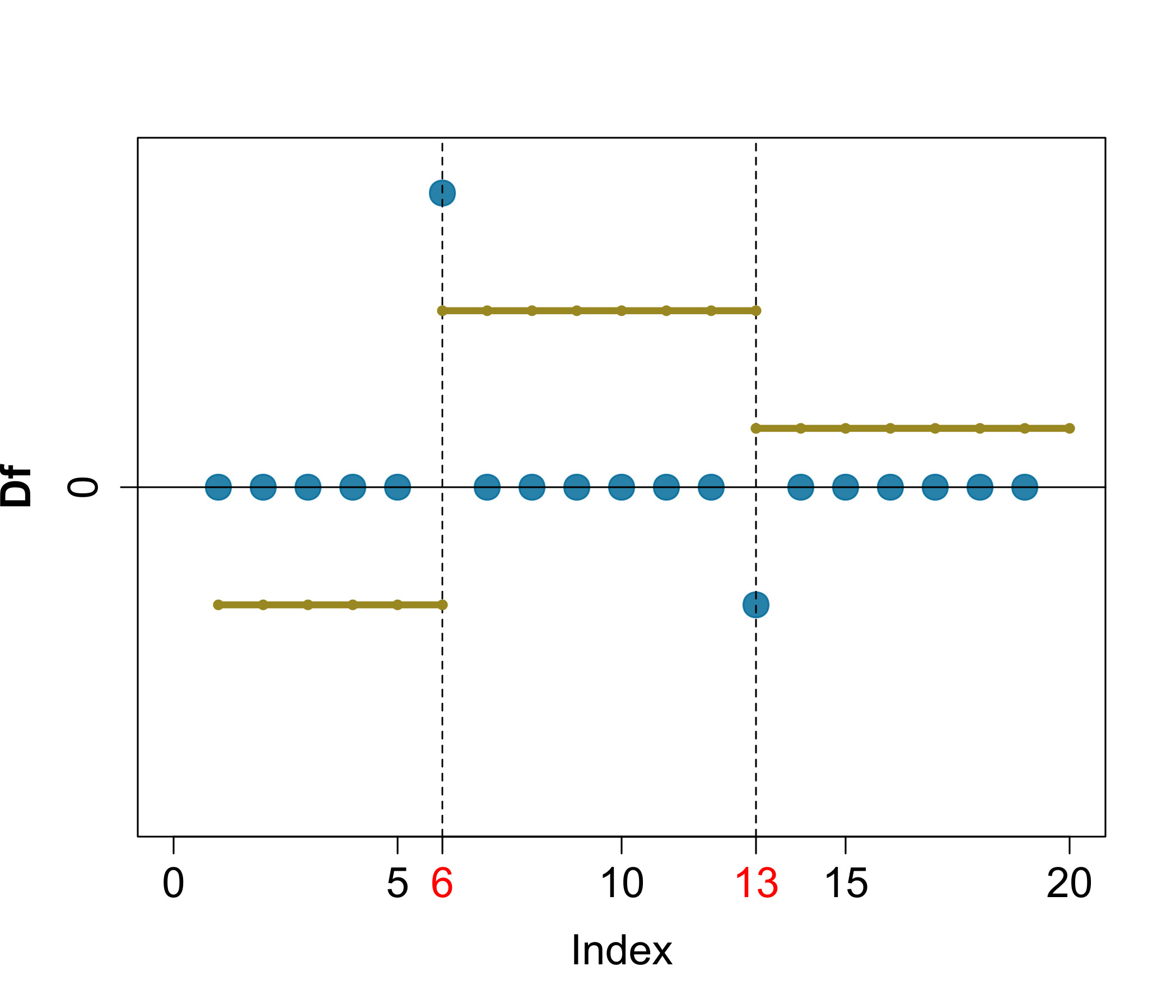}
  \caption{Piecewise constant, $r=0$.}
\end{subfigure}
\begin{subfigure}{.43\textwidth}
  \centering
  \includegraphics[width=1\linewidth]{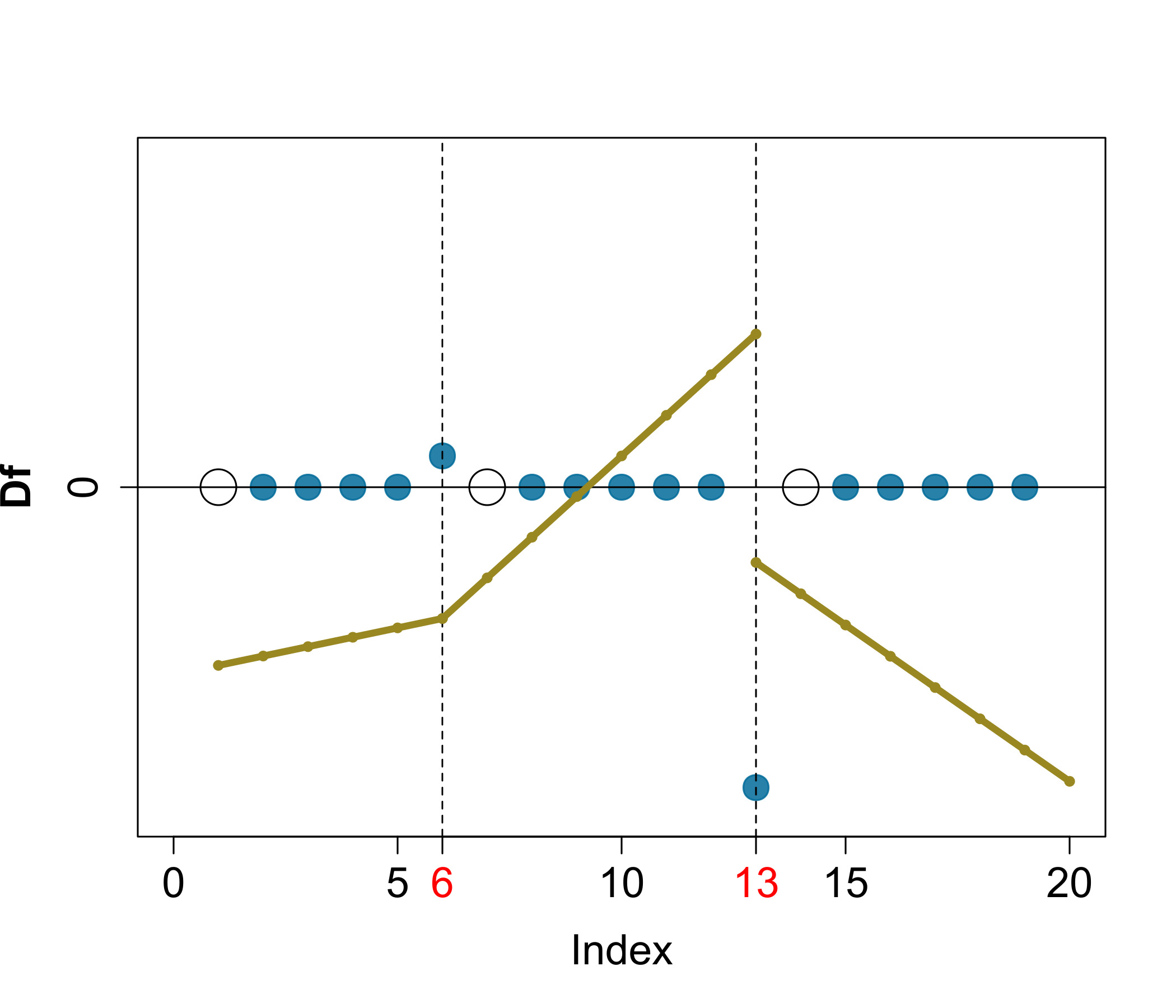}
  \caption{Piecewise linear, $r=1$.}
\end{subfigure}
\\
\begin{subfigure}{.43\textwidth}
  \centering
  \includegraphics[width=1\linewidth]{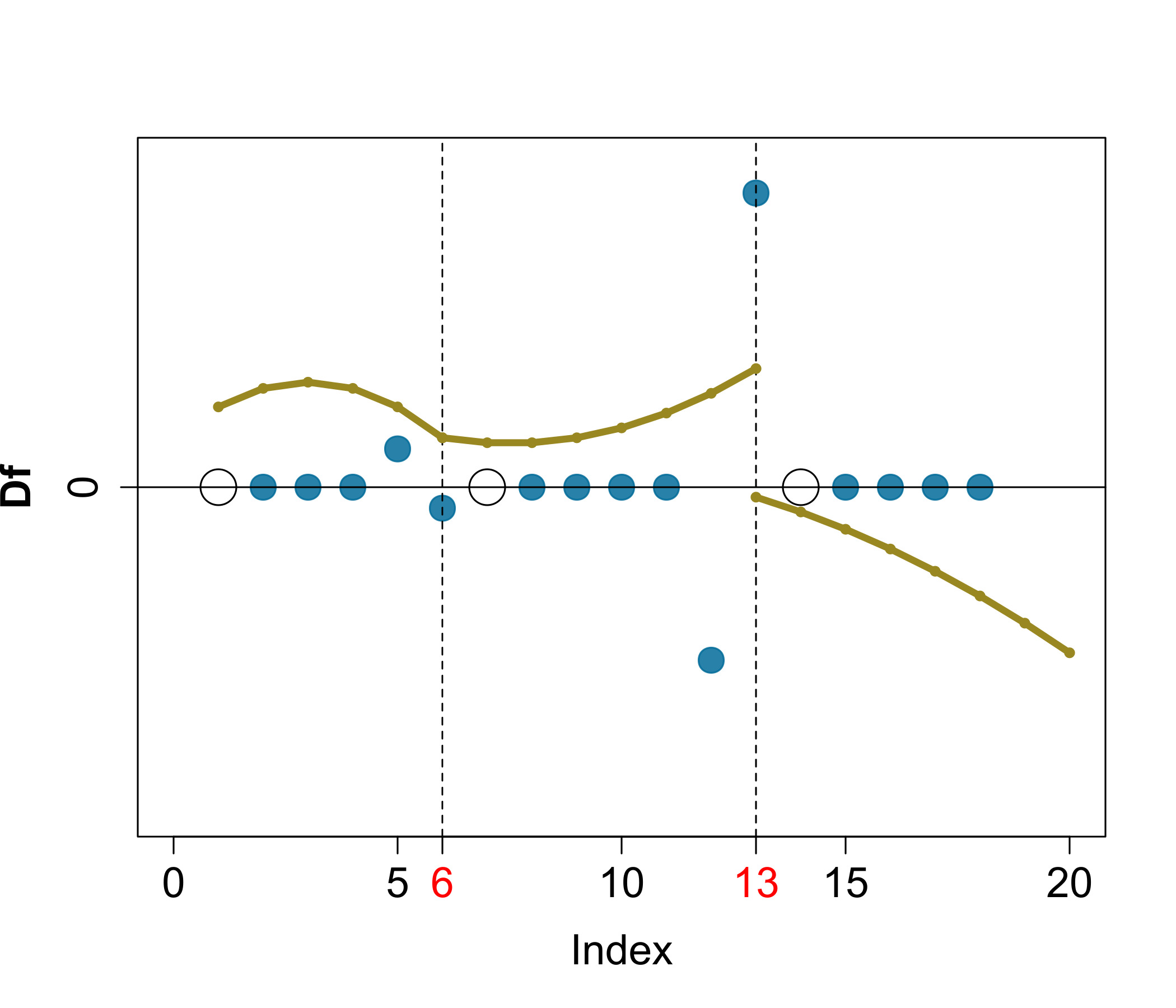}
  \caption{Piecewise quadratic, $r=2$.}
\end{subfigure}
\begin{subfigure}{.43\textwidth}
  \centering
  \includegraphics[width=1\linewidth]{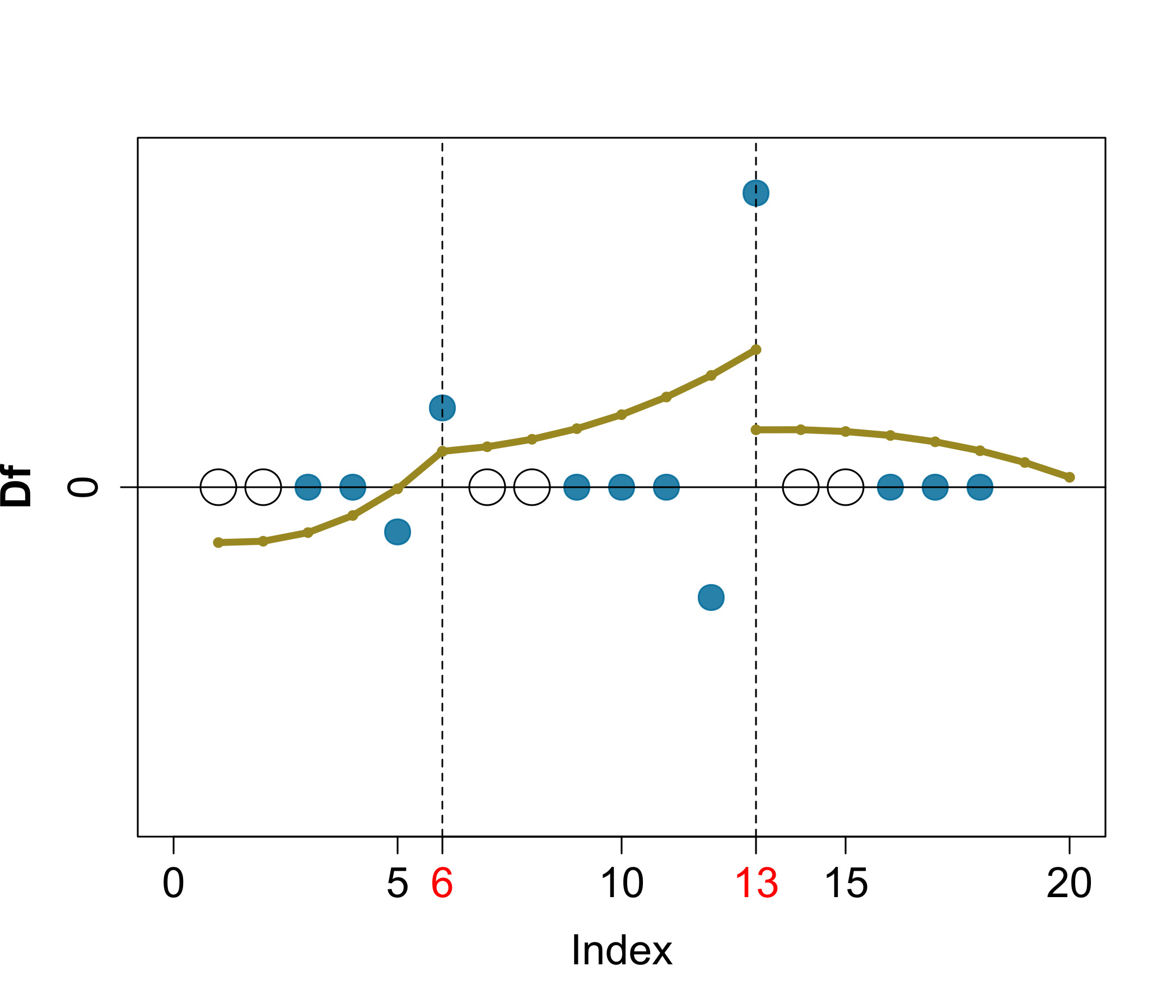}
  \caption{Piecewise cubic, $r=3$.}
\end{subfigure}
\caption[Structure of $\mbf D\mbf f$ For Piecewise Polynomial Signals]{Structure of $\mbf D\mbf f$ for piecewise polynomial signals with various orders $r=0,\,1,\,2,\,3$. The olive lines display the true signals with two change points at the locations $6$ and $13$. Empty circles represent the indices that $\mbf D\mbf f$ does not exist.}
\label{fig:coor-removal}
\end{center}
\end{figure}

In the following, we discuss the need for the augmented boundary set $\mca A$. We begin by studying the structure of the dual vector $\mbf u=\mbf D\mbf f$ in a piecewise polynomial signal of order $r$, where the signal is partitioned into a number of blocks defined by the position of the change points. Because the signal $f$ is a piecewise polynomial of order $r$, to compute the $i$-th coordinate of the vector $\mbf{u}$, we need $r_{_b}=\lceil (r+1)/2\rceil-1$ points directly before the $i$-th element of $\mbf{f}$ as well as $r_{_a}=\lfloor (r+1)/2\rfloor$ points immediately after that. Consequently, the first $r_{_a}$ elements of $\mbf D\mbf f$ within each block cannot be computed. Moreover, within each block, the last $r_{_b}+1$ elements of $\mbf D\mbf f$ are all nonzero due to the existence of a change point. This observation is depicted in Figure \ref{fig:coor-removal} for $r=0,\, 1,\, 2,\, 3$. To explain this point clearly, consider the case of $r=2$ in Figure \ref{fig:coor-removal} in which the structure of $\mbf {Df}$ is shown, where the true change points are at $6$ and $13$. As can be seen, the points on the boundary -- the nonzero coordinates of $\mbf{Df}$ -- are $\mca B(\lambda)= \{ 5,\,6,\,12,\,13\}$ with their respective signs $\mbf s_{\mca B}(\lambda)=\{ 1,\,1,\,-1,\,-1 \}$. Notice that $\mbf{Df}$ does not exist at points 7 and 14. The augmented boundary set contains these points as well as the boundary points; that is $\mca A(\lambda)= \{5,\,6,\,7,\,12,\,13,\,14\}$. The respective signs of the coordinates in the augmented boundary set $\mca A(\lambda)$ are given by $\mbf s_{\mca A}(\lambda)=\{1,\,1,\,1,\,-1,\,-1,\,-1\}$. At each value of $\lambda$, we call the coordinates that belong to the augmented boundary set $\mca A(\lambda)$ the augmented boundary coordinates, and the rest, the interior coordinates.

At the $j$-th iteration with $\lambda=\lambda_{_j}$, we assume that the boundary set and its corresponding sign vector are $\mca B=\mca B(\lambda)$ and $\mbf s_{\mca B}=\mbf s_{\mca B}(\lambda)$, respectively. Furthermore, we assume the augmented boundary set and its sign vector are $\mca A=\mca A(\lambda)$ and $\mbf s_{\mca A}=\mbf s_{\mca A}(\lambda)$, respectively. Dual coordinates can be split into augmented boundary coordinates $\widehat{\mathbf{u}}_{_{\lambda_j,\,\mathcal{A}}}$ and interior coordinates $\widehat{\mathbf{u}}_{_{\lambda_j,\,-\mathcal{A}}}$. Recall from Section \ref{sec:notations.proj1} that $\widehat{\mathbf{u}}_{_{\lambda_j,\,\mathcal{A}}}$ represents the subvector of $\widehat{\mbf u}_{_\lambda}$ with the coordinate labels in the set $\mca A$ and $\widehat{\mathbf{u}}_{_{\lambda_j,\,-\mathcal{A}}}$ represents the subvector of $\widehat{\mbf u}_{_\lambda}$ with the coordinate labels in the set $\lbrace 1,\,2,\, \cdots, \,m\rbrace \backslash \mca A$. It is apparent from the definition of the boundary coordinates that
\begin{align}\label{u.boundary}
    \widehat{\mathbf{u}}_{_{\lambda_j,\,\mathcal{A}}}=\lambda_{_j}\, \mathbf{s}_{\mathcal{A}}\,.
\end{align}
Replacing the boundary coordinate with $\lambda_{_j}\,\mbf s_{\mca A}$ in \eqref{tf.dual.obj.proj1} and solving the resulting quadratic problem with respect to the interior coordinates, lead to their least square estimates, given by
\begin{align}\label{u.interior}
    \widehat{\mathbf{u}}_{_{\lambda_j,\mathcal{-A}}}&=\left(\mathbf{D}_{_{-\mathcal{A}}}\mathbf{D}_{_{-\mathcal{A}}}^T\right)^{-1}\mathbf{D}_{_{-\mathcal{A}}}\left(\mathbf{y}-\lambda_{_j} \mathbf{D}_{_{\mathcal{A}}}^T~\mathbf{s}_{\mathcal{A}}\right).
\end{align}

It should be noted that for the purpose of simplicity, we denote $(\mbf D_{\mca A})^T$ and $(\mbf D_{_{-\mca A}})^T$ with $\mbf D_{\mca A}^T$ and  $\mbf D_{_{-\mca A}}^T$, respectively. Notice that in \eqref{u.interior}, the first term $\big( \mathbf{D}_{_{-\mathcal{A}}} \mathbf{D}_{_{-\mathcal{A}}}^T \big)^{-1} \mathbf{D}_{_{-\mathcal{A}}}\, \mathbf{y}$ simply yields the least square estimate of regressing the response vector $\mbf y$ on the design matrix $\mathbf{D}_{_{-\mathcal{A}}}$. The second term $-\lambda_{_j}\, \big( \mathbf{D}_{_{-\mathcal{A}}} \mathbf{D}_{_{-\mathcal{A}}}^T \big)^{-1} \mathbf{D}_{_{-\mathcal{A}}}\, \mathbf{D}_{_{\mathcal{A}}}^T~ \mathbf{s}_{\mathcal{A}}$ can be interpreted as a shrinkage term due to the condition $\|\mbf u\|_{_\infty}\leq \lambda$.
The expression \eqref{u.interior} is true for $\lambda\leq \lambda_{_j}$ until either an interior coordinate joins to the boundary or a coordinate in the boundary set leaves the boundary. The following argument explains how to specify values of $\lambda$ while the interior coordinates change.

We define the joining time associated with the interior coordinate $i\in \lbrace 1,\,2,\, \cdots, \,m\rbrace \backslash \mca A$ as the time at which this interior coordinate joins the boundary. To determine the next joining time, we reduce the value of $\lambda$ in a linear direction starting from $\lambda_{_j}$ and solve $\widehat{\mathbf{u}}_{_{\lambda,\mathcal{-A}}}=(\pm\lambda,\,\cdots,\,\pm\lambda)^T$. Note that the right-hand side of \eqref{u.interior} can be expressed as $\bsy a-\lambda_{_j}\,\mbf b$, where 
\begin{align}\label{ab}
    \bsy a&=\big(\mathbf{D}_{_{-\mca A}}\mathbf{D}_{_{-\mca A}}^T\big)^{-1}\mathbf{D}_{_{-\mca A}}\,\mathbf{y}\,,\\[8pt]
    \mathbf{b}&=\big(\mathbf{D}_{_{-\mca A}}\mathbf{D}_{_{-\mca A}}^T\big)^{-1}\mathbf{D}_{_{-\mca A}}\mathbf{D}_{_{\mca A}}^T\,\mathbf{s}_{_{\mca A}}\,.
\end{align}
The joining time for every $i\in \lbrace 1,\,2,\, \cdots, \,m\rbrace \backslash \mca A\,$  is hence the solution of the equation $a_{_i}-\lambda \,b_{_i}=\pm\lambda$ with respect to $\lambda$, which is given by
\begin{align*}
    \lambda_{_i}^{^\textrm{join}}=\frac{a_{_i}}{b_{_i}\pm 1}\,, \qquad\qquad i\in\lbrace 1,\,2,\, \cdots, \,m\rbrace \backslash \mca A\,.
\end{align*}
Note that $\lambda_{_i}^{^\textrm{join}}$ is uniquely defined because only one of the signs $-1$ or $+1$ yields $\lambda_{_i}\in [0,\, \lambda_{_j}]$.

Now we turn the attention to the characterization of a coordinate which leaves the boundary set $\mca B$. For $i\in \mca B$, the leaving time is defined as the time that the coordinate $i$ leaves the boundary set $\mca B$. Since $\mbf s_{\mca B}$ is the sign vector of changes captured by $\big[ \mbf D\, \widehat{\mbf f} \big]_{\mca B}$, then $\diag{\mbf s_{\mca B}}\, \big[ \mbf D\, \widehat{\mbf f} \big]_{\mca B}> \mbf 0$, which in turn, along with Equation \eqref{dual.to.primal}, implies $\diag{\mbf s_{\mca B}} \big[ \mbf D\, \big(\mbf y-\mbf D^T \widehat{\mbf u}_{_\lambda}\big)\, \big]_{\mca B}> \mbf 0$. Here, for any vector $\bsy{\eta}$, $\diag{\bsy{\eta}}$ denotes the diagonal matrix with the diagonal elements given by $\bsy{\eta}$, and $\bsy{\eta} > \mbf{0}$ holds element-wise. Therefore, a coordinate $i\in\mca B$ leaves the boundary set $\mca B$ if $\diag{\mbf s_{\mca B}} \big[ \mbf D\, \big(\mbf y-\mbf D^T \widehat{\mbf u}_{_\lambda}\big)\, \big]_{\mca B}> \mbf 0$ is violated. Using the relation 
\begin{align*}
    \Big[ \mbf D\, \big(\mbf y-\mbf D^T \widehat{\mbf u}_{_\lambda}\big)\, \big]_{\mca B}=\mbf D_{\mca B}\, \big( \mbf y-\mbf D^T \widehat{\mbf u}_{_\lambda} \big)\,,
\end{align*}
and the decomposition $\mbf D^T \widehat{\mbf u}_{_\lambda}=\mbf D_{\mca A}^T\, \widehat{\mbf u}_{_{\lambda,\,\mca A}}+\mbf D_{_{-\mca A}}^T\, \widehat{\mbf u}_{_{\lambda,-\mca A}}$, we obtain
\begin{align}
    \diag{\mbf s_{\mca B}} \Big[ \mbf D\, \big(\mbf y-\mbf D^T \widehat{\mbf u}_{_\lambda}\big) \Big]_{\mca B}=\mbf c-\lambda\,\mbf d\,,
\end{align}
where
\begin{align}\label{cd}
    \mathbf{c}&=\mathrm{diag}(\mathbf{s}_{\mathcal{B}})\,\mathbf{D}_{\mathcal{B}}\big(\mathbf{y}-\mathbf{D}_{_{-\mathcal{A}}}^T\,\bsy{a}\big)\,,\\[8pt]
    \mathbf{d}&=\mathrm{diag}(\mathbf{s}_{\mathcal{B}})\,\mathbf{D}_{\mathcal{B}}\big(\mathbf{D}_{\mathcal{A}}^T\,\mathbf{s}_{\mathcal{A}}-\mathbf{D}_{_{-\mathcal{A}}}^T\,\mathbf{b}\big).
\end{align}
Hence, a leaving time is obtained from the equation $c_{_i}-\lambda\, d_{_i} > 0$ as
\begin{align*}
    \lambda_{_i}^{^\textrm{leave}}=\left\{\begin{array}{lll}
    \dfrac{c_{_i}}{d_{_i}},     && \textrm{if}~ c_{_i}<0~ \textrm{~and~} ~ d_{_i}<0\,, \\[8pt]
    0,     && \textrm{otherwise}\,.
    \end{array}\right.
\end{align*}
The conditions in the aforementioned equation is due to the fact that at the $j$-th iteration with $\lambda\leq \lambda_{_j}$, the expression $c_{_i}-\lambda\, d_{_i} > 0$ fails for $i\in \mca B$, if both $c_{_i}$ and $d_{_i}$ are negative.
An alternative way to determine the next leaving time is to use the KKT optimality conditions of \eqref{tf.dual.obj.proj1}. We refer the reader to the supplementary materials of \cite{tibshirani2011solution}.

The following algorithm, PRUTF, describes the process of constructing the entire dual solution path of trend filtering.

\begin{algorithm}[PRUTF Algorithm] \label{tf.path.alg}
\begin{enumerate}
     \item[]
     \item Initialize the set of change points locations as $\bsy\tau_{_0}=\emptyset$, the empty set.
    \item At step $j=1$, initialize the boundary set $\mathcal{B}_{_1}=\big\{\tau_{_1}-r_{_b},\,\tau_{_1}-r_{_b}+1,\,
    \ldots,\tau_{_1}\big\}$ and its associated sign vector $s_{_{\mathcal{B}_1}}=\{s_{_1},\ldots,s_{_1}\}$, both with cardinality of $r_{_b}+1$, where $\tau_{_1}$ is obtained by
    \begin{align}\label{firstjoin}
        \tau_{_1}=\underset{i=1,\,\ldots,\,m}{\rm argmax} \mid \widehat{u}_{_i}\mid \,,
    \end{align}
    and $s_{_1}=\sign{ \widehat{u}_{_{ \tau_{_1}}}}$, where $\widehat{u}_{_i}$ is the $i$-th element of the vector $\widehat{\mathbf{u}}=\left(\mathbf{DD}^T\right)^{-1}\mathbf{D}\,\mathbf{y}$. The updated set of change points locations is now $\bsy\tau_{_1}=\{\tau_{_1}\}$. We also record the first joining time $\lambda_{_1}= \mid\widehat{u}_{\tau_{_1}}\mid$ and keep track of the augmented boundary set $\mathcal{A}_{_1}=\{\tau_{_1}-r_{_b},\ldots,\tau_{_1}+r_{_a}\}$ and its corresponding sign vector $\mathbf{s}_{_{\mathcal{A}_{_1}}}=\{s_{_1},\,\ldots,\,s_{_1}\}$ of length $r+1$. The dual solution is regarded as $\widehat{\mbf u}(\lambda)=\left(\mathbf{DD}^T\right)^{-1}\mathbf{D}\,\mathbf{y}$, for $\lambda\geq \lambda_{_1}$.
    
    \item For step $j=2,\,3,\,\ldots\,,$
    \begin{enumerate}
    \item Obtain the pair $\big( \tau_{_j}^{^\mathrm{join}},s_{_j}^{^\mathrm{join}} \big)$ from
    \begin{align}\label{joinpair}
        \big(\tau_{_j}^{^\mathrm{join}},s_{_j}^{^\mathrm{join}}\big)=~ \underset{i\notin \mathcal{A}_{_{j-1}},\,s\in\{-1,\,1\}}{\rm argmax}~~ \frac{a_{_i}}{s+b_{_i}}\cdot \mathbbm{1} \left\{0\leq \dfrac{a_{_i}}{s+b_{_i}} \leq \lambda_{_{j-1}}\right\},
    \end{align}
    and set the next joining time $\lambda_{_j}^{^\mathrm{join}}$ as the value of $\frac{a_{_i}}{s+b_{_i}}$, for $i=\tau_{_j}^{^\mathrm{join}}$ and $s= s_{_j}^{^\mathrm{join}}$.
    
    \item Obtain the pair $\big( \tau_{_j}^{^\mathrm{leave}},s_{_j}^{^\mathrm{leave}}\big)$ from 
        \begin{align}\label{leavepair}
        \big(\tau_{_j}^{^\mathrm{leave}},\,s_{_j}^{^\mathrm{leave}}\big)=~ \underset{i\in \mathcal{B}_{_{j-1}},\,s\in\{-1,\,1\}}{\rm argmax}~~ \dfrac{c_{_i}}{d_{_i}}\cdot\, \mathbbm{1} \Big\{c_{_i} < 0~,~ d_{_i}< 0\Big\},
    \end{align}
    and assign  the next leaving time $\lambda_{_j}^{^\mathrm{leave}}$ as the value of $\dfrac{c_{_i}}{d_{_i}}$, for $i=\tau_{_j}^{^\mathrm{leave}}$ and $s=s_{_j}^{^\mathrm{leave}}$.
    
    \item Let $\lambda_{_j}=\max \big\{\lambda_{_j}^{^\mathrm{join}}\, ,\, \lambda_{_j}^{^\mathrm{leave}}\big\}$,
    then the boundary set $\mathcal{B}_{_j}$ and its sign vector $\mathbf{s}_{_{\mathcal{B}_{j}}}$ are updated in the following fashion:
    \begin{itemize}
    \item[-- ] Either append $\big\{\tau_{_j}^{^\mathrm{join}}-r_{_b},\,\tau_{_j}^{^\mathrm{join}}-r_{_b}+1,\,\ldots,\tau_{_j}^{^\mathrm{join}}\big\}$ and the corresponding signs $\big\{s_{_j}^{^\mathrm{join}},\,\ldots,\,s_{_j}^{^\mathrm{join}}\big\}$ to $\mathcal{B}_{_{j-1}}$ and $\mathbf{s}_{_{\mathcal{B}_{j-1}}}$, respectively, provided that $\lambda_{_j}=\lambda_{_j}^{^\mathrm{join}}$. Also, add $\tau_{_j}^{^\mathrm{join}}$ to $\bsy\tau_{_{j-1}}$. 
    \item[-- ] Or remove $\big\{\tau_{_j}^{^\mathrm{leave}},\, \tau_{_j}^{^\mathrm{leave}}+1,\,
    \ldots,\,\tau_{_j}^{^\mathrm{leave}}+r_{_b}\big\}$ and the corresponding signs $\big\{s_{_j}^{^\mathrm{leave}}, \,\ldots$, $ \,s_{_j}^{^\mathrm{leave}}\big\}$ from $\mathcal{B}_{_{j-1}}$ and $\mathbf{s}_{_{\mathcal{B}_{j-1}}}$, respectively, provided that $\lambda_{_j}=\lambda_{_j}^{^\mathrm{leave}}$. Also, remove $\tau_{_j}^{^\mathrm{leave}}$ from $\bsy\tau_{_{j-1}}$.
    \end{itemize}
    In the same manner, the augmented boundary set, $\mathcal{A}_{_j}$ and its sign, $\mathbf{s}_{_{\mathcal{A}_{_j}}}$ are formed by adding $\big\{\tau_{_j}^{^\mathrm{join}}-r_{_b},\,\ldots,\, \tau_{_j}^{^\mathrm{join}}+r_{_a}\big\}$  and $\big\{s_{_j}^{^\mathrm{join}},\,\ldots,\,s_{_j}^{^\mathrm{join}}\big\}$ to $\mathcal{A}_{_{j-1}}$ and $\mathbf{s}_{_{\mathcal{A}_{j-1}}}$, respectively, if $\lambda_{_j}=\lambda_{_j}^{^\mathrm{leave}}$ or, otherwise, by removing the associated set $\big\{\tau_{_j}^{^\mathrm{leave}},\,\ldots,\, \tau_{_j}^{^\mathrm{leave}}+r\big\}$  and $\big\{s_{_j}^{^\mathrm{leave}},\,\ldots,\,s_{_j}^{^\mathrm{leave}}\big\}$ from $\mathcal{A}_{_{j-1}}$ and $\mathbf{s}_{_{\mathcal{A}_{_{j-1}}}}$. Thus, the dual solution is computed as $\widehat{\mbf u}_{_{\mca A_{_j}}}(\lambda)=\bsy a-\lambda\, \mbf b$ for interior coordinates and $\widehat{\mbf u}_{_{-\mca A_{_j}}}(\lambda)=\lambda\,\mbf s_{_{\mca A_{_j}}}$ for boundary coordinates over $\lambda_{_j}\leq\lambda\leq\lambda_{_{j-1}}$. 
    \end{enumerate}
    \item Repeat step 3 until $\lambda_{_j}> 0$.
    \end{enumerate}
\end{algorithm}

The critical points $\lambda_{_1}\, \geq\, \lambda_{_2}\, \geq\,  \ldots\, \geq\, 0$ indicate the values of the regularization parameter at which the boundary set changes.

\begin{remark}
Notice that the vector $\bsy\tau$ derived by the PRUTF algorithm represents the locations of change points for the dual variables. In order to obtain the locations of change points in primal variables, we must add $r_{_a}$ to any element of $\bsy\tau$, that is, $\big\{ \tau_{_1}+r_{_a}, \, \tau_{_2}+r_{_a},\, \ldots\, \big\}$. This relationship between the primal and dual change point sets is visible from Figure \ref{fig:coor-removal}.
\end{remark}

\begin{remark}
For fused lasso, $r=0$, Lemma 1 of \cite{tibshirani2011solution}, known as the boundary lemma, is satisfied since the matrix $\mbf D\mbf D^T$ is diagonally dominant, meaning that $\big[\,\mbf D \mbf D^T\,\big]_{i,i} \geq \sum_{j\neq i}\big| \big[\,\mbf D \mbf D^T\, \big]_{i,j} \big|$, for $i=1,\ldots,m$. This lemma states that when a coordinate joins the boundary, it will stay on the boundary for the rest of the path. Consequently, part (b) of step 3 in Algorithm \ref{tf.path.alg} is unnecessary, and hence the next leaving time in part (c) is set to zero, i.e., $\lambda_{_j}^{^\mathrm{leave}}=0$, for every step $j$. However, the boundary lemma is not satisfied for $r=1,\,2,\,3,\,\ldots$. 
\end{remark}

\begin{remark}\label{rem:tibshirani_difference:proj1}
There is a subtle and important distinction between our proposed algorithm, PRUTF, and the one presented in \cite{tibshirani2011solution}. 
The latter work studies the generalized lasso problem for any arbitrary penalty matrix $\mbf D$ (unlike $\mbf D$ used in trend filtering, which must have a certain structure). The proposed algorithm in \cite{tibshirani2011solution} relies on adding or removing only one coordinate to or from the boundary set at every step. The key attribute of our algorithm is to add or remove $r+1$ coordinates to or from the augmented boundary set, an approach inspired by the argument presented at the beginning of this section. Essentially, this attribute makes PRUTF, presented in Algorithm \ref{tf.path.alg}, well-suited for change point analysis. It is important to mention that PRUTF requires at least $r+1$ data points between neighbouring change points.
\end{remark}

\begin{remark}
For a given $\lambda$, equations \eqref{u.boundary} and \eqref{u.interior} give the values of the dual variables in $\widehat{\mbf u}_{_\lambda}$. The equations demonstrate that the dual solution path is a linear function of $\lambda$ with change in the slopes at joining or leaving times $\lambda_1\geq\lambda_2\geq\ldots\geq 0$. 
\end{remark}

\begin{remark}
The number of iterations required for PRUTF, presented in Algorithm \ref{tf.path.alg}, is at most $(3^{\,p}+1)/2$, where $p=\lceil \frac{m}{r+1}\rceil$, see \cite{tibshirani2013lasso}, Lemma 6. However, this upper bound for the number of iterations is usually very loose. The upper bound comes from the following realization discovered by \cite{osborne2000lasso} and later improved by \cite{mairal2012complexity}. Any pair $\big(\mca A\,,\,\mbf s_{\mca A}\big)$ appears at most once throughout the solution path. In other words, if $\big(\mca A\,,\,\mbf s_{\mca A}\big)$ is visited in one iteration of the algorithm, the pair $\big(\mca A\,,\,-\mbf s_{\mca A}\big)$ as well as $\big(\mca A\,,\,\mbf s_{\mca A}\big)$ cannot reappear again for the rest of the algorithm. Interestingly, this fact says that once a coordinate enters the boundary set, it cannot immediately leave the boundary set at the next step.

\noindent Moreover, note that at one iteration of the PRUTF algorithm with the augmented boundary set $\mca A$, the dominant computation is in solving the least square problem
\begin{align}\label{lsproblem}
    \min\limits_{\mathbf{u}\, \in\, \mbb R^{m}}~ \frac{1}{2}\, \big\| \,\mathbf{y}-\mathbf{D}_{_{\mca A}}^T\, \mathbf{u}\, \big\|_{2}^2 \,.
\end{align}
One can apply QR decomposition
of $\mathbf{D}_{_{\mca A}}^T$ to solve the least square problem, and then update the decomposition as set $\mca A$ changes. However, since $\mathbf{D}_{_{-\mathcal{A}}} \mathbf{D}_{_{-\mathcal{A}}}^T$ is a banded Toeplitz matrix (see Section \ref{sec:property.solution.path.proj1}), a solution of \eqref{lsproblem} always exists and can be computed using a banded Cholesky decomposition. Thus, the computational complexity for the iteration is of order $O \big((m-|\mca A|)\, r^2 \big)$, which is linear in the number of interior coordinates as $r$ is fixed and usually small. Overall, if $K$ is the total number of steps run by the PRUTF algorithm, then the total computational complexity is $O \big(K(m-|\mca A|)\, r^2 \big)$. See \cite{tibshirani2011solution} and \cite{arnold2016efficient}.
\end{remark}

\section{Statistical Properties of the Solution Path}
\label{sec:property.solution.path.proj1}

An important component of the methodology that we develop in this work involves computing algebraic expressions based on the matrix $\mbf D=\mbf D^{(r+1)}$. In this section, we describe the properties of such expressions. To begin with, let $\mathcal{A}=\{A_{_1},\,\ldots,\,A_{_J}\}$ and $\mathbf{s}_{\mathcal{A}}=\{\mbf s_{_1},\,\ldots,\,\mbf s_{_J}\}$ be the augmented boundary set and its corresponding sign vector, respectively, after a number of iterations of Algorithm \ref{tf.path.alg}, where $A_{_j}=\big\{\tau_{_j}-r_{_b},\,\tau_{_j}-r_{_b}+1,\,\ldots,\,\tau_{_j}+r_{_a}\big\}$ and $\mathbf{s}_{_j}=\{s_{_j},\,\ldots,\,s_{_j}\}$ for $j=1,\,\ldots,\,J$. This augmented boundary set corresponds to $J$ change points $\{\tau_{_1},\,\ldots,\,\tau_{_J}\}$ that partition all the dual variables into $J+1$ blocks $B_{_j}=\big\{\tau_{_j}+1,\,\ldots,\tau_{_{j+1}}\big\}$ for $j=0,\,1,\,\ldots,\,J$, with the conventions that $\tau_{_0}=0$ and $\tau_{_{J+1}}=m$. In the following, we list some properties of matrix multiplications involving $\mbf D$. 
 
\begin{itemize}
    \item It follows from the definition of the matrix $\mbf D$ that it is a banded Toeplitz matrix with bandwidth $r+1$. It tuns out that the matrix $\mbf D\mbf D^T$ reveals the same property, meaning that it is a square banded Toeplitz matrix. Moreover, its $r+1$ nonzero row elements are consecutive binomial coefficients of order $2\,r+2$ with alternating signs. In other words, $(i\, ,\, j)$-th element of $\mbf D\mbf D^T$ for $i\geq j$ is $(-1)^{\, i-j}{2\,r+2 \choose r+1+i-j}$. An example, for $r=1$, is given in panel (a) of Figure \ref{fig.Dstruct}. Note that $\mbf D\mbf D^T$ is a symmetric, nonsingular and positive definite matrix \citep{demetriou2001certain}.
    
    \item The matrix $\mathbf{D}_{_{-\mathcal{A}}}\mathbf{D}_{_{-\mathcal{A}}}^T$ is a block diagonal matrix whose diagonal submatrices correspond to $J+1$ blocks. More precisely, the $j$-th submatrix on the diagonal of  $\mathbf{D}_{_{-\mathcal{A}}}\mathbf{D}_{_{-\mathcal{A}}}^T$ is a matrix with the first $(\tau_{_{j+1}}-\tau_{_j}-r)$ rows and columns of $\mbf D\mbf D^T$, see panel (b) of Figure \ref{fig.Dstruct}. Notice that, due to its non-singularity, $\mathbf{D}_{_{-\mathcal{A}}}\mathbf{D}_{_{-\mathcal{A}}}^T$ is always invertible. In fact, both $\big(\mathbf{D}_{_{-\mathcal{A}}}\mathbf{D}_{_{-\mathcal{A}}}^T\big)^{-1}$ and $\big(\mathbf{D}_{_{-\mathcal{A}}}\mathbf{D}_{_{-\mathcal{A}}}^T\big)^{-1}\mathbf{D}_{_{-\mathcal{A}}}$ are block diagonal matrices. Another interesting result is that every row of the matrix $\big(\mathbf{D}_{_{-\mathcal{A}}}\, \mathbf{D}_{_{-\mathcal{A}}}^T\big)^{-1} \mathbf{D}_{_{-\mathcal{A}}}$ is a contrast vector, meaning that for any $t=1,\,\ldots,\,m$,
    \begin{align*}
        \ssum{1}{n}\left[\big(\mathbf{D}_{_{-\mathcal{A}}}\mathbf{D}_{_{-\mathcal{A}}}^T\big)^{-1}\mathbf{D}_{_{-\mathcal{A}}}\right]_{t,\,i}=0\,.
    \end{align*}

    \begin{figure}[!t]
        \centering
        \begin{subfigure}[b]{0.3\textwidth}
        \centering
        {\small    
        \begin{align*}
            \left(\begin{array}{rrrrrrrrrrr}
            6 & -4 & 1 & 0 & 0 & 0 & 0 & 0 & 0 \\ 
            -4 & 6 & -4 & 1 & 0 & 0 & 0 & 0 & 0 \\ 
            1 & -4 & 6 & -4 & 1 & 0 & 0 & 0 & 0 \\ 
            0 & 1 & -4 & 6 & -4 & 1 & 0 & 0 & 0 \\ 
            0 & 0 & 1 & -4 & 6 & -4 & 1 & 0 & 0 \\ 
            0 & 0 & 0 & 1 & -4 & 6 & -4 & 1 & 0 \\ 
            0 & 0 & 0 & 0 & 1 & -4 & 6 & -4 & 1 \\ 
            0 & 0 & 0 & 0 & 0 & 1 & -4 & 6 & -4 \\ 
            0 & 0 & 0 & 0 & 0 & 0 & 1 & -4 & 6 \\ 
            \end{array}\right)
        \end{align*}}
            \caption{Structure of $\mbf D \mbf D^T$.}
            \label{fig.Dstruct.a}
        \end{subfigure}
        \quad
        \begin{subfigure}[b]{0.3\textwidth}
        \centering
        {\small 
    \begin{align*}
    \left(\begin{array}{ccc|cccc}
      6 & -4 & 1 & 0 & 0 & 0 & 0 \\ 
      -4 & 6 & -4 & 0 & 0 & 0 & 0 \\ 
      1 & -4 & 6 & 0 & 0 & 0 & 0 \\\hline 
      0 & 0 & 0 & 6 & -4 & 1 & 0 \\ 
      0 & 0 & 0 & -4 & 6 & -4 & 1 \\ 
      0 & 0 & 0 & 1 & -4 & 6 & -4 \\ 
      0 & 0 & 0 & 0 & 1 & -4 & 6 \\ 
     \end{array}\right)
    \end{align*}}
            \caption{The structure of $\mbf D_{_{-\mca A}}\, \mbf D^T_{_{-\mca A}}$.}
            \label{fig.Dstruct.b}
        \end{subfigure}
        \caption[Structure of Quadratic Forms of Matrix $\mbf D$]{Structure of quadratic forms of matrix $\mbf D$.}
        \label{fig.Dstruct}
    \end{figure}

    \item Another interesting term in analyzing the behaviour of the dual variables is $\mbf D_{\mca A}^T\,\mbf s_{\mca A}$. It can be shown that the vector $\mbf D_{\mca A}^T\,\mbf s_{\mca A}$ can be partitioned into $J+1$ subvectors associated with the change points $\tau_{_j},~j=1,\,\ldots,\,J$. The subvector associated with $\tau_{_j},\, \,j=2,\, \ldots, \,J-1$, is $\mbf D_{_{A_j}}^T\,\mbf s_{_{A_j}}$, whose elements are zero, except the first consecutive $r+1$ as well as the last consecutive $r+1$ elements. The  first $r+1$ nonzero elements of $\mbf D_{_{A_j}}^T\,\mbf s_{_{A_j}}$ are the binomial coefficients in the expansion of $s_{_j}\,(x-1)^r$, and its last $r+1$ elements are the binomial coefficients in the expansion of $-s_{_{j+1}}\,(x-1)^r$. Furthermore, the first $r+1$ elements of the first subvector and the last $r+1$ elements of the last subvector are also equal to zero. For example, for a piecewise cubic signal, $r=3$, with two change points $\big(\tau_{_1}\, ,\, \tau_{_2}\big)$ and signs $\big(-1\, ,\, 1\big)$, the vector $\mathbf{D}_{_{\mathcal{A}}}^T\, \mathbf{s}_{_{\mathcal{A}}}$ becomes
    {\small
    \begin{align*}
        \left(\underbrace{0,\ldots,\,0,\,1,\, -3,\, 3,\,-1}_{1\, :\, (\tau_{_1}+r_{_a})}\, ,\,  \underbrace{-1,\, 3,\, -3,\, 1,\, 0,\, \ldots, \,0,\, -1,\, 3,\, -3,\, 1}_{(\tau_{_1}+r_{_a}+1)\, :\, (\tau_{_2}+r_{_a})}\, ,\, \underbrace{1,\, -3,\, 3,\, -1,\, 0,\,\ldots,\,0 }_{(\tau_{_2}+r_{_a}+1)\, :\, m}\right).
    \end{align*}}
    Consequently, the structure of $\mbf D_{_{A_j}}^T\,\mbf s_{_{A_j}}$ allows us to write $\mathbf{D}_{_{\mathcal{A}}}^T\,\mathbf{s}_{_{\mathcal{A}}}=\sum_{j=0}^J \mathbf{D}_{_{{A}_j}}^T\,\mathbf{s}_{_j}$. Additionally, if the signs of two consecutive change points $\tau_{_{j}}$ and $\tau_{_{j+1}}$ are the same, then
    \begin{align}\label{drift.term.staircase.proj1}
        \left[ \big(\mathbf{D}_{_{-\mca A}} \mathbf{D}_{_{-\mca A}}^T \big)^{-1}\mathbf{D}_{_{-\mca A}}\right]_t\,\left(\mathbf{D}_{_{A_{j+1}}}^T \mathbf{s}_{_{j+1}}+\mathbf{D}_{_{A_j}}^T \mathbf{s}_{_j}\right)= -s_j,
    \end{align}
    for $t=\tau_{_j}+r_{_a}\, ,\, \ldots\, ,\, \tau_{_{j+1}}+r_{_b}$.
    
    \item 
    Let $\mbf P_D=\mbf D_{_{-\mca A}}^T \big( \mbf D_{_{-\mca A}} \mbf D_{_{-\mca A}}^T \big)^{-1}\mbf D_{_{-\mca A}}$ be the projection matrix onto the row space of the matrix $\mbf D_{-\mca A}$. Moreover, let $\mbf X_j$ be the design matrix of the $r$-th polynomial regression on the indices of
    $j$-th segment $\big\{ \tau_{j}+1\, ,\, \ldots\, ,\, \tau_{j+1} \big\}$, that is, 
    \begin{align*}
        \mbf X_{j}=\begin{pmatrix}
          1 & \frac{\tau_{_j}+1}{n} & \left(\frac{\tau_{_j}+1}{n}\right)^2 & \cdots & \left(\frac{\tau_{_j}+1}{n}\right)^r 
          \\[10pt]
          1 & \frac{\tau_{_j}+2}{n} & \left(\frac{\tau_{_j}+2}{n}\right)^2 & \cdots & \left(\frac{\tau_{_j}+2}{n}\right)^r \\
          \vdots & \vdots & \vdots &  & \vdots \\
          1 & \frac{\tau_{_{j+1}}}{n} & \left(\frac{\tau_{_{j+1}}}{n}\right)^2 & \cdots & \left(\frac{\tau_{_{j+1}}}{n}\right)^r \\
        \end{pmatrix}.
    \end{align*}
    Therefore, the orthogonal projection matrix $\mbf I - \mbf P_{D}$ is a block diagonal matrix whose $j$-th block associated with the segment $\big\{ \tau_{j}+1\, ,\, \ldots\, ,\, \tau_{j+1} \big\}$ is equal to the projection map onto the column space of $\mbf X_j$, i.e.,
    \begin{align}\label{Dprojection.map.proj1}
        \mbf I - \mbf P_{D} = \mbf X_j \big(\mbf X_j^T\, \mbf X_j \big)^{-1}\, \mbf X_j^T.
    \end{align}
\end{itemize}

Equation \eqref{u.boundary} says that the absolute values of the boundary coordinates are $\lambda$, that is,
\begin{align}
    \widehat{u}(t;\lambda)=\lambda\,s_{_j}\qquad\qquad\textrm{for }t\in A_j.
\end{align}
On the other hand, the values of the interior coordinates are given by
{\small
\begin{align}\label{usegment}
    \widehat{u}(t;\lambda)=\left\{\begin{array}{ll}
    \left[\big(\mathbf{D}_{_{-\mathcal{A}}}\mathbf{D}_{_{-\mathcal{A}}}^T\big)^{-1}\mathbf{D}_{_{-\mathcal{A}}}\right]_t\,\left(\mathbf{y}-\lambda \,\mathbf{D}_{_{{A}_1}}^T \mathbf{s}_{_1}\right),     &  1\leq t<\tau_{_1}-r_{_b}\\\\
    \left[\big(\mathbf{D}_{_{-\mathcal{A}}}\mathbf{D}_{_{-\mathcal{A}}}^T\big)^{-1}\mathbf{D}_{_{-\mathcal{A}}}\right]_t\,\left(\mathbf{y}-\lambda\, \left(\mathbf{D}_{_{A_{j+1}}}^T \mathbf{s}_{_{j+1}}+\mathbf{D}_{_{A_j}}^T \mathbf{s}_{_j}\right)\right),     &   \tau_{_j}+r_{_a}< t<\tau_{_{j+1}}-r_{_b}\\\\
    \left[\big(\mathbf{D}_{_{-\mathcal{A}}}\mathbf{D}_{_{-\mathcal{A}}}^T\big)^{-1}\mathbf{D}_{_{-\mathcal{A}}}\right]_t\,\left(\mathbf{y}-\lambda \,\mathbf{D}_{_{A_{_J}}}^T \mathbf{s}_{_J}\right),     &   \tau_{_J}+r_{_a} < t \leq m\,.
    \end{array}\right.
\end{align}}

For a given $\lambda$, the dual variables $\widehat{u}(t;\,\lambda)$ for $t=0,\,\ldots, \,m$ can be collectively viewed as a random bridge, that is, a conditioned random walk with drift whose end points are set to zero. Moreover, $\widehat{u}(t;\,\lambda)$ is bounded between $-\lambda$ and $\lambda$.
The quantity $\widehat{u}(t;\,\lambda)$ can also be decomposed into a sum of several smaller random bridges which are formed by blocks created from the change points. Recall that the last consecutive $r_{_b}+1$ elements of the block $B_{_j}$ are $\lambda\, s_{_j}$, for any $j=0,\,1,\,\cdots, \,J$.
Hence, for $t=\tau_{_j}+r_{_a},\,\ldots,\,\tau_{_{j+1}}-r_{_b}$, the random bridge associated with the $j$-th block is given by 
\begin{align}\label{z.dual.seg}
    \widehat{u}_{_j}(t;\, \lambda)=\left[\big(\mathbf{D}_{_{-\mathcal{A}}}\mathbf{D}_{_{-\mathcal{A}}}^T\big)^{-1}\mathbf{D}_{_{-\mathcal{A}}}\right]_t\, \left(\mathbf{y}-\lambda \,\big(\mathbf{D}_{_{A_{j+1}}}^T \mathbf{s}_{_{j+1}}+\mathbf{D}_{_{A_j}}^T \mathbf{s}_{_j}\big)\right),\quad j=0,\,\ldots,\,J\,,
\end{align}
with the conventions $\mbf s_{_0}=\mbf s_{_{J+1}}=\mbf 0\in \mathbb{R}^{^{r+1}}$. It is important to note that similar to $\widehat{u}(t;\,\lambda)$, the process $\widehat{u}_{_j}(t;\,\lambda)$ satisfies the conditions $\widehat{u}_{_j}(\tau_{_j}+r_{_a};\,\lambda)=\lambda\,s_{_j}$ and $\widehat{u}_{_j}(\tau_{_{j+1}}-r_{_b};\lambda)=\lambda\,s_{_{j+1}}$.
From \eqref{z.dual.seg}, the process $\widehat{u}_{_j}(t;\,\lambda)$ is composed of the stochastic term
\begin{align}\label{u.stoch}
   \widehat{u}_{_j}^{\,\textrm{st}}(t)=\left[\big(\mathbf{D}_{_{-\mathcal{A}}}\mathbf{D}_{_{-\mathcal{A}}}^T\big)^{-1}\mathbf{D}_{_{-\mathcal{A}}}\right]_t\, \mathbf{y},
\end{align}
and the drift term
\begin{align}\label{u.drift}
    \widehat{u}_{_j}^{\,\textrm{dr}}(t;\,\lambda)=-\lambda\left[\big(\mathbf{D}_{_{-\mathcal{A}}}\mathbf{D}_{_{-\mathcal{A}}}^T\big)^{-1}\mathbf{D}_{_{-\mathcal{A}}}\right]_t\,\left(\mathbf{D}_{_{A_{j+1}}}^T \mathbf{s}_{_{j+1}}+\mathbf{D}_{_{A_j}}^T \mathbf{s}_{_j}\right).
\end{align}

According to model \eqref{fmodel.proj1} with Gaussian noises, it turns out that the discrete time stochastic process term $\widehat{u}_{_j}^{\,\textrm{st}}(t)$ can be embedded in a continuous time Gaussian bridge process. The following theorem describes the characteristics of this process.

\begin{theorem}\label{thm:gaussian.bridge.proj1}
Suppose the observation vector $\mbf y$ is drawn from the model \eqref{fmodel.proj1}, where the error vector $\bsy\varepsilon$ has a Gaussian distribution with mean zero and covariance matrix $\sigma^2\mbf\, \mbf I$. For given $\mbf D$ and $\mca A$,
\begin{enumerate}[label=(\alph*)]
    \item  Define
    \begin{align}\label{wj.process.proj1}
        W_j(t)= \big( \tau_{j+1}-\tau_j-r \big)^{-(2r+1)/2} \left[ \big(\mathbf{D}_{_{-\mathcal{A}}} \mathbf{D}_{_{-\mathcal{A}}}^T \big)^{-1} \mathbf{D}_{_{-\mathcal{A}}} \right]_{\lfloor mt\rfloor}\mathbf{y},
    \end{align}
    for $(\tau_{_j}+r_{_a})/m ~\leq~ t ~\leq~ (\tau_{_{j+1}}-r_{_b})/m$,  where
    \begin{align}\label{wbridge.tails.proj1}
    W_{_{j}} \Big(\, \frac{\tau_{_j}+r_{_a}}{m}\, \Big)= W_{_{j}} \Big(\, \frac{\tau_{_{j+1}}-r_{_b}}{m}\, \Big)=0,
    \end{align}
    and, for $j=0\, ,\, \ldots\, ,\, J$. Then the stochastic process $\mbf W_j=\big\{ W_j(t):~(\tau_{_j}+r_{_a})/m\leq t\leq (\tau_{_{j+1}}-r_{_b})/m\big\}$ is a Gaussian bridge process with mean vector zero and covariance function 
    \begin{align}\label{w.cov}
        {\rm Cov} \Big( W_j(t)\, ,\, W_j(t')\Big)= \sigma^2 \left[\big(\mathbf{D}_{_{-\mathcal{A}}}\mathbf{D}_{_{-\mathcal{A}}}^T\big)^{-1}\right]_{\lfloor mt\rfloor,\lfloor mt'\rfloor},
    \end{align}
     for any $(\tau_{_j}+r_{_a})/m ~\leq~ t\, ,\,    t' ~\leq~ (\tau_{_{j+1}}-r_{_b})/m$.
     
    \item The processes $\mbf W_j$ and $\mbf W_{j'}$ are independent, for $j'\neq j$.
    
\end{enumerate}
\end{theorem}
A proof is given in Appendix \ref{prf:gaussian.bridge.proj1}.

This theorem could be extended to the case of non-Gaussian random variables and therefore establishes a Donsker type Central Limit Theorem for $\mbf W_j$.  Theorem \ref{thm:gaussian.bridge.proj1} guarantees that the dual variable process associated with the $j$-th block, i.e. 
\begin{align*}
    \mbf u_j= \Big\{\, \widehat{u} \big( \lfloor mt\rfloor;\,\lambda \big):~ (\tau_{_j}+r_{_a})/m\leq t\leq (\tau_{_{j+1}}-r_{_b})/m \Big\}
\end{align*}
is a Gaussian bridge process with the drift term 
\begin{align}\label{W.drift.simple}
    -\lambda\left[\big(\mathbf{D}_{_{-\mathcal{A}}}\mathbf{D}_{_{-\mathcal{A}}}^T\big)^{-1}\mathbf{D}_{_{-\mathcal{A}}}\right]_{\lfloor mt\rfloor}\,\left(\mathbf{D}_{_{A_{j+1}}}^T \mathbf{s}_{_{j+1}}+\mathbf{D}_{_{A_j}}^T \mathbf{s}_{_j}\right),
\end{align}
and the covariance matrix stated in \eqref{w.cov}. 

Recall that a standard Brownian bridge process defined on the interval $[a,\,b]$ is a standard Brownian motion $B(t)$ conditioned on the event $B(a)=B(b)=0$. It is often characterized from a Brownian motion $B(t)$ with $B(a)=0$, by setting $$B_{_0}(t)=B(t)-\frac{t-a}{b-a}\,B(b)\,.$$
The mean and covariance functions of the Brownian bridge $B_{_0}(t)$ are given by ${\rm E} \big(B_{_0}(t)\big)=0$ and ${\rm Cov} \big( B_{_0}(s),\,B_{_0}(t) \big)= \min\lbrace s-a,\,t-a\rbrace -(b-a)^{-1}(s-a)(t-a)$ for any $s,\, t\in[a,\,b]$, respectively. A Gaussian bridge process is an extension of the Brownian bridge process when the Brownian motion $B(t)$, in the definition of the Brownian bridge $B_{_0}(t)$, is replaced by a more general Gaussian process $G(t)$. See, for example, \cite{gasbarra2007gaussian}. 
\begin{remark}
The celebrated Donsker theorem \cite{donsker1951invariance} states that the partial sum process of a sequence of i.i.d. random variables, with mean zero and variance 1, converges weakly to a Brownian bridge process. See \cite{van1996weak} or \cite{billingsley2013convergence}. A version of Theorem \ref{thm:gaussian.bridge.proj1} involving non-Gaussian random variables would extend this result to weighted partial sum processes and show that the limiting process is a Gaussian bridge with a certain covariance structure. So the Gaussian assumption in Theorem \ref{thm:gaussian.bridge.proj1} is not restrictive. It is also interesting to show that for $r=0, \,1$, the process $\widehat{u}_{_j}^{\textrm{\,st}} \big( \lfloor mt\rfloor \big)$ boils down to its respective CUSUM processes. To show this, consider the interval $\big[(\tau_{_j}+r_{_a})/m\, , (\tau_{_{j+1}}-r_{_b})/m\big]$,
\begin{itemize}
    \item For the piecewise constant signals, $r=0$, the quantity $\left[\big(\mathbf{D}_{_{-\mathcal{A}}} \mbf  D_{_{-\mathcal{A}}}^T\big)^{-1}\mbf D_{_{-\mathcal{A}}}\right]_{\lfloor mt\rfloor}\mbf y$ can be written as
    {\small
    \begin{align*}
        \bigg(0,\ldots,\underbrace{0}_{ \tau_{_j}},1-\frac{\lfloor mt\rfloor}{\tau_{_{j+1}}-\tau_{_j}}, \ldots, \underbrace{1-\frac{\lfloor mt\rfloor}{\tau_{_{j+1}}-\tau_{_j}}}_{\lfloor mt\rfloor}, -\frac{\lfloor mt\rfloor}{\tau_{_{j+1}}-\tau_{_j}}, \ldots,-\frac{\lfloor mt\rfloor}{\tau_{_{j+1}}-\tau_{_j}},\, \underbrace{0}_{\tau_{_{j+1}}},\ldots,0 \bigg)\, \mbf y.
    \end{align*}}
Notice that the above statement is the CUSUM statistic for the $j$-th segment, that is
\begin{align}\label{CUSUMr0}
    \sum\limits_{k=\tau_{_j}+1}^{\lfloor mt \rfloor}\, \Big( y_{_k}-\overline{y}_{_{(\tau_{_j}+1):\tau_{_{j+1}}}}\Big)\,,
\end{align}
where $\overline{y}_{_{(\tau_{_j}+1):\tau_{_{j+1}}}}$ is the sample average of $\big( y_{_{\tau_{_j}+1}}, \,\ldots,\, y_{_{\tau_{_{j+1}}}} \big)$. It is well known that the CUSUM statistic \eqref{CUSUMr0} converges weakly to the Brownian bridge. In addition, for any $(\tau_{_j}+r_{_a})/m\leq t' \leq t\leq (\tau_{_{j+1}}-r_{_b})/m$, the covariance function becomes
\begin{align*}
    \left[\big(\mathbf{D}_{_{-\mathcal{A}}}\mathbf{D}_{_{-\mathcal{A}}}^T\big)^{-1}\right]_{(\lfloor mt'\rfloor,\lfloor mt\rfloor)}= (\lfloor mt'\rfloor-\tau_{_j})- \frac{(\lfloor mt'\rfloor-\tau_{_j}) (\lfloor mt\rfloor-\tau_{_j})}{\tau_{_{j+1}}-\tau_{_j}},
    \end{align*}
which is identical to the covariance function of the Brownian bridge.

    \item For the piecewise linear signals $r=1$, the quantity $\left[\big(\mathbf{D}_{_{-\mathcal{A}}} \mbf  D_{_{-\mathcal{A}}}^T\big)^{-1}\mbf D_{-\mathcal{A}}\right]_{\lfloor mt\rfloor}\mbf y$ reduces to
    \begin{align}\label{CUSUMr1}
      \sum\limits_{k=\tau_{_j}+1}^{\lfloor mt \rfloor} \, k\, \big( y_{_k}-\widehat{ f}_{_k} \big)\,,
    \end{align}
where $\widehat{f}$ is the least square fit of the simple linear regression of $\big(y_{_{\tau_{_j}+1}}, \,\ldots,\, y_{_{\tau_{_{j+1}}}}\big)$ onto $\big(\tau_{_j}+1, \,\ldots,\, \tau_{_{j+1}}\big)$. As proved in  Theorem \ref{thm:gaussian.bridge.proj1}, the preceding statistic \eqref{CUSUMr1} is also a Gaussian bridge process. Furthermore, using the results in \cite{hoskins1972some}, for any $(\tau_{_j}+r_{_a})/m\leq t' \leq t\leq (\tau_{_{j+1}}-r_{_b})/m$, the covariance function of this Gaussian bridge process is given by 
{\small
\begin{align*}
    \left[\big(\mathbf{D}_{_{-\mathcal{A}}}\mathbf{D}_{_{-\mathcal{A}}}^T \big)^{-1} \right]_{(\lfloor mt'\rfloor,\lfloor mt\rfloor)}&= \frac{\big(\Delta_{_j}- \lfloor mt\rfloor+\tau_{_j}\big)\big(\Delta_{_j}- \lfloor mt\rfloor+\tau_{_j}+1\big)}{3\, \Delta_{_j} \big(\Delta_{_j}+1\big) \big(\Delta_{_j}+2\big)}\\\\
    &\hspace{-3cm}\times \big(\lfloor mt'\rfloor-\tau_{_j}\big) \big(\lfloor mt'\rfloor-\tau_{_j}+1\big)\\\\
    &\hspace{-3cm}\times \Big [ \big(\lfloor mt\rfloor-\tau_{_j}+1\big) \big(\lfloor mt'\rfloor-\tau_{_j}-1\big) \big(\Delta_{_j}+2\big)-\big(\lfloor mt\rfloor-\tau_{_j}\big)\big(\lfloor mt'\rfloor-\tau_{_j}+2\big)\Delta_{_j} \Big ],
\end{align*}}
where $\Delta_{_j}=\tau_{_{j+1}}-\tau_{_j}$.
\end{itemize}
\end{remark}

\section{Stopping Criterion}
\label{sec:stop.rule.proj1}

\begin{figure}[!t]
\begin{subfigure}{.45\textwidth}
  \centering
  \includegraphics[width=1\linewidth]{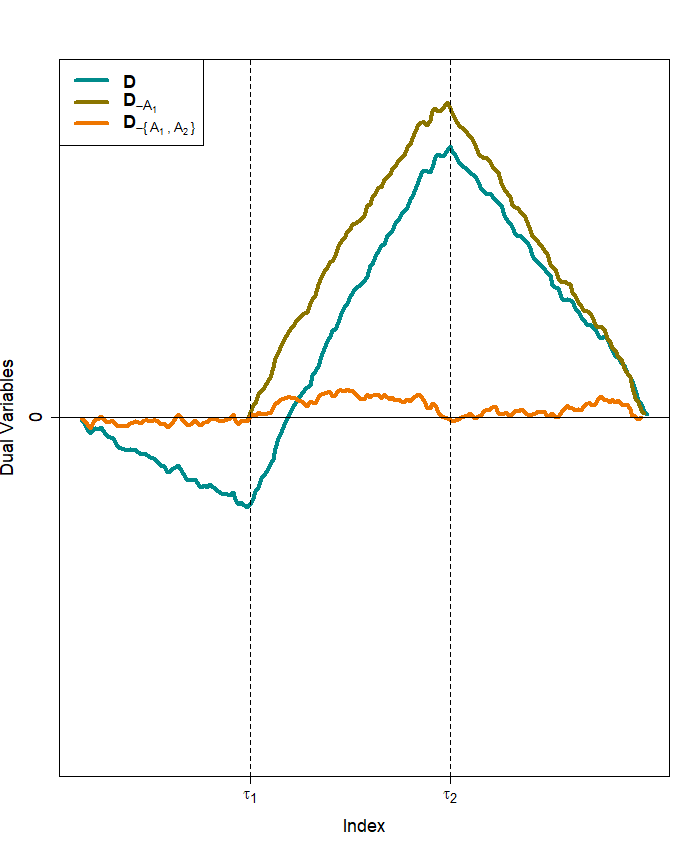}
  \caption{Piecewise constant with $r=0$}
\end{subfigure}
\qquad
\begin{subfigure}{.45\textwidth}
  \centering
  \includegraphics[width=1\linewidth]{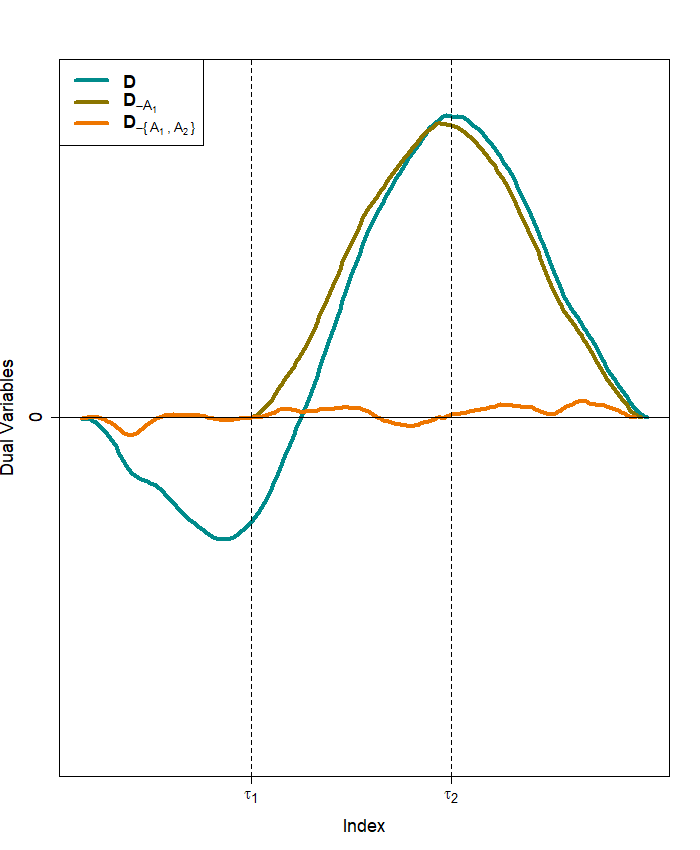}
  \caption{Piecewise linear with $r=1$}
\end{subfigure}
\caption[Structure of $\widehat{u}^{\,\trm{st}}(t)$  With Removed Change Points]{The cyan-colored lines
show the dual variables for the full matrix $\mbf D$. Dual variables computed after removing rows of the matrix $\mbf D$ associated with $\tau_{_1}$, that is $\mbf D_{_{-\mca A_{_1}}}$, are displayed by the olive-colored lines.
The augmented boundary set $\mca A_{_2}$ corresponding to $\tau_{_1}$ and $\tau_{_2}$ results to the dual variables shown by orange-colored lines. 
}
\label{fig:stopping-rule}
\end{figure}

This section concerns developing a stopping criterion for the PRUTF algorithm. We provide tools for deriving a threshold value at which the PRUTF algorithm terminates the search if no values of dual variables exceed this threshold. Consider the dual variables at the first step of the algorithm, i.e. $\widehat{u}^{\,\trm{st}}(t)= \big[\left(\mathbf{D}\mathbf{D}^T\right)^{-1}\mathbf{D}\big]_t\,\mathbf{y}$, for $t=0,\ldots,m$, which correspond to $\widehat{u}^{\,\trm{st}}(t)$ in \eqref{u.stoch} with $\mca A=\emptyset$. It turns out that $\widehat{u}^{\,\trm{st}}(t)$ is a stochastic process with local minima and maxima attained at the  change points. This structure is displayed with cyan-colored lines (\tikz\draw [color=cyan, thick, solid] (0,0) -- (.5,0);) in Figure \ref{fig:stopping-rule} for both piecewise constant $r=0$ and piecewise linear $r=1$ signals. As the PRUTF algorithm detects more change points and forms the augmented boundary set $\mca A$, the local minima or maxima corresponding to these change points are removed  from the stochastic process
\begin{align}\label{Uhatst}
    \widehat{u}_{_{-\mca A}}^{\,\trm{st}}(t)= \left[\big(\mathbf{D}_{_{-\mathcal{A}}}\mathbf{D}_{_{-\mathcal{A}}}^T\big)^{-1}\mathbf{D}_{_{-\mathcal{A}}}\right]_t\mathbf{y}= \ssum[j]{0}{J}  ~\widehat{u}_{_j}^{\,\trm{st}}(t)\, \mathbbm 1 \big\{ t \in B_j \big\},
\end{align}

\noindent for $t=1,\,\ldots,\, m-|\mca A|$. This fact is shown by olive-colored lines (\tikz\draw [color=olive,thick,solid] (0,0) -- (.5,0);) in Figure \ref{fig:stopping-rule}. The last equality in \eqref{Uhatst} expresses that the $\widehat{u}_{_{-\mca A}}^{\,\trm{st}}(t)$ is the stochastic term of the dual variables for all the interior coordinates and is derived by stacking the stochastic terms of the dual variables associated with $j$-th block, $\widehat{u}_{_j}^{\,\trm{st}}(t)$, as defined in \eqref{u.stoch}, for $j=0,\,\ldots,\,J$. This behaviour suggests a way to introduce a stopping rule for the PRUTF algorithm. As can be viewed from the orange-colored lines (\tikz\draw [color=orange,thick,solid] (0,0) -- (.5,0);) of Figure \ref{fig:stopping-rule}, if all true change points are captured by the algorithm and stored in the augmented set $\mca A_0$, the resulting process
\begin{align*}
    \widehat{u}_{_{-\mca A_0}}^{\,\trm{st}}(t)=\left[\big(\mathbf{D}_{_{-\mca A_0}}\mathbf{D}_{_{-\mca A_0}}^T\big)^{-1}\mathbf{D}_{_{-\mca A_0}}\right]_t\,\mathbf{y}
    \qquad \text{ for }\quad t=0,\,\ldots,\,m-|\mca A_0|\,,
\end{align*}
contains no noticeable optimum points and tends to fluctuate close to the zero line (x-axis).  

We terminate the search in Algorithm \ref{tf.path.alg} at step $j$ by checking whether the maximum of $\big|\,\widehat{ u}_{_{-\mca A_j}}^{\,\trm{st}} (t)\,\big|$, for $t=0,\,\ldots,\,m-|\mathcal{A}_{_j}|$, is smaller than a certain threshold. To exactly specify this threshold, as suggested by Theorem \ref{thm:gaussian.bridge.proj1}, we need to calculate the {\it excursion probabilities} of a Gaussian bridge process. As stated in \cite{adler2009random}, analytic formulas for the excursion probabilities are known to be available only for a small number of Gaussian processes. One of such Gaussian processes is the Brownian bridge process. It is well known that for the Brownian bridge process $B_{_0}(t)$ defined on the interval $[a,\,b]$ 
\begin{align}\label{BBmax}
    \Pr\Big(\sup_{a\leq t\leq b} \big| \,B_{_0}(t)\, \big| \geq x \Big)=2\,\sum\limits_{i=1}^{\infty}\, (-1)^{i+1}\exp\left(\frac{-2\,i^2\,x^2}{b-a}\right)\,.
\end{align}
See, for example, \cite{adler2009random}, and \cite{shorack2009empirical}.   
Hence for the piecewise constant signals, the required threshold for stopping Algorithm \ref{tf.path.alg} can be obtained from \eqref{BBmax}, for a  suitably chosen interval $[a,\,b]$. That is, for a given value $\alpha$, we choose $x_{_\alpha}$ such that $\Pr\big( \sup_{a\, \leq\,  t\, \leq\, b} |\,B_{_0}(t)\,| \geq x_{_\alpha} \big)=1-\alpha$.  Therefore, for $r=0$ and $a=0$, $b=1$, we stop Algorithm \ref{tf.path.alg} at the iteration $j_{_0}$ if 
\begin{align*}
    \max_{0\, \leq\,  t\, \leq\, 1}~ \Big|\widehat{\mbf u}_{_{-\mca A_{j_{_0}}}}^{\,\trm{st}}\big( \lfloor\, kt\,\rfloor \big) \Big| ~\leq~ \sigma\, x_{_\alpha}\,\sqrt{k}\,, \qquad\qquad \text{ for   } \quad t=0,\,\ldots,\,m-|\mca A_{_{j_0}}|\,, 
\end{align*}
and $k=m-|\mca A_{j_0}|$.

For $r\geq 1$, the threshold is obtained in a similar fashion. Although the excursion probabilities for the Gaussian bridge processes are not known, we notice that by adopting the steps for the proof of \eqref{BBmax} in \cite{beghin1999maximum}, we can establish a similar formula for the Gaussian bridge process $G_{_0}(t)$ in Theorem \ref{thm:gaussian.bridge.proj1} as  
\begin{align}\label{GBmax}
    \Pr\Big(\sup_{a\, \leq\,  t\, \leq\, b}\, \big|\,G_{_0}(t)\, \big| \geq x \Big)=2\, \sum\limits_{i=1}^{\infty}~(-1)^{i+1}\, \exp\left(\frac{-2\,i^2\,x^2}{S_r^2(k)}\right)\,,
\end{align}
where $k=m-|\mca A_{_{j_{_0}}}|$, and the quantity $S_r^2(k)$ is the $k$-th diagonal element of the matrix $$\Big(\mathbf{D}_{_{-\mathcal{A}_{j_{_0}}}}\mathbf{D}_{_{-\mathcal{A}_{j_{_0}}}}^T\Big)^{-1}\,.$$
Hence, we stop Algorithm \ref{tf.path.alg} at the iteration $j_{_0}$ if
\begin{align}\label{stop.rule.proj1}
    \max_{0\, \leq\, t\, \leq\, 1}~ \Big|\widehat{\mbf u}_{_{-\mca A_{j_{_0}}}}^{\,\trm{st}}\big( \lfloor\, kt\,\rfloor\big) \Big| \leq \sigma x_{_\alpha}\left(k-r\right)^{(2\,r+1)/2}\,, \qquad \text{ for   }\quad t=0,\,\ldots,\,m-|\mca A_{_{j_0}},
\end{align}
where $x_{_\alpha}$ is derived from the equation
\begin{align}\label{thresh.stop.rule}
    \sum\limits_{i=1}^{\infty}(-1)^{i+1}\exp\left(\frac{-2\,i^2\,x_{_\alpha}^2}{S_r^2(k)}\right)=\frac{\alpha}{2}\,.
\end{align}

\section{Pattern Recovery and Theories}
\label{sec:pattern.recovery.proj1}

The main purpose of this section is to investigate whether the PRUTF algorithm can recover features of the true signal $\mbf f$. We also demonstrate conditions under which the structure of the estimated signal $\widehat{\mbf f}$ matches the true signal $\mbf f$. To verify the performance of PRUTF in the discovery the true signal, we first define what we mean by pattern recovery. 
\begin{definition}{(Pattern Recovery):}
A trend filtering estimate $\widehat{\mbf f}$  recovers the pattern of the true signal $\mbf f$ if
\begin{align}\label{pat.rec}
    \trm{sign} \big( \big[\,\mbf D\widehat{\mbf f}\, \big]_i \big)=\trm{sign}\big(\big[\,\mbf D\mbf f\,\big]_i\big), \qquad\qquad \textrm{for} \quad i=1,\ldots,m,
\end{align}
where $m=n-r-1$ is the number of rows of matrix $\mbf D$. We use the notation $\widehat{\mbf f} \stackrel{pr}{=}\mbf f$ to briefly denote the pattern recovery feature of $\widehat{\mbf f}$.
\end{definition}
In the asymptotic framework, a trend filtering estimate is called pattern consistent if 
\begin{align}
    \Pr \big(\, \widehat{\mbf f}\, \stackrel{pr}{=} \, \mbf f \, \big) ~\longrightarrow 1\qquad\qquad \textrm{as} \quad n\longrightarrow \infty,
\end{align}
where $\widehat{\mbf f}=\widehat{\mbf f}_n$, to denote its dependency to the sample size $n$. Pattern recovery is very similar to the concept of sign recovery in lasso \citep{zhao2006model,wainwright2009sharp} as it deals with the specification of both locations of non-zero coefficients and their signs.

The problem of pattern recovery is studied for the special case of the fused lasso in several papers.  \cite{rinaldo2009properties} derived conditions under which fused lasso consistently identifies the true pattern. This was contradicted by  \cite{rojas2014change}, who argued that fused lasso does not always succeed in discovering the exact change points.  \cite{rojas2014change} showed that fused lasso can be reformulated as the usual lasso, for which the necessary conditions for exact sign recovery have been established in the literature. Then, they proved that one such necessary condition, known as the irrepresentable condition, is not satisfied for the transformed lasso when there is a specific pattern called a staircase (Definition \ref{staircase.def}). Corrections to \cite{rinaldo2009properties} were appeared in \cite{rinaldo2014corrections}. Later on, \cite{qian2016stepwise} proposed a method called puffer transformation, which is shown to be consistent in specifying the exact change points, including in the presence of staircases. 

In the remaining part of this section, we use the dual variables to demonstrate the situations in which PRUTF can correctly recover the pattern of the true signal. Exact pattern recovery implies that the dual variables are comprised of $J_{_0}+1$ consecutive bounded processes whose endpoints correspond to the true change points. The following lemma describes the situations in which exact pattern recovery can be attained. A particular case of this result in the context of piecewise constant signals was established in \cite{rinaldo2014corrections}. 

\begin{theorem}\label{thm:consistency.constraints.proj1}
Exact pattern recovery in PRUTF occurs when the discrete time processes $\big\{ \widehat{u}_{_j}^{\,\mathrm{st}}(t)\, ,$   $t=\tau_{_j}+r_{_a}\, ,\, \ldots\, ,\, \tau_{_{j+1}}-r_{_b} \big\}$, for $j=0,\,\ldots,\,J_{_0}$, satisfy the following conditions simultaneously with probability one:
\begin{enumerate}[label=(\alph*)]
    \item {\bf First block constraint:} for $t=1\, ,\,\ldots,\,\tau_{_1}-r_{_b}$,
    {\footnotesize
    \begin{align}\label{block.const.first}
        -\lambda\left(1-\left[\big( \mathbf{D}_{_{-\mca A}}\mathbf{D}_{_{-\mca A}}^T \big)^{-1}\mathbf{D}_{_{-\mca A}}\right]_t\,\mbf D_{_{A_1}}^T\mbf 1\right)& ~\leq~ \widehat{u}_{_0}^{\,\mathrm{st}}(t) ~\leq~ \lambda\left(1+\left[\big(\mathbf{D}_{_{-\mca A}}\mathbf{D}_{_{-\mca A}}^T \big)^{-1}\mathbf{D}_{_{-\mca A}}\right]_t\,\mbf D_{_{A_1}}^T\mbf 1\right)\,.
    \end{align}}
    \item {\bf Last Block constraint:} for $t=\tau_{_{J_0}}+r_{_a}\, ,\, \ldots\, ,\, m$,
     {\footnotesize
     \begin{align}\label{block.const.last}
        \hspace{-.4cm}-\lambda\left(1+\left[ \big(\mathbf{D}_{_{-\mca A}}\mathbf{D}_{_{-\mca A}}^T\big)^{-1}\mathbf{D}_{_{-\mca A}}\right]_t\,\mbf D_{_{{A}_{_{J_0}}}}^T\mbf 1\right)& ~\leq~ \widehat{u}_{_{J_0}}^{\,\mathrm{st}}(t) ~\leq~ \lambda \left(1-\left[\big(\mathbf{D}_{_{-\mca A}}\mathbf{D}_{_{-\mca A}}^T \big)^{-1}\mathbf{D}_{_{-\mca A}}\right]_t\,\mbf D_{_{{A}_{_{J_0}}}}^T\mbf 1\right)\,.
    \end{align}}
    \item {\bf Interior Block constraints:} for $t=\tau_{_j}+r_{_a}\, ,\, \ldots\, ,\, \tau_{_{j+1}}-r_{_b}$, if $s_{_{j}}\neq s_{_{j+1}}$
    {\footnotesize
    \begin{align}\label{block.const.inter}
        &-\lambda\left(1-\left[ \big(\mathbf{D}_{_{-\mca A}}\mathbf{D}_{_{-\mca A}}^T \big)^{-1}\mathbf{D}_{_{-\mca A}}\right]_t\, \big(\mbf D_{_{A_{j+1}}}^T\mbf 1-\mbf D_{_{A_j}}^T\mbf 1 \big)\right)\, \leq \, \widehat{u}_{_j}^{\,\mathrm{st}}(t) \\[10pt]
        &\hspace{5cm}\leq\lambda\left(1+\left[\big(\mathbf{D}_{_{-\mca A}}\mathbf{D}_{_{-\mca A}}^T \big)^{-1}\mathbf{D}_{_{-\mca A}}\right]_t\, \big(\mbf D_{_{A_{j+1}}}^T\mbf 1-\mbf D_{_{A_j}}^T\mbf 1 \big)\right)\,,
    \end{align}}
    and if $s_{_{j}}\neq s_{_{j+1}}$, which corresponds to a staircase block, $\widehat{u}_{_j}^{\,\mathrm{st}}(t) \,\leq\, 0$ or $\widehat{u}_{_j}^{\,\mathrm{st}}(t) \,\geq\, 0$.
\end{enumerate}
\end{theorem}

\noindent
In the foregoing equations, $\mbf 1 \in \mbb R^{r+1}$ is a vector of size $r+1$ whose elements are all 1. A proof of the theorem is given in Appendix \ref{prf:consistency.constraints.proj1}.

We analyze the performance of the PRUTF algorithm in pattern recovery in two different scenarios; signals with staircase patterns and
signals without staircase patterns. To our knowledge, \cite{rojas2014change} was the first paper to carefully investigate the staircase pattern for the piecewise constant signals in the change points analysis setting. In \cite{rojas2014change}, a staircase pattern for a piecewise constant signal refers to the phenomenon of equal signs in two consecutive changes. We extend this concept to the general case, which covers any piecewise polynomial signals of order $r$, by applying the penalty matrix $\mbf D=\mathbf{D}^{(r+1)}$.

\begin{definition}\label{staircase.def}
Suppose that the true signal $\mathbf{f}$ is a piecewise polynomial of order $r$ with change points at the locations $\boldsymbol\tau= \big\{\tau_{_1},\,\ldots,\,\tau_{_{J_{_0}}} \big\}$. Moreover, let $\mathbf{B}= \big\{B_{_0},\,\ldots,\,B_{_{J_{_0}}}\big\}$ be blocks created by the change points in $\bsy \tau$. A staircase occurs in block $B_{_j},~ j=1,\,\ldots,\,J_{_0}-1$ if
\begin{align}\label{stair}
    \sgn\big( \big[\,\mathbf{D}\mbf f\,\big]_{\tau_{_j}}\big)=\sgn\big(\big[\,\mathbf{D}\mbf f\,\big]_{\tau_{_{j+1}}}\big).
\end{align}
\end{definition}

The following theorem investigates the consistency of PRUTF in pattern recovery, in both with and without staircases. Specifically, it shows that for a signal without a staircase, the exact pattern recovery conditions are satisfied with probability one. On the other hand, in the presence of staircases in the signal, the probability of these conditions holding, which is equivalent to the probability of a Gaussian bridge process never crossing the zero line, converges to zero.  

In the literature, the consistency of a change point method is usually characterized by the signal size $n$, the number of change points $J_0$, the noise variance $\sigma_n^2$, the minimal spacing between change points,
\begin{align*}
    \underline{L}_n= \min\limits_{j=0,\,\ldots,\,J_{_0}}~  \big|L_{n,\, j} \big|= \min\limits_{j=0,\,\ldots,\,J_{_0}}~  \big|\tau_{_{j+1}}-\tau_{_j}\big|,
\end{align*}
and the minimum magnitude of jumps between change points,
\begin{align*}
    \delta_n= \min\limits_{j=1,\,\ldots,\,J_{_0}}~ \big| \mbf D_{\tau_j} \mbf f \big|.
\end{align*}
All the above quantities are allowed to change as $n$ grows.

In the following, we present our main theorem providing conditions under which the output of the PRUTF algorithm consistently recovers the pattern of the true signal $\mbf f$.

\begin{theorem}\label{thm:consistency.proj1}
Suppose that $\mbf y$ follows the model in \eqref{fmodel.proj1}. Let $\bsy\tau$ be the set of $J_0$ change points for the true signal $\mbf f$. Additionally, assume that $\widehat{\bsy\tau}_n$ and $\widehat{\mbf f}_n$ are the set of change points estimates and the corresponding signal estimate obtained by the PRUTF algorithm, respectively. The followings hold for the PRUTF algorithm.
\begin{enumerate}[label=(\alph*)]
    \item {\bf Non-staircase Blocks:} Suppose there is no staircase block in the true signal $\mbf f$. For some $\xi>0$ and with 
    \begin{align*}
        \lambda_n < \dfrac{\delta_n\, \underline{L}_n^{2r+1}}{n^{2r}\, 2^{\,r+2}},
    \end{align*}
  if the conditions
    \begin{itemize}
        \item \begin{minipage}{0.875\textwidth}
        \begin{align}\label{consistent.conditions1.proj1}
            \frac{\delta_{n}\, \underline{L}_{n}^{r+1/2}}{n^r\, \sigma_n} \longrightarrow \infty
            \qquad\quad \trm{and}  \qquad\quad
            \frac{ \delta_{n}\, \underline{L}_{n}^{r+1/2}}{ 2^{\,r/2+2}\,n^r\, \sigma_n\, \sqrt{\log (J_0)}} ~>~  (1+\xi),
        \end{align}
        \end{minipage}
        
        \item \begin{minipage}{0.875\textwidth}
        \centering 
        \begin{align}\label{consistent.conditions2.proj1}
            \frac{\lambda_n\,\, \underline{L}_{n}^{r+1/2}}{n^r\, \sigma_n} \longrightarrow \infty
            \qquad\quad \trm{and}  \qquad\quad
            \frac{ 2^{\,r/2+1}\, \lambda_n\,\, \underline{L}_{n}^{r+1/2}}{n^r\, \sigma_n\, \sqrt{\log (n-J_0)}} ~>~  (1+ \xi),
        \end{align}
        \end{minipage}
    \end{itemize}
    hold, then the PRUTF algorithm guarantees exact pattern recovery with probability approaching one. That is,
    \begin{align*}
        \Pr \big(\, \widehat{\mbf f}_n\, \stackrel{pr}{=}\, \mbf f \big)\longrightarrow 1
        \qquad\qquad\qquad
        \trm{as}
        \qquad
        n \longrightarrow \infty.
    \end{align*}

    \item {\bf Staircase Blocks:} On the other hand, if the true signal $\mbf f$ contains at least one staircase block, then the probability of exact pattern recovery by the PRUTF algorithm converges to zero. That is, \begin{align*}
        \Pr \big(\, \widehat{\mbf f}_n\, \stackrel{pr}{=}\, \mbf f \big)\longrightarrow 0
        \qquad\qquad\qquad
        \trm{as}
        \qquad
        n \longrightarrow \infty.
    \end{align*}
   
\end{enumerate}
\end{theorem}
A proof is given in Appendix \ref{prf:consistency.proj1}.

\begin{remark}
The performance of PRUTF in terms of consistent pattern recovery relies on the quantity $\delta_n\, \underline{L}_{n}^{r+1/2}/\sigma_n$ and the choice of $\lambda_n$. In the piecewise constant case, the former quantity reduces to the well-known signal-to-noise-ratio quantity, which is crucial for a consistent change point estimation \citep{fryzlewicz2014wild,wang2020univariate}. The statements in \ref{consistent.conditions1.proj1} illustrate that the consistency of PRUTF in non-staircase blocks is achievable if the quantity $\delta_n\,\underline{L}_{n}^{\,r+1/2}/\sigma_n$ is of order $O(n^{r+c})$, for some $c>0$. In addition, the number of the change points $J_0$ is allowed to diverge, provided
\begin{align*}
    \log \big( J_0 \big) ~\lesssim~ \frac{ \delta_{n}^2\,\, \underline{L}_{n}^{2r+1}}{ n^{2r}\, \sigma_n^2}\,. 
\end{align*}
\end{remark}

\begin{figure}[!ht]
\begin{center}
\begin{subfigure}{.38\textwidth}
  \centering
  \includegraphics[width=1\linewidth]{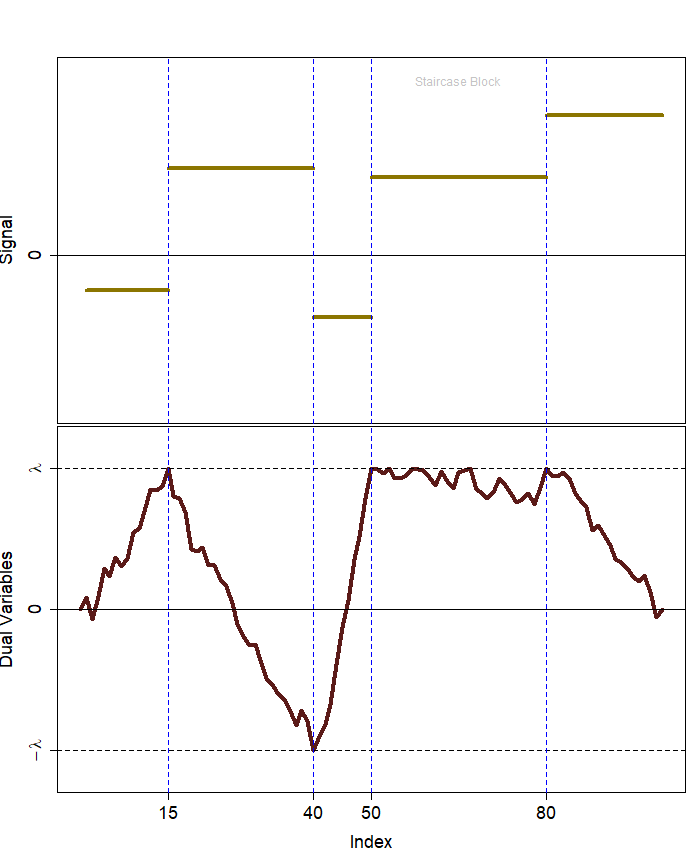}
  \caption{Piecewise constant signal with staircase block (50, 80].}
\end{subfigure}
\qquad\qquad
\begin{subfigure}{.38\textwidth}
  \centering
  \includegraphics[width=1\linewidth]{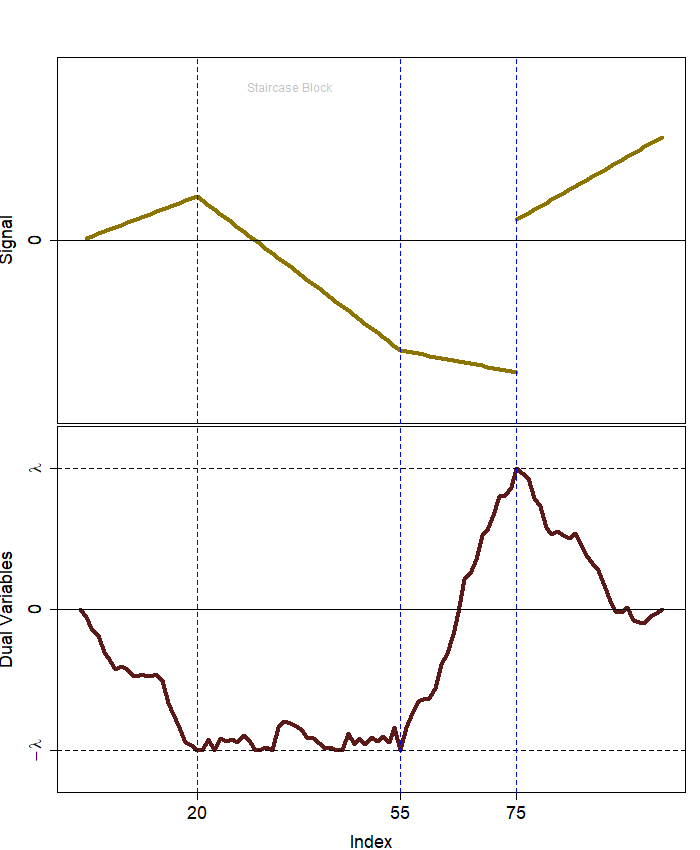}
  \caption{Piecewise linear signal with staircase block (20, 55].}
\end{subfigure}

\caption[Piecewise Constant and Piecewise Linear Signals With Staircase Patterns]{Piecewise constant and piecewise linear signals with staircase pattern at blocks (50, 80] and (20, 55] and their corresponding dual variables. }
\label{fig:stair-dual}
\end{center}
\end{figure}

The drift term \eqref{u.drift} plays a key role in assessing the performance of PRUTF in pattern recovery. From \eqref{drift.term.staircase.proj1}, this drift for a staircase block $B_{_j}$ becomes $\lambda \,s_{_j}$, which is constant in $t$ for the entire block. Consequently, the interior dual variables $\widehat{u}_{_j}(t;\lambda)$ for the staircase block $B_{_j}$ contain only the stochastic term $\widehat{u}_{_j}^{\,\textrm{st}}(t)=\left[\big(\mathbf{D}_{_{-\mca A}}\mathbf{D}_{_{-\mca A}}^T\big)^{-1}\mathbf{D}_{_{-\mca A}}\right]_t\, \mathbf{y}$, which fluctuates around the line $\lambda\, s_{_j}$. Recall that the KKT conditions for the dual problem of trend filtering require $\widehat{u}_{_j}(t;\,\lambda)$ to stay within the lines $-\lambda$ and $\lambda$. Thus, for a signal with staircase patterns, the PRUTF algorithm is sensitive to the variability of random noises and identifies change points once $\widehat{u}_{_{j}}^{\,\textrm{st}}(t)$ touches the $\pm\lambda$ boundaries.
Examples of piecewise constant and piecewise linear signals, along with their corresponding dual variables, are depicted in Figure \ref{fig:stair-dual}, in which the above argument can be clearly seen.


According to Theorem \ref{thm:consistency.proj1}, if there is no staircase pattern in the underlying signal, the PRUTF algorithm consistently estimates the true signal, and fails to do so, otherwise. Given the results in Theorem \ref{thm:consistency.proj1}, the natural question is whether Algorithm \ref{tf.path.alg} could be modified to enjoy the consistent pattern recovery in any case. In the next section, we will present an effective remedy based on altering the sign of a change associated with a staircase block.

\section{Modified PRUTF Algorithm}
\label{sec:modified.trend.filtering.proj1}
\begin{figure}[!t]
\begin{center}
\begin{subfigure}[b]{.36\textwidth}
  \centering
  \includegraphics[width=1\linewidth]{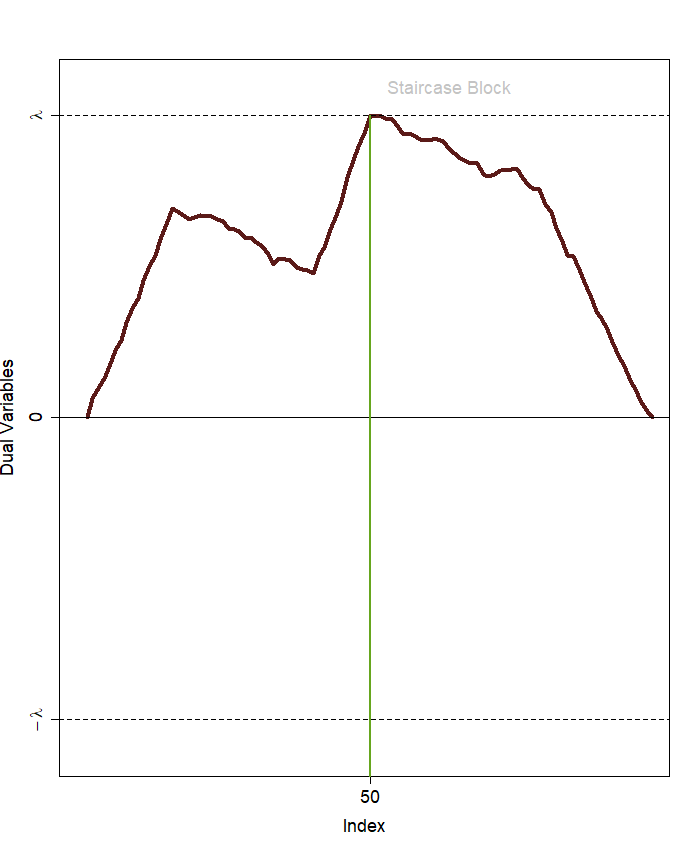}
  \caption{First change point at $\tau=50$.}
\end{subfigure}
\qquad\quad
\begin{subfigure}[b]{.36\textwidth}
  \centering
  \includegraphics[width=1\linewidth]{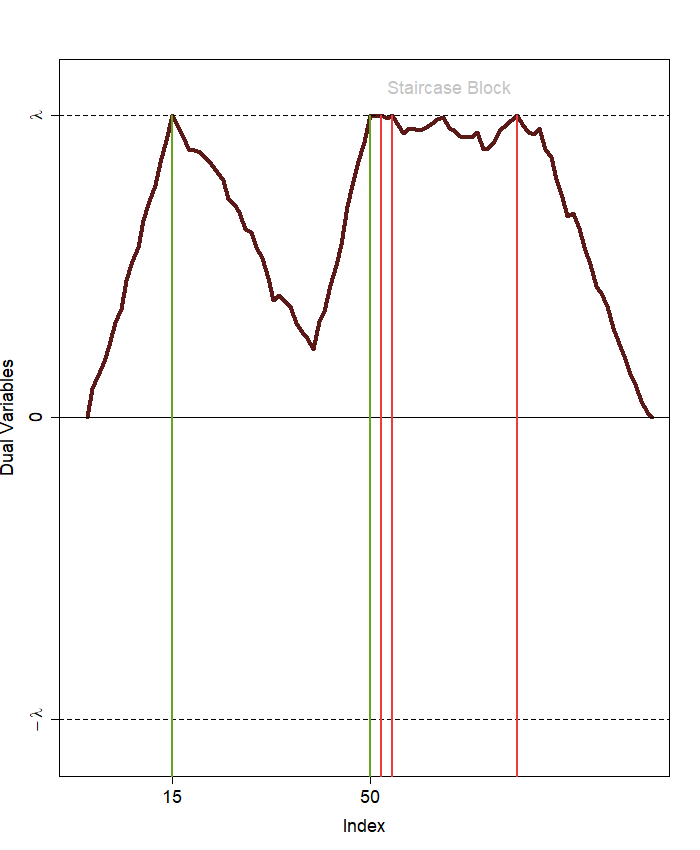}
  \caption{Second change point at $\tau=15$.}
\end{subfigure}
\\
\begin{subfigure}[b]{.36\textwidth}
  \centering
  \includegraphics[width=1\linewidth]{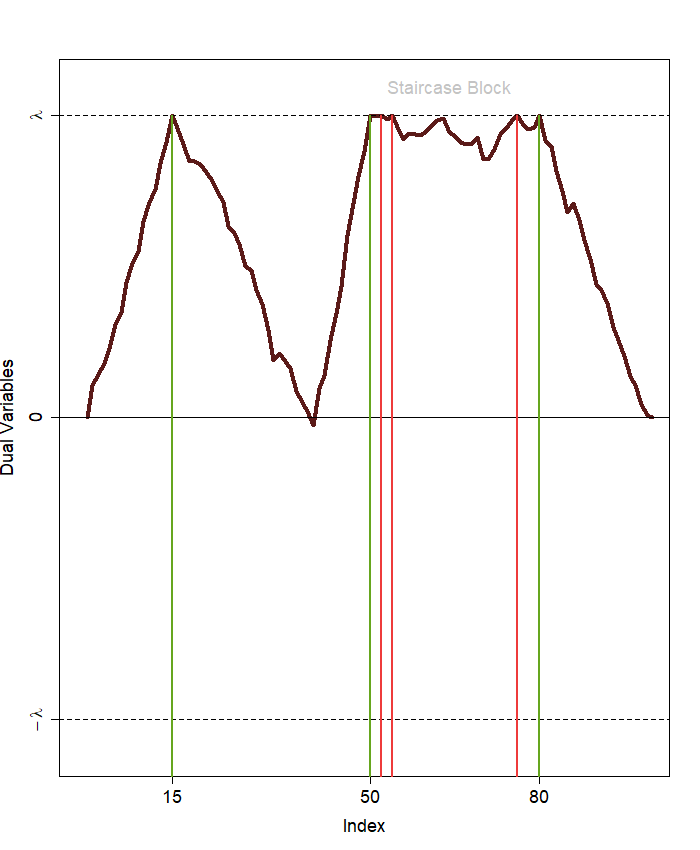}
  \caption{Third change point at $\tau=80$.}
\end{subfigure}
\qquad\quad
\begin{subfigure}[b]{.36\textwidth}
  \centering
  \includegraphics[width=1\linewidth]{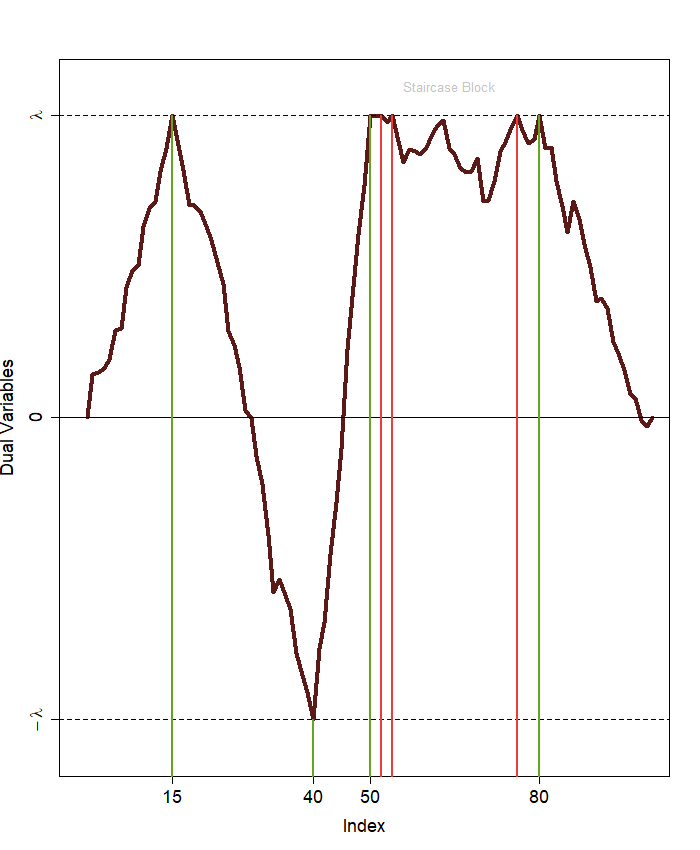}
  \caption{Fourth change point at $\tau=40$.}
\end{subfigure}
\caption[Impact of Staircase Patterns in Change Point False Discovery] {The process of detecting change points using PRUTF for a signal with a staircase pattern. In panel (b), there are three falsely detected change points $\{52, \,54,\, 76\}$ which is due to the staircase block $(50, \,80]$.}
\label{fig:stair.false.discovery}
\end{center}
\end{figure}

In this section, we attempt to modify the PRUTF algorithm in such a way that it produces consistent estimates of the number and locations of change points even in the presence of staircase patterns. As previously mentioned, for a staircase block, the drift term \eqref{u.drift} is constant and leads to false discoveries in change points. This is shown in Figure \ref{fig:stair.false.discovery} with a piecewise constant signal of size $n=100$ and the true change points at $\bsy\tau= \big\{15,\, 40,\, 50,\, 80 \big\}$. The figure reveals that the staircase block $(50,\, 80]$ leads to three false discoveries at the locations 52, 54 and 76.


\begin{figure}[!hb]
\begin{center}
\begin{subfigure}{.36\textwidth}
  \centering
  \includegraphics[width=1\linewidth]{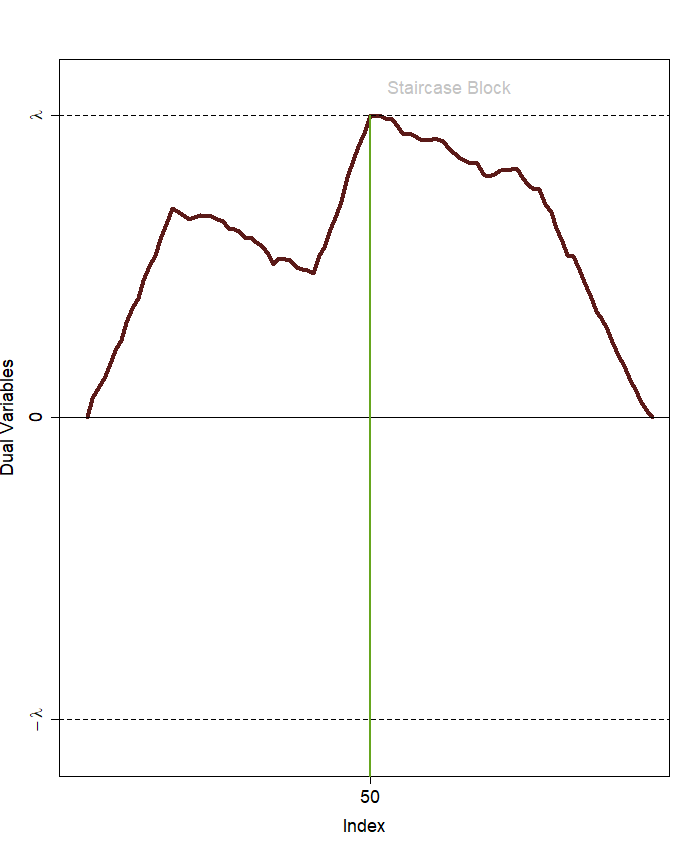}
  \caption{First change point at $\tau=50$.}
\end{subfigure}
\qquad\quad
\begin{subfigure}{.36\textwidth}
  \centering
  \includegraphics[width=1\linewidth]{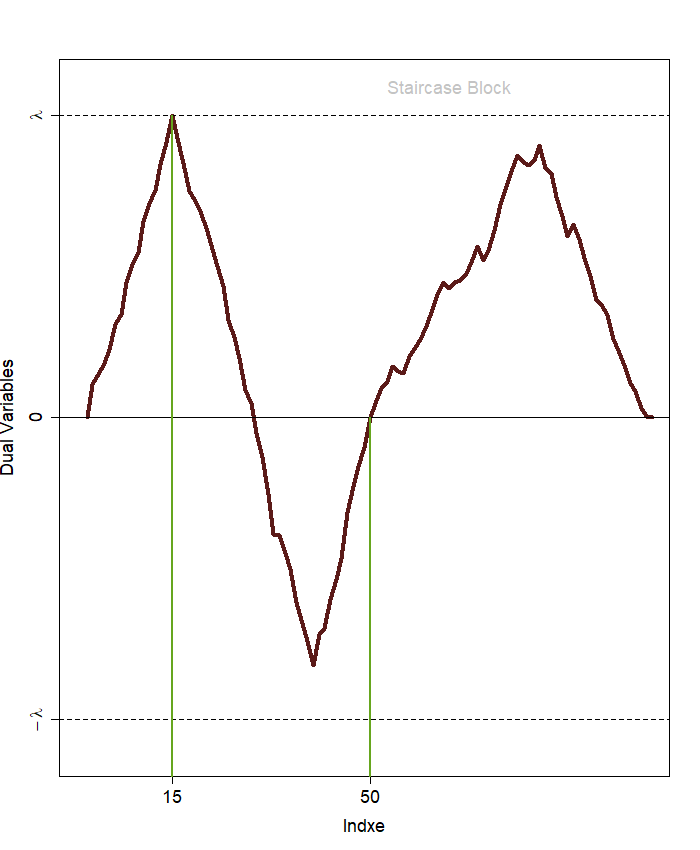}
  \caption{Second change point at $\tau=15$.}
\end{subfigure}
\\
\begin{subfigure}{.36\textwidth}
  \centering
  \includegraphics[width=1\linewidth]{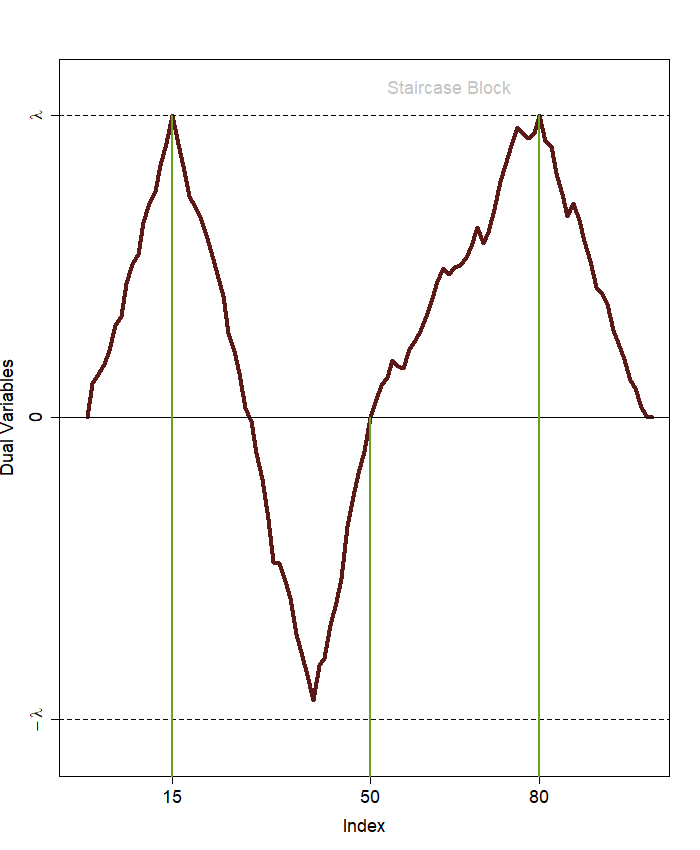}
  \caption{Third change point at $\tau=80$.}
\end{subfigure}
\qquad\quad
\begin{subfigure}{.36\textwidth}
  \centering
  \includegraphics[width=1\linewidth]{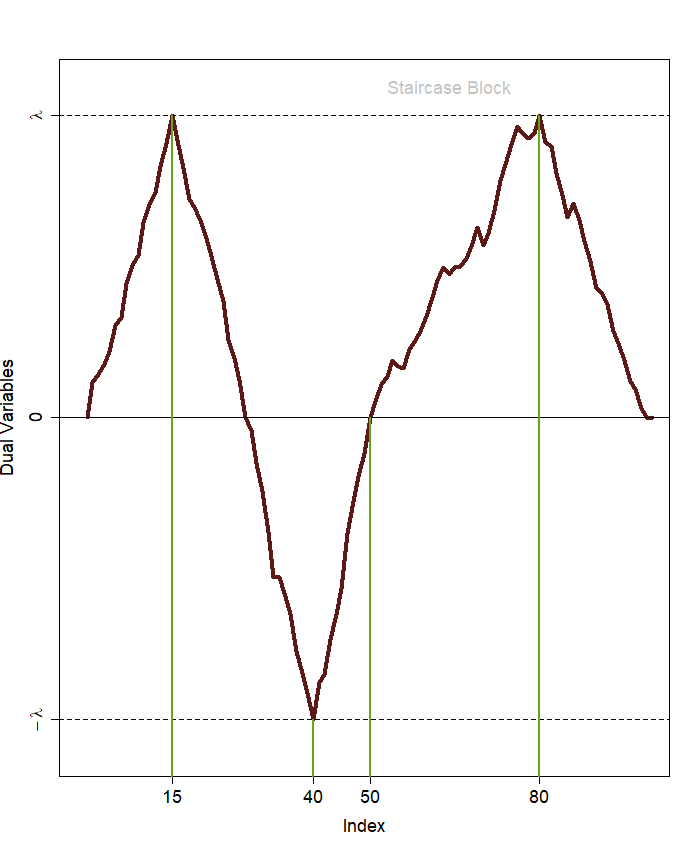}
  \caption{Fourth change point at $\tau=40$.}
\end{subfigure}
\caption[Steps of the mPRUTF Algorithm]{Steps of the mPRUTF algorithm until all four true change points are identified.}
\label{fig:modified-tf}
\end{center}
\end{figure}

The inconsistency of PRUTF in the presence of a staircase as established in Theorem \ref{thm:consistency.proj1}, stems from the fact that the change signs of the two consecutive change points at both ends of the staircase block are identical. That is, for the staircase block $B_{_j}$, $\sgn \big( [\,\mathbf{D} \mbf f\,]_{\tau_{_j}} \big)= \sgn \big( [\,\mathbf{D} \mbf f\,]_{\tau_{_{j+1}}} \big)$. Therefore, a question arises: can we modify Algorithm \ref{tf.path.alg} in such a way that the change signs of two neighbouring change points never become equal but still yield the solution path of trend filtering? We suggest a simple but very efficient solution to the above question.

Once a new change point is identified, we check whether its $r$-th order difference sign is the same as that of the change points right before and after. If these change signs are not identical, then the procedure continues to search for the next change point. Otherwise, we replace the sign of the neighbouring change point with zero. This replacement of the sign prevents the drift term \eqref{u.drift} from becoming zero. This idea is implemented for the above signal, and the result is displayed in Figure \ref{fig:modified-tf}. As shown in panel (b), the sign of the first change point at location 50 is set to zero since its sign is identical to the sign of the second change point at 15. This sign replacement vanishes false discoveries appeared in panel (b) of Figure \ref{fig:stair.false.discovery}.

Based on the above argument, PRUTF presented in Algorithm \ref{tf.path.alg} can be modified as follows to avoid false discovery and to produce consistent pattern recovery.

    

\begin{algorithm}[mPRUTF] \label{tf.modified.path.alg}

\begin{enumerate}
    \item[]
    
    \item Execute steps 1 and 2 of Algorithm \ref{tf.path.alg}.
    
    \item 
    \begin{enumerate}
        \item Execute part (a) of step 3 in Algorithm \ref{tf.path.alg} to obtain $\tau_{_j}^{^\mathrm{join}}$ and its sign $s_{_j}^{^\mathrm{join}}$. At this point, the algorithm checks whether $s_{_j}^{^\mathrm{join}}$ is identical to the signs of the change points just before and after $\tau_{_j}^{^\mathrm{join}}$. If so, set the sign of change point which is identical to $s_{_j}^{^\mathrm{join}}$ to zero.
        Then, repeat part (a) of step 3 again to obtain new $\tau_{_j}^{^\mathrm{join}}$ and $s_{_j}^{^\mathrm{join}}$ and update the sets $\mca A_{_j}$ and $\mca B_{_j}$.
        
        \item Execute parts $(b)$ and $(c)$ of step 3 in Algorithm \ref{tf.path.alg}.
        
    \end{enumerate}
    \item Repeat step 3 until either $\lambda_{_j}>0$ or a stopping rule is met.
\end{enumerate}
\end{algorithm}

The modified PRUTF (mPRUTF) algorithm produces consistent change point estimations, even in the presence of staircase patterns. This consistency has been achieved by converting the staircase patterns to non-staircase patterns that avoid false change point detection. In other words, running mPRUTF on an arbitrary signal (with or without staircases) is equivalent to running PRUTF on a signal without any staircase; see Figures \ref{fig:stair.false.discovery} and \ref{fig:modified-tf}. Thus, from part (a) of Theorem \ref{thm:consistency.proj1}, the mPRUTF algorithm is consistent in pattern recovery.

\begin{remark}
In step 2, part (a) of the mPRUTF algorithm, presented in Algorithm \ref{tf.modified.path.alg}, it is impossible for the sign $s_{_j}^{^\mathrm{join}}$ of the new change point to be identical to the sign of both of its immediate neighbouring change points, because the algorithm has already checked the equality of signs at previous steps. If they are equal, the sign of the immediate neighbouring change point will be set to zero.
\end{remark}

Recall that the KKT optimality conditions for solutions of the trend filtering problem in \eqref{tf.dual.obj.proj1} requires the dual variables $\widehat{\mbf u}_{_\lambda}$ to be less than or equal to $\lambda$ in absolute values, i.e., $|\widehat{\mbf u}_{_\lambda}|\leq \lambda$. This condition still holds when we replace the sign values ($+1$ or $-1$) with 0. Consequently, we have the following theorem.
\begin{theorem}\label{lem:modif.alg}
The mPRUTF algorithm presented in Algorithm \ref{tf.modified.path.alg} is a solution path of trend filtering.
\end{theorem}
For brevity, we do not provide the proof of Theorem \ref{lem:modif.alg} here. We refer the reader to the similar arguments for the LARS algorithm of lasso in \cite{tibshirani2013lasso}.

\begin{figure}[!h]
\begin{center}
\begin{subfigure}{.4\textwidth}
  \centering
  \includegraphics[width=\linewidth]{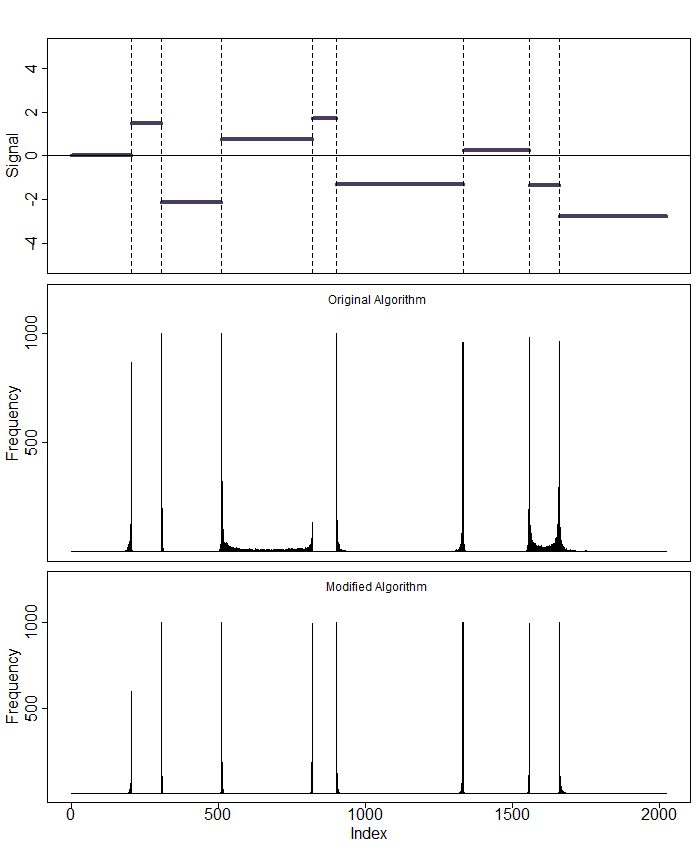}  
  \caption{A piecewise constant signal with blocks 4 and 8 as staircase blocks.}
\end{subfigure}
\quad
\begin{subfigure}{.4\textwidth}
  \centering
  \includegraphics[width=\linewidth]{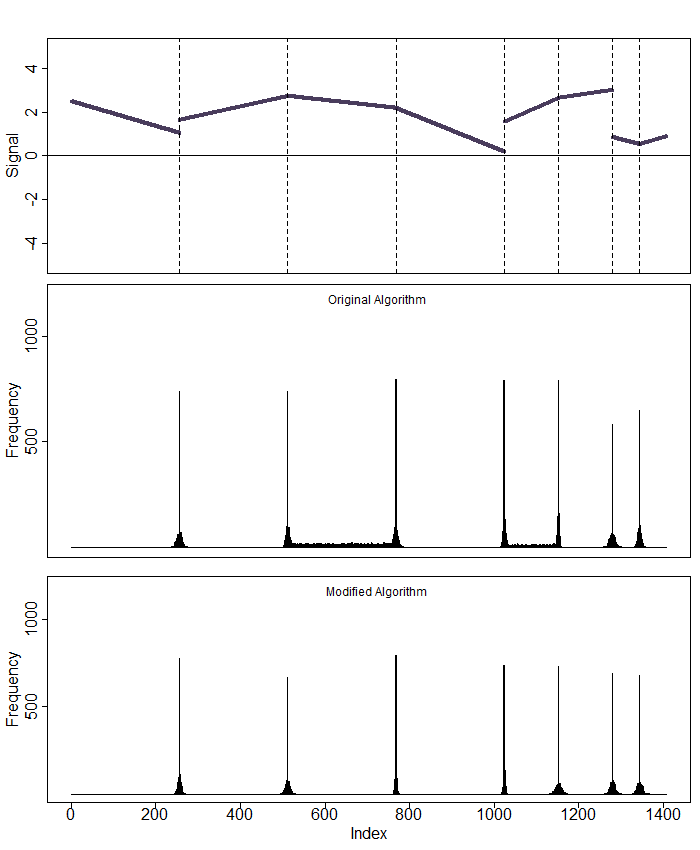}  
  \caption{A piecewise linear signal with blocks 3 and 5 as staircase blocks.}
\end{subfigure}
\caption[Change Point Frequency Plots For the PRUTF and mPRUTF Algorithms]{The frequency plots of estimated change points using the PRUTF and mPRUTF algorithms.}
\label{fig:simul-mod-vs-orig}
\end{center}
\end{figure}

It is worth pointing out that the mPRUTF algorithm requires slightly more computation than the original PRUTF algorithm. The increase in computation time directly depends on the number of staircase blocks in the underlying signal. To show how mPRUTF resolves the problem of false discovery in signals with staircases, we ran both algorithms for 1000 generated datasets from a piecewise constant and piecewise linear signals. The frequency plot of the estimated change points for both algorithms are represented in Figure \ref{fig:simul-mod-vs-orig}. The figure reveals that the original algorithm produces false discoveries within staircase blocks for both signals, whereas mPRUTF resolves this issue.


\section{Numerical Studies}
\label{sec:simulation.proj1}
In this section, we provide numerical studies to demonstrate the effectiveness and performance of our proposed algorithm, mPRUTF . We begin with a simulation study and then provide real data analyses.

\subsection{Simulation Study}
In this section, we investigate the performance of our proposed method, mPRUTF, by a simulation study. We consider two scenarios, namely piecewise constant and piecewise linear signals with staircase patterns. We compare our method to some powerful state-of-the-art approaches in change point analysis. These methods, a list of their available packages on CRAN, and their applicability for different scenarios are listed in Table \ref{tab:CPmethods}. 

\begin{table}[!t]
    \centering
    {\small
    \begin{tabular}{ll ll ll llll}
    \hline
    Method && Reference && R Package && \multicolumn{3}{c}{Signal} \\
    \cline{7-9}
     &&  &&  && PWC && PWL \\ \cline{1-1}\cline{3-3}\cline{5-5}\cline{7-9}
         PELT  && \cite{killick2012optimal} && {\bf changepoint} && \checkmark && \ding{53} \\
         WBS && \cite{fryzlewicz2014wild} && {\bf wbs}  && \checkmark && \ding{53} \\
         SMUCE  && \cite{frick2014multiscale} && {\bf stepR} && \checkmark && \ding{53} \\
         NOT  && \cite{baranowski2019narrowest} && {\bf not} && \checkmark && \checkmark \\    
         ID  && \cite{anastasiou2019detecting} && {\bf IDetect} && \checkmark && \checkmark \\
    \hline
    \end{tabular}}
    \caption[A List of Change Point Detection Methods With Their Packages in CRAN]{A list of change point detection and estimation methods with their packages in CRAN. The last two columns indicate which methods can be applied to piecewise constant  or/and piecewise linear  signals.}
    \label{tab:CPmethods}
\end{table}

We have adopted the simulation setting of \cite{baranowski2019narrowest}, and consider piecewise constant and piecewise linear signals as follows.
\begin{enumerate}[label=(\roman*)]
    \item A piecewise constant signal (PWC) of size $n=2024$ with the number of change points $J_{_0}=8$.  The locations of the true change points are $\bsy\tau= \big\{$205, 308, 512, 820, 902, 1332, 1557, 1659$\big\}$ with jump sizes 1.464, -0.656, 0.098, 1.830, 0.537, 0.768, -0.574, -3.335. We set the starting intercept to 0.
    
    \item A piecewise linear signal (PWL) of size $n=1408$ and the number of change points $J_{_0}=7$. The true change points are located at $\bsy\tau=\big\{$256, 512, 768, 1024, 1152, 1280, 1344$\big\}$. The corresponding intercepts and slopes for 8 created blocks by $\bsy\tau$ are 0.111, 0.553, -0.481, 3.002, -7.169, -0.030, 7.217, -0.958 and -8, 6, -3, -11, 12, 4, -7, 8, respectively.
\end{enumerate}

\begin{figure}[!b]
\begin{center}
\begin{subfigure}{.475\textwidth}
  \centering
  \includegraphics[width=1\linewidth]{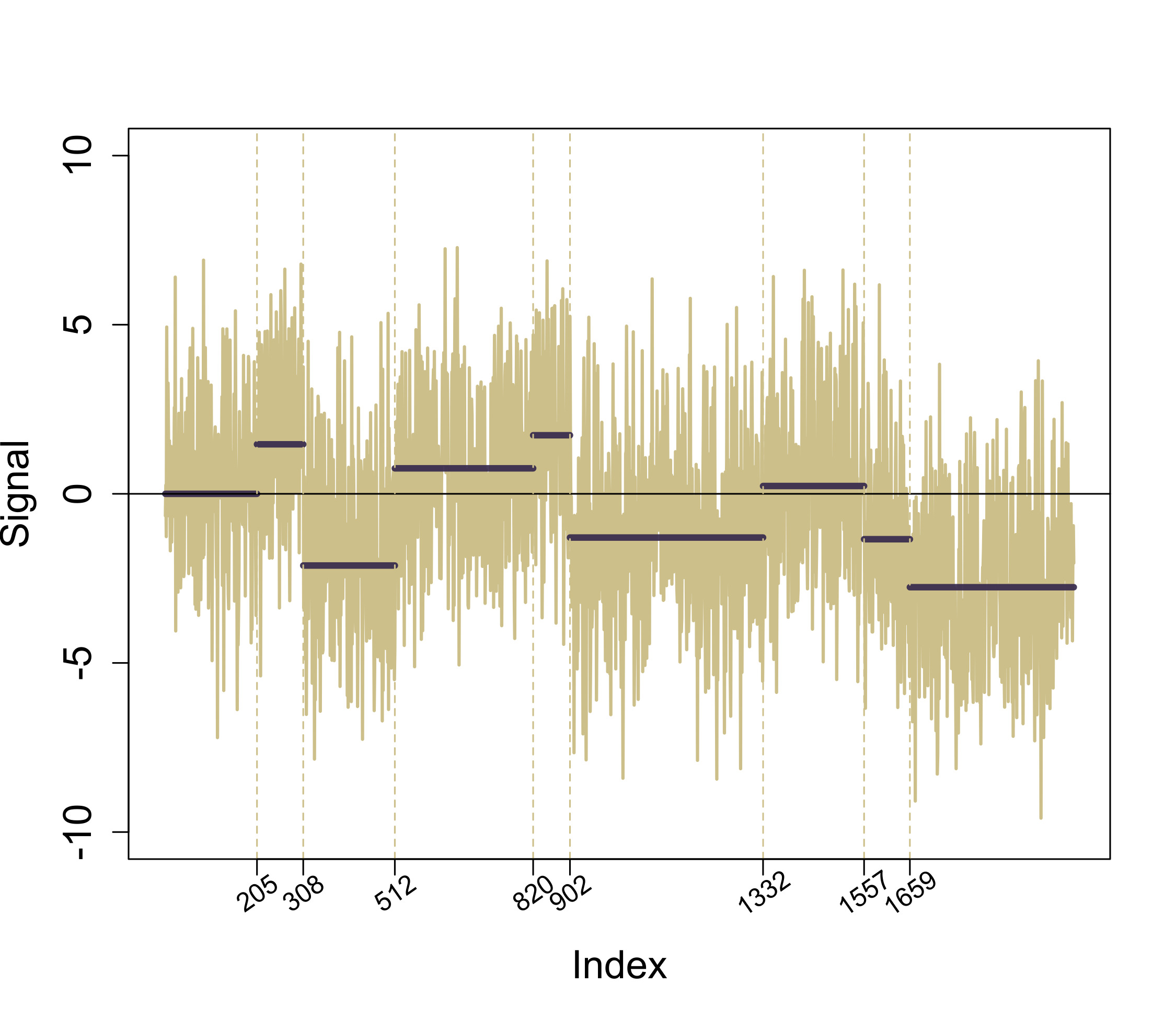}
  \caption{PWC signal with staircases at blocks $(512\,,\, 820]$ and $(1557\,,\, 1659]$.}
\end{subfigure}
\quad
\begin{subfigure}{.475\textwidth}
  \centering
  \includegraphics[width=1\linewidth]{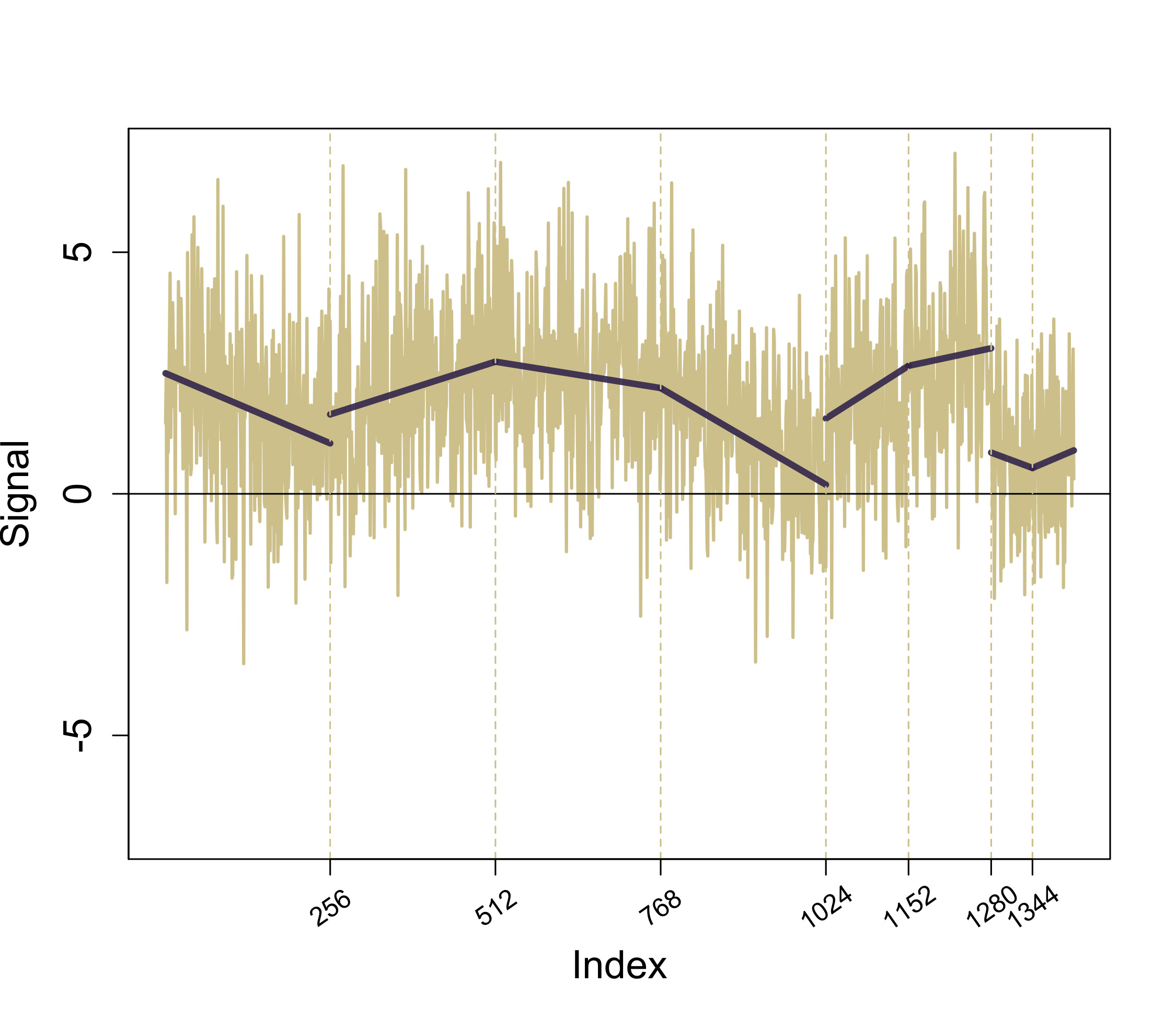}
  \caption{PWL signal with staircases at blocks $(512\,,\,768]$ and $(1024\,,\,1152]$.}
\end{subfigure}
\caption[Simulated Piecewise Constant and Piecewise Linear Signals]{The piecewise constant (PWC) and piecewise linear (PWL) signals with the generated samples used in the simulation study.}
\label{fig:signal-scenarios}
\end{center}
\end{figure}

\noindent Figure \ref{fig:signal-scenarios} displays the true PWC and PWL signals, with their representative datasets generated using model \eqref{fmodel.proj1}. We note that both PWC and PWL signals contain two staircase blocks. These blocks for the PWC signal are $(512\,,\, 820]$, $(1557\,,\, 1659]$ and for PWL signal are $(512\,,\,768]$ and $(1024\,,\,1152]$.

We apply mPRUTF presented in Algorithm \ref{tf.modified.path.alg} to estimate the number and the locations of the change points for the PWC and PWL signals. In each iteration of the simulation study, we simulate a dataset according to model \eqref{fmodel.proj1} under the assumption that the error terms are independently and identically distributed as $N(0\,,\,\sigma^2)$. Moreover, we set the significance level to $\alpha=0.05$ for the stopping rule in \eqref{stop.rule.proj1}.

\begin{figure}[!b]
\begin{center}
\begin{subfigure}{.47\textwidth}
  \centering
  \includegraphics[width=\linewidth]{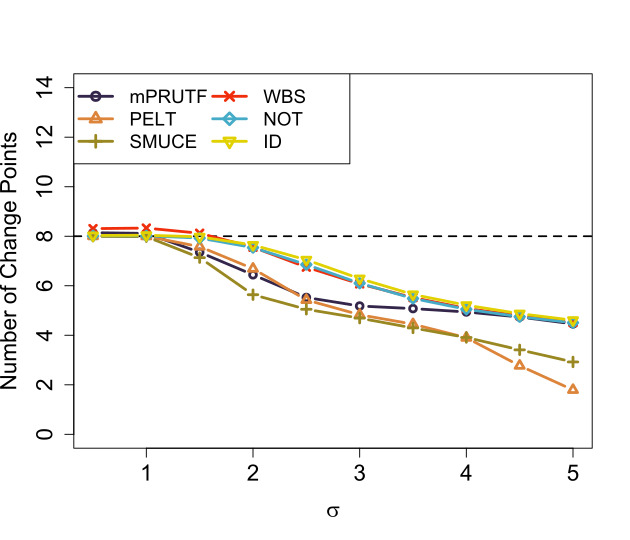}
  \caption{Average number of change points.}
  \label{fig:simul-pc-a}
\end{subfigure}
\quad
\begin{subfigure}{.47\textwidth}
  \centering
  \includegraphics[width=\linewidth]{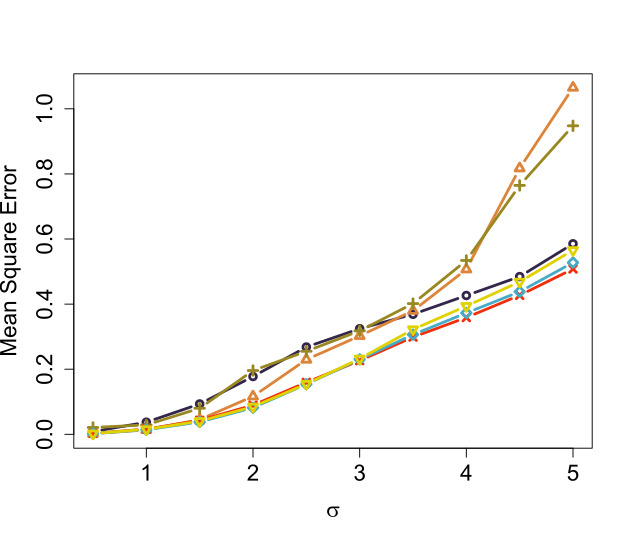}
  \caption{MSE estimations.}
  \label{fig:simul-pc-b}
\end{subfigure}
\\
\begin{subfigure}{.47\textwidth}
  \centering
  \includegraphics[width=\linewidth]{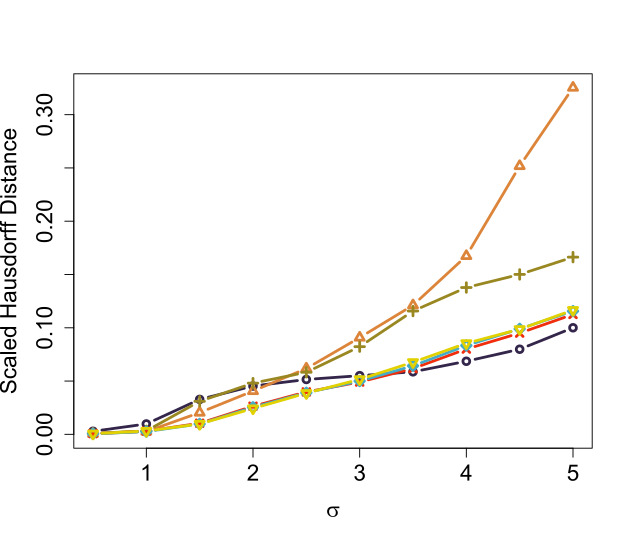}
  \label{fig:simul-pc-c}
\end{subfigure}
\quad
\begin{subfigure}{.47\textwidth}
  \centering
  \includegraphics[width=\linewidth]{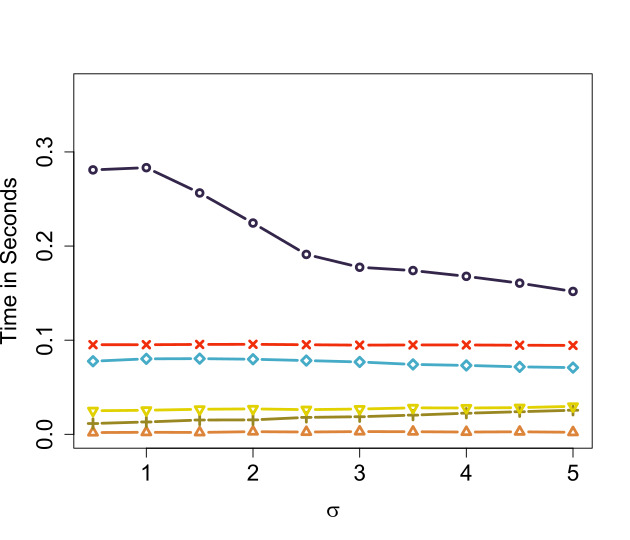}
  \caption{Computation time.}
  \label{fig:simul-pc-d}
\end{subfigure}
\caption[Comparison of mPRUTF With the State-of-the-Art Change Point Methods: Piecewise Constant Signal]{The estimated average number of change points, MSE and Hausdorff distance,  as well as the computation time of various methods for PWC signal. The results are provided for different values of the noise variability $\sigma$.}
\label{fig:simul-pc}
\end{center}
\end{figure}

\begin{figure}[!b]
\begin{center}
\begin{subfigure}{.47\textwidth}
  \centering
  \includegraphics[width=\linewidth]{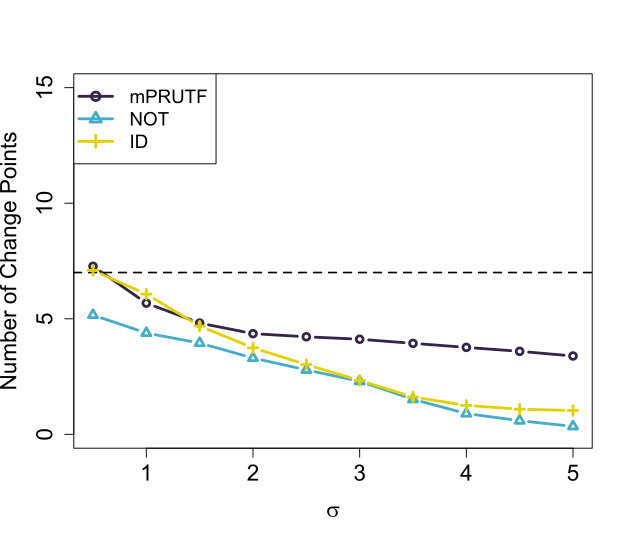}
  \caption{Average number of change points.}
  \label{fig:simul-pl-a}
\end{subfigure}
\quad
\begin{subfigure}{.47\textwidth}
  \centering
  \includegraphics[width=\linewidth]{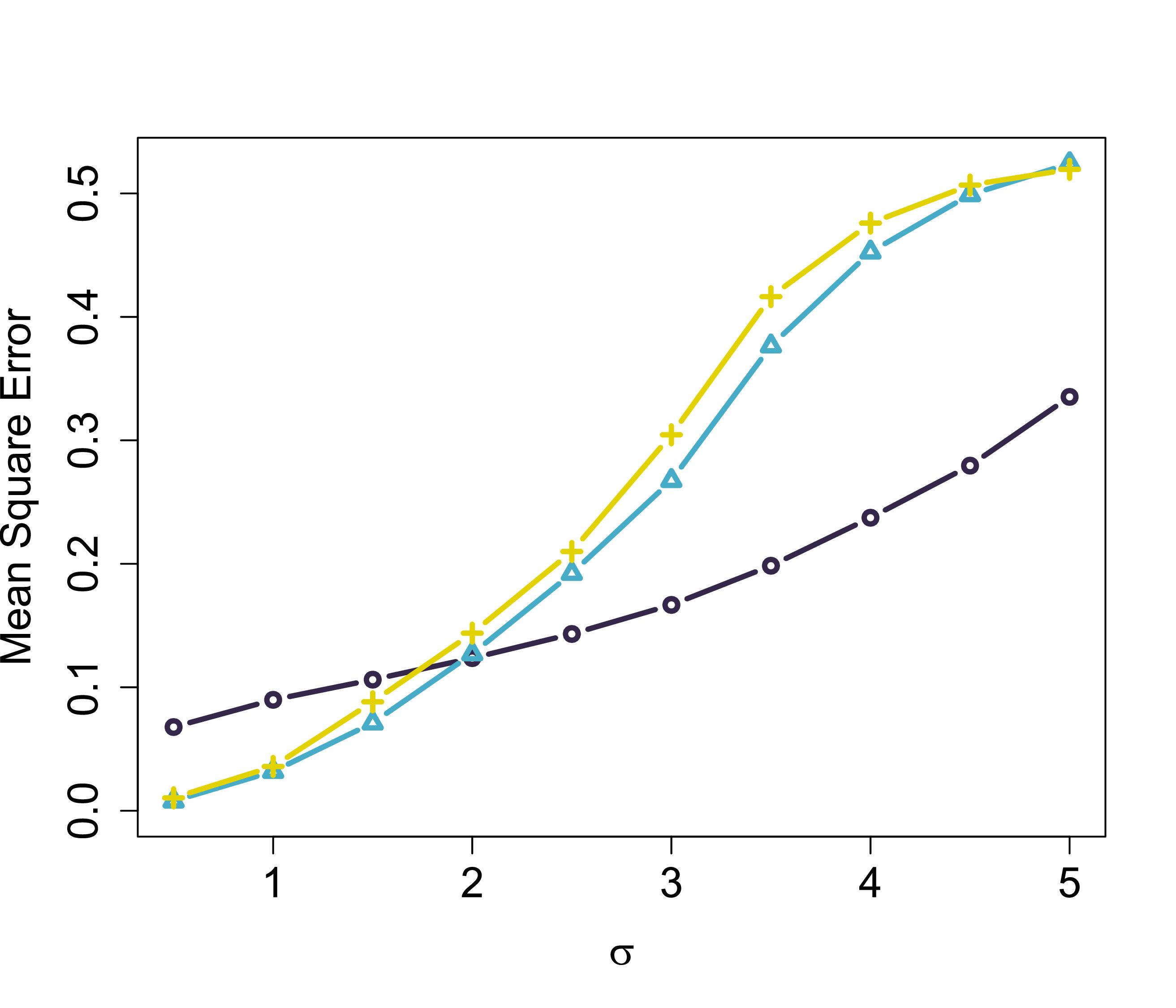}
  \caption{MSE estimations}
  \label{fig:simul-pl-b}
\end{subfigure}
\newline
\begin{subfigure}{.47\textwidth}
  \centering
  \includegraphics[width=\linewidth]{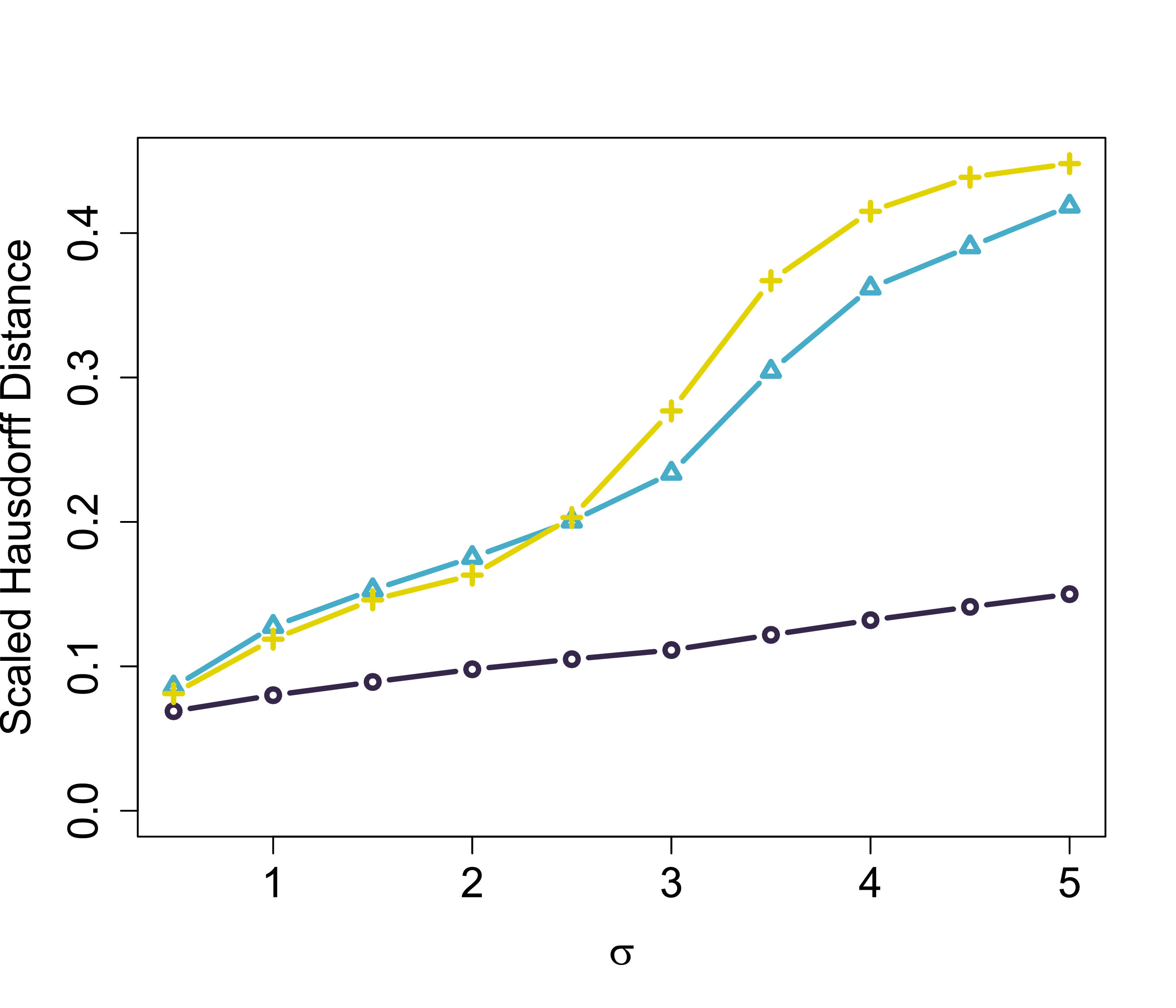}
  \caption{Hausdorff distance.}
  \label{fig:simul-pl-c}
\end{subfigure}
\quad
\begin{subfigure}{.47\textwidth}
  \centering
  \includegraphics[width=\linewidth]{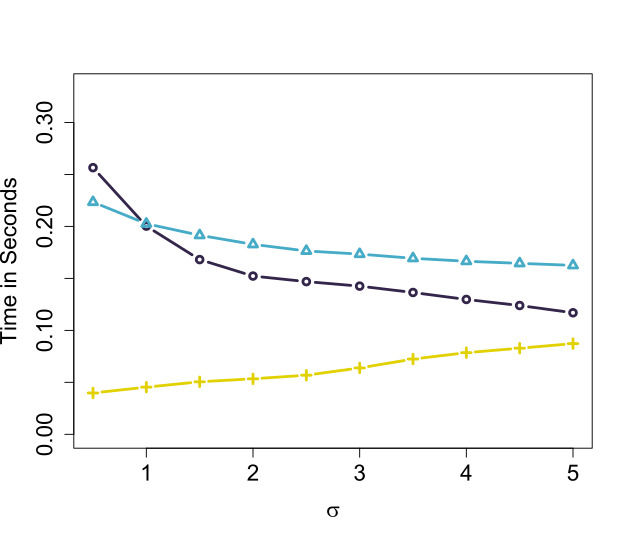}
  \caption{Computation time.}
  \label{fig:simul-pl-d}
\end{subfigure}
\caption [Comparison of mPRUTF With the State-of-the-Art Change Point Methods: Piecewise Linear Signal]{The estimated average number of change points, MSE and Hausdorff distance,  as well as the computation time of various methods for PWL signal. The results are provided for different values of the noise variability $\sigma$.}
\label{fig:simul-pl}
\end{center}
\end{figure}


In order to explore the impact of different noise levels on the change point methods, we run each simulation for various values of $\sigma$ in $\big\{ 0.5 , \,1 , \,1.5 ,\, \ldots\, ,\, 4.5 , \,5 \big\}$. We run the simulation $N=5000$ times and report the results for each change point technique in terms of estimates of the number of change points, estimates of the mean square error given by $\textrm{MSE}=N^{-1} \sum_{\, i=1}^{N} \big(\widehat{f}_{_i}-f_{_i} \big)^2$, estimates of the scaled Hausdorff distance given by
\begin{align*}
    d_H=\frac{1}{N}\,\max\left\{\max_{j=0,\,\ldots,\,J_{_0}}\,\,\min_{i=0,\,\ldots,\,\widehat{J}_{_0}}\,\big|\widehat{\tau}_{_i}-\tau_{_j} \big|~~,~~\max_{i=0,\,\ldots,\,\widehat{J}_{_0}}\,\min_{j=0,\,\ldots,\,J_{_0}}\, \big|\widehat{\tau}_{_i}-\tau_{_j} \big|\right\},
\end{align*}
and the computation time in seconds. These quantities are frequently used to assess the performance of a change point detection technique in the literature, for example, see \cite{baranowski2019narrowest}, \cite{anastasiou2019detecting}. The signal estimate, $\widehat{f}$, is computed by the least square fit of a polynomial of order $r$ to the observations within segments created by each change point method. We also remark that the tuning parameters and stopping criteria for the methods listed in Table \ref{tab:CPmethods} are set to the default values by the packages.

The results for the PWC and PWL signals are presented in Figures \ref{fig:simul-pc} and \ref{fig:simul-pl}, respectively. In the case of piecewise constant signal, as in Figure \ref{fig:simul-pc}, mPRUTF performs comparable to PELT and SMUCE in terms of the average number of change points, MSE and the scaled Hausdorff distance up to $\sigma=3$, and outperforms them as $\sigma$ increases. For $\sigma \geq 4$, similar performance to WBS, NOT and ID is viewed from these measurements. As indicated by the average number of change points, MSE and the scaled Hausdorff distance, WBS, NOT and ID outperform the other methods in almost all noise levels. From a computational point of view, mPRUTF takes a slightly longer time, mainly due to the matrix $\mbf D$ multiplications, however, this computation time decreases as noise level $\sigma$ increases. As in panel (d) of Figure \ref{fig:simul-pc}, the methods PELT, SMUCE and ID are the fastest ones.

In the case of piecewise linear signal, mPRUTF is only compared to NOT and ID methods, which are applicable to the piecewise polynomials of order $r \geq 1$. As in Figure \ref{fig:simul-pl} mPRUTF outperforms both NOT and ID in terms of the average number of change points and the scaled Hausdorff distance for all noise levels. In terms of MES, mPRUTF outperforms the other two for $\sigma \geq 2$. As shown in Panel (d) of Figure \ref{fig:simul-pl}, the computation time of mPRUTF ranks second after ID.

The mPRUTF method performs well in terms of the estimation of the number of change points, their locations, as well as the true signals. In fact, simulation results for most of the scenarios indicate that mPRUTF is among the most competitive change point detection approaches in the literature.

\subsection{Real Data Analysis}
In this section, we have analyzed UK HPI and GISTEMP and COVID-19 datasets, using our proposed algorithm. Because $\sigma^2$ is unknown for these real datasets, we applied median absolute deviation (MAD) proposed by \cite{hampel1974influence}, to robustly estimate $\sigma^2$. More specifically, a MAD estimate of $\sigma$ for piecewise constant signals is given by $\widehat{\sigma}=\textrm{Median}\big(\mbf D^{(1)}\,\mbf y \big) \big/ \big[\sqrt{2}\, \Phi^{-1}(0.75) \big]$ and for piecewise linear signals by $\widehat{\sigma}=\textrm{Median} \big( \mbf D^{(2)}\,\mbf y \big) \big/ \big[\sqrt{6}\,\Phi^{-1}(0.75) \big]$, where $\Phi^{-1}(\cdot)$ represents the inverse cumulative density function of the standard normal distribution.

\begin{figure}[!t]
\begin{subfigure}{.47\textwidth}
  \centering
  \includegraphics[width=1\linewidth]{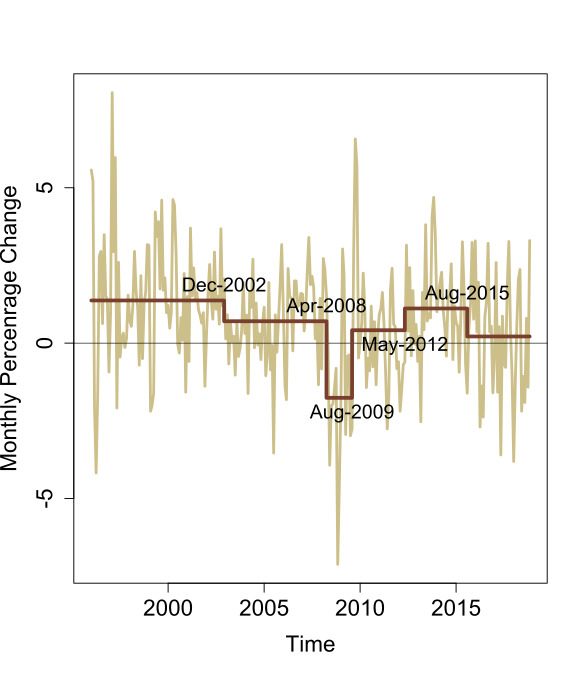}
  \caption{UK HPI dataset and its piecewise constant fit.}
\end{subfigure}
\quad
\begin{subfigure}{.47\textwidth}
  \centering
  \includegraphics[width=1\linewidth]{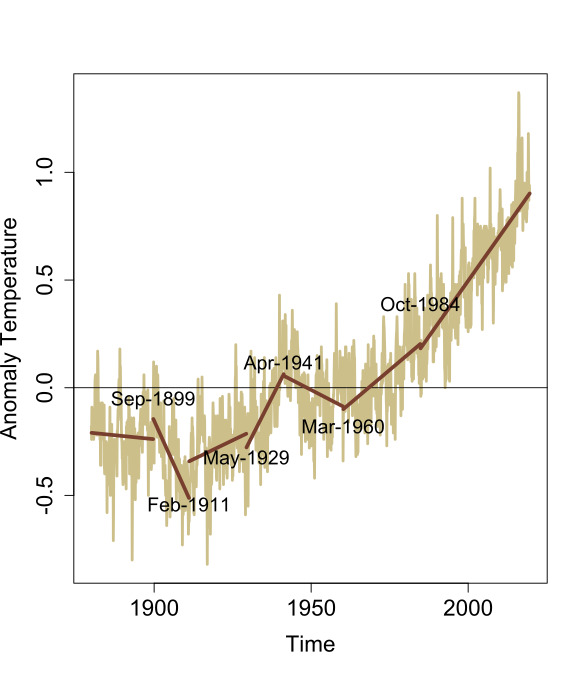}
  \caption{GISTEMP dataset and its piecewise linear fit.}
\end{subfigure}
\caption[Detected Change Points Using mPRUTF For UK HPI and GISTEMP Datasets]{The time series and fitted signals for both Tower Hamlet HPI and GISTEMP datasets presented in examples }
\label{fig:HPI-GISTEMP}
\end{figure}

\begin{example}[UK HPI Data] \label{HPIdata.example}

The UK House Price Index (HPI) is a National Statistic that shows changes in the value of residential properties in England, Scotland, Wales and Northern Ireland. The HPI measures the price changes of residential housing by calculating the price of completed houses sale transactions as a percentage change from some specific start date. The UK HPI uses the hedonic regression model as a statistical approach to produce estimates of the change in house prices for each period. For more details, see \url{https://landregistry.data.gov.uk/app/ukhpi}.Many researchers, including \cite{fryzlewicz2018detecting}, \cite{fryzlewicz2018tail} and \cite{anastasiou2019detecting}, have studied the UK HPI dataset in carrying out change point analysis.
In the current study, we consider monthly percentage changes in the UK HPI at Tower Hamlets (an east borough of London) from January 1996 to November 2018.

We have applied the mPRUTF algorithm to the dataset. The algorithm have found five change points located at the dates December 2002, April 2008 and August 2009 (may be attributed to the Credit Crunch and Financial Crises), May 2012 (may be attributed to The London 2012 Summer Olympics) and August 2015 (may be attributed to regulatory and tax changes, and also by lower net migration from the EU). The dataset, the change points derived by mPRUTF and its piecewise constant fit are presented in panel (a) of Figure \ref{fig:HPI-GISTEMP}.  
\end{example}

\begin{example} [GISTEMP Data] \label{GISTEMPdata.example}

The Goddard Institute for Space Studies (GISS) monitors broad global changes around the world. The GISS Surface Temperature Analysis (GISTEMP) is an estimate of the global surface temperature changes (see \url{https://www.giss.nasa.gov}). In the analysis of GISTEMP data, the temperature anomalies are used rather than the actual temperatures. A temperature anomaly is a difference from an average or baseline temperature. The baseline temperature is typically computed by averaging thirty or more years of temperature data (1951 to 1980 in the current dataset). A positive anomaly indicates the observed temperature was warmer than the baseline, while a negative anomaly indicates the observed temperature was cooler than the baseline. For more details see \cite{hansen2010global} and \cite{lenssen2019improvements}.

The GISTEMP dataset has been frequently explored in change point literature, for example see \cite{ruggieri2013bayesian}, \cite{james2015change} and \cite{baranowski2019narrowest}.
Panel (b) of Figure \ref{fig:HPI-GISTEMP} displays the monthly land-ocean temperature anomalies recorded from January 1880 to August 2019 (see \url{https://data.giss.nasa.gov/gistemp}). The plot reveals the presence of a linear trend with several change points in the dataset. For this dataset, we have identified six change points using mPRUTF  located in September 1899, February 1911, May 1929, April 1941, March 1960, October 1984. The locations of change points and an estimate of the piecewise linear signal are presented in panel (b) of Figure \ref{fig:HPI-GISTEMP}.  

\end{example}

\begin{example}[COVID-19 Data] \label{example:covid19.proj1}
Since the initial outbreak of Novel Coronavirus Disease 2019 (COVID-19) in Wuhan, China, in mid-November 2019, the virus has rapidly spread throughout the world. The pandemic has infected 21.26 million people and killed more than 761,000 \url{https://covid19.who.int/}, greatly stressing public health systems and adversely influencing global society and economies. Therefore, every country has attempted to slow down the transmission rate by various regional and national policies such as the declaration of national emergencies, quarantines and mass testing. Of vital interest to governments is understanding the pattern of the epidemic growth and assessing the effectiveness of policies undertaken. A scientist can investigate these matters by analyzing the sequence of infection data for COVID-19. Changepoint detection is one possible framework for studying the behaviour of COVID-19 infection curves. By detecting the locations of alterations in the curves, change point analysis gives us insights into changes in the transmission rate or efficiency of interventions. It also enables us to raise warning signals if the disease pattern changes.


For this example, we consider the log-scale of the cumulative number of confirmed cases for Australia, Canada, the United Kingdom and the United States, during the period March 10, 2020 through April 30, 2021. We have applied mPRUTF to detect change points that have occurred in the data for each country. We then fitted a piecewise linear model to the data using the selected change points, which provides a more direct perception of how the growth rate changes over time.

\begin{figure}[!t]
\begin{subfigure}{.47\textwidth}
  \centering
  \includegraphics[width=1\linewidth]{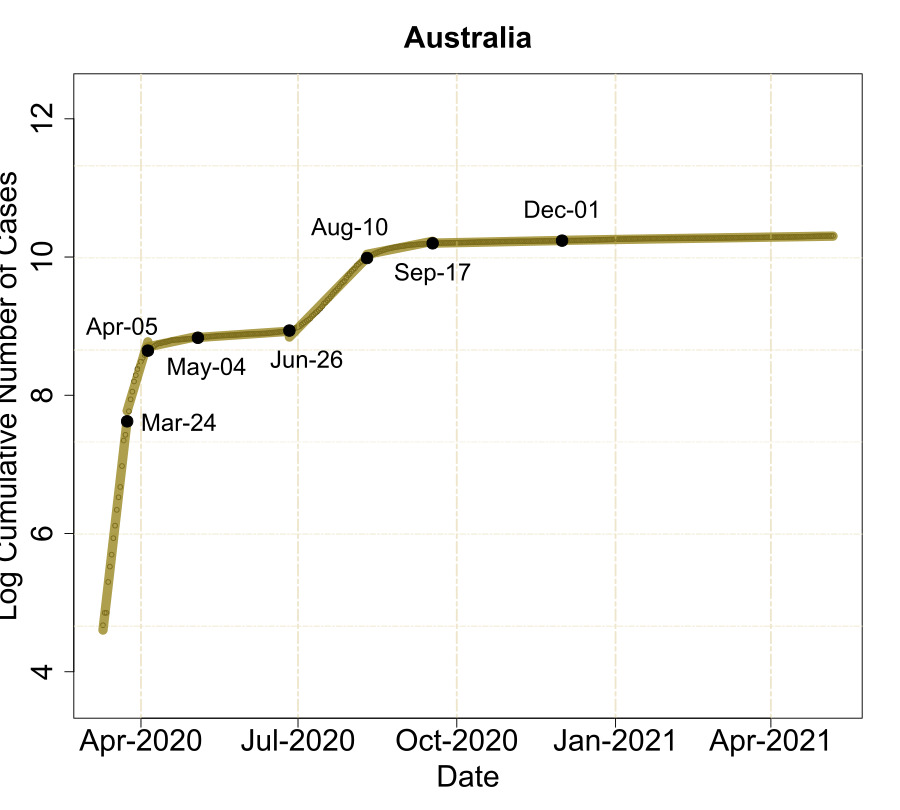}
\end{subfigure}
\quad
\begin{subfigure}{.47\textwidth}
  \centering
  \includegraphics[width=1\linewidth]{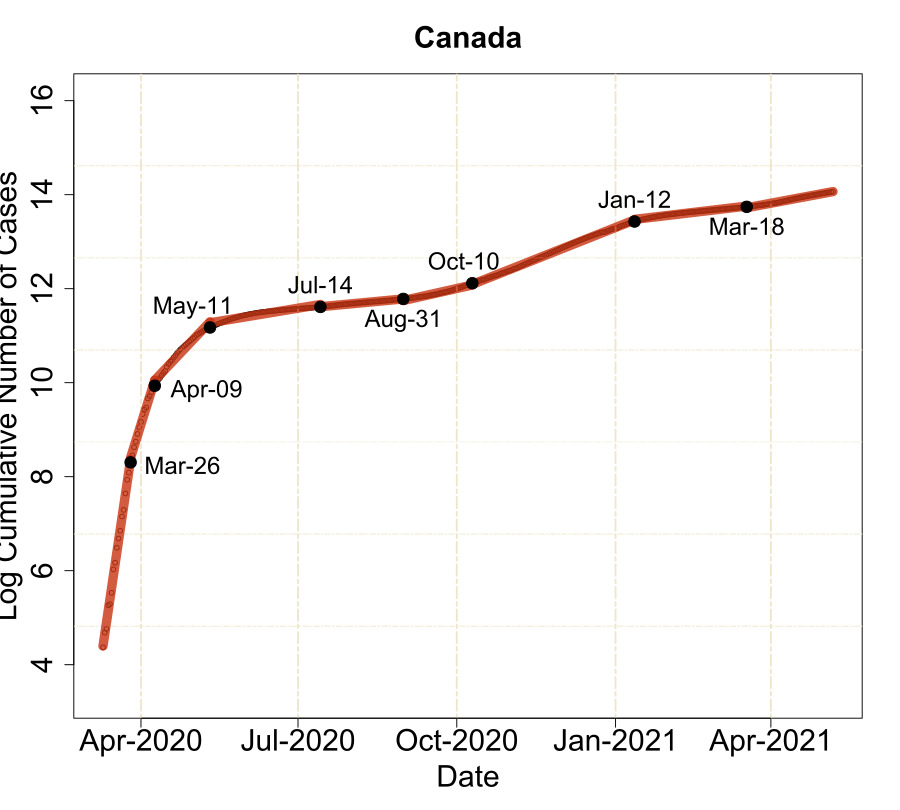}
\end{subfigure}
\\
\vspace{.6cm}
\\
\begin{subfigure}{.47\textwidth}
  \centering
  \includegraphics[width=1\linewidth]{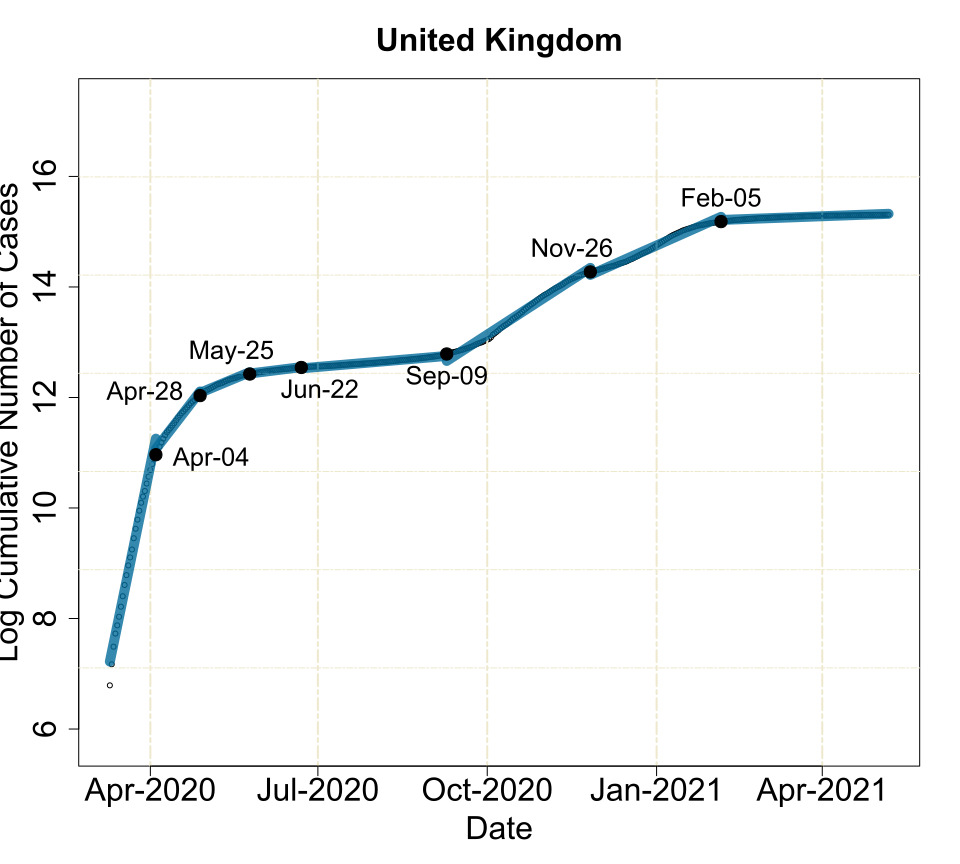}
\end{subfigure}
\quad
\begin{subfigure}{.47\textwidth}
  \centering
  \includegraphics[width=1\linewidth]{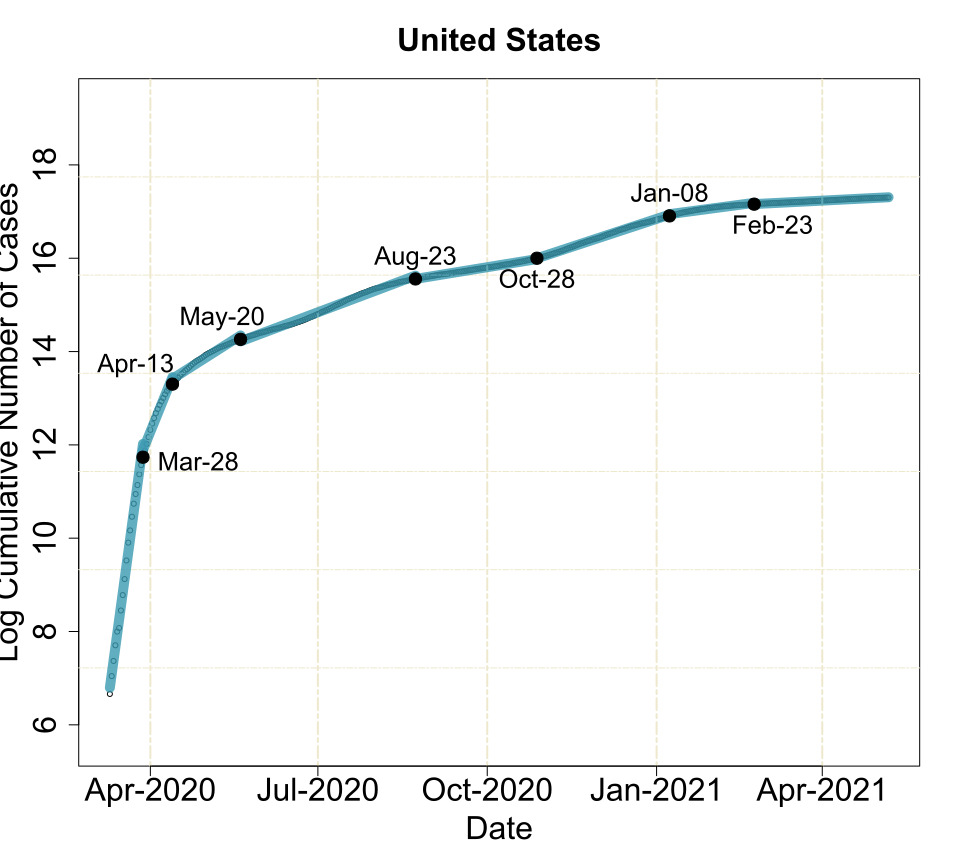}
\end{subfigure}
\caption[Detected Change Points Using mPRUTF For COVID-19 Datasets]{The change point locations and estimated linear trend for the transformed COVID-19 datasets in Example \ref{example:covid19.proj1}.  The dates indicated on each plot show the detected change points.}
\label{fig:covid19}
\end{figure}

Figure \ref{fig:covid19} displays the locations of change points detected by the mPRUTF algorithm as well as the estimated linear trends for the four countries. For example, our algorithm has identified eight change points for Canada, on March 26, 2020; April 9, 2020; May 11, 2020; July 14, 2020; August 31, 2020; October 10, 2020; January 12, 2021 and March 18, 2021. The figure shows segments created by the estimated change points as well as their growth rate. The growth rate for the first segment (from March 10, 2020 to March 26, 2020) is remarkably high, but starts to slightly decline after the first change point on March 26, 2020. This mild decline may be linked to the declaration of the the state of emergency, quarantine and international travel ban declared by the Government of Canada. The third segment (from April 9, 2020 to May 11, 2020), the fourth segment (from May 11, 2020 to July 14, 2020) and the fifth segment (from July 14, 2020 to August 31, 2020) have witnessed noticeable decreases in the growth rate. The decrease can perhaps/probably be explained by the mandatory use of face-coverings and the border closure with the United States for the third segment, and the use of COVID-19 serological tests and the national contact tracing for the fourth and fifth segments. An upward trend in the growth rate observed from August 31, 2020 to October 10, 2020 could have resulted from the opening of businesses and public spaces. It seems that the second wave started on October 10, 2020, with a remarkable increase in the rate that continued until January 12, 2021. After this date, the rate again declined until March 18, 2021, which could be the result of provincial states of emergency and lockdowns. The last segment witnessed another surge in the rate, perhaps due to new variants of Coronavirus.

The mPRUTF algorithm has also detected seven change points for the United Kingdom on the following dates: April 4, 2020; April 28, 2020; May 25, 2020; June 22, 2020; September 9, 2020; November 26, 2020 and February 5, 2021. As can be viewed from the figure, there are remarkable declines in the growth rates for the second segment (perhaps due to the nationwide lockdown), the third segment (perhaps due to the international travel ban) and the segments from May 25, 2020 to September 9, 2020 (perhaps due to mandatory use of face masks and comprehensive contact tracing). The country witnessed a significant increase in the growth rate starting from September 9, 2020, which aligns with the reopening of businesses, schools and universities. The second national lockdown could be linked to the very small decrease in the slope of the segment from November 26, 2020 to February 5, 2021. Finally, the growth rate in the last segment seemed to be under control, which could be the result of COVID vaccinations.

\end{example}

\section{More on Models With Frequent Change Points or With Dependent Errors}
\label{sec:model_misspecification.proj1}

This section empirically investigates the performance of mPRUTF in models with frequent change points as well as models with dependent random errors.
\subsection{mPRUTF in Signals With Frequent Change Points}
\label{discussion:frequent.changepoint.proj1}
\begin{figure}[!b]
\begin{center}
\begin{subfigure}{.45\textwidth}
  \centering
  \includegraphics[width=\linewidth]{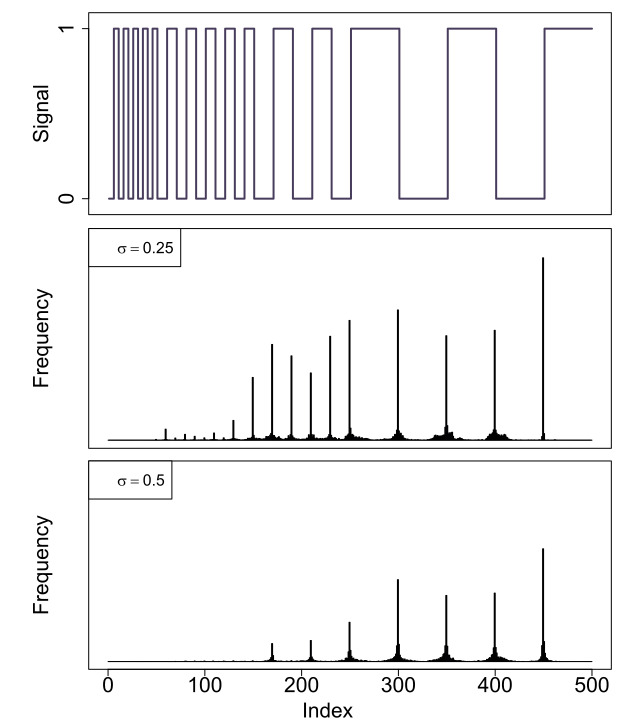}  
  \caption{Teeth signal}
\end{subfigure}
\quad
\begin{subfigure}{.45\textwidth}
  \centering
  \includegraphics[width=\linewidth]{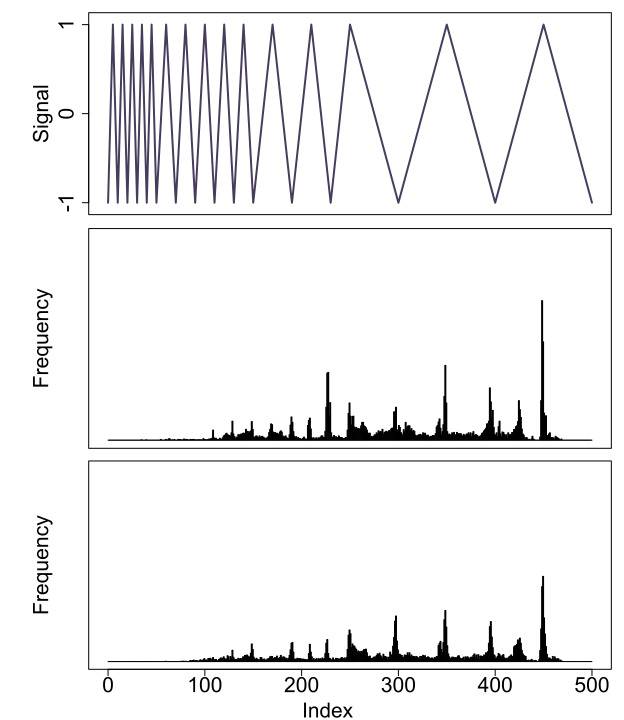}  
  \caption{Wave signal}
\end{subfigure}
\caption[Histograms of the Locations of Change Points For the Teeth and Wave Signals]{Histograms of the locations of change points for the teeth and wave signals. The histograms show the frequencies of the change points detected using mPRUTF in both signals. The result are displayed for two different $\sigma$ values.}
\label{fig:cpts_frequent_changepoint.proj1}
\end{center}
\end{figure}

In order to evaluate the detection power of mPRUTF in signals with frequent change points, we employ a teeth signal for the piecewise constant case and a wave signal for the piecewise linear case. For the teeth signal, we consider a signal with 29 change points and varying segment lengths defined as follows: \begin{itemize}
    \item for $1 \leq t \leq 50$, $f_t=0$ if $(t~ \trm{mod}~ 10) \in \{1, \ldots, 5\}$; $f_t=1$, otherwise,
    \item for $51 \leq t \leq 150$, $f_t=0$ if $(t~ \trm{mod}~ 20) \in \{1, \ldots, 10\}$; $f_t=1$, otherwise,
    \item for $151 \leq t \leq 250$, $f_t=0$ if $(t~ \trm{mod}~ 40) \in \{1, \ldots, 20\}$; $f_t=1$, otherwise,
    \item for $251 \leq t \leq 500$, $f_t=0$ if $(t~ \trm{mod}~ 100) \in \{1, \ldots, 50\}$; $f_t=1$, otherwise.
\end{itemize}
The signal is displayed in the top-left panel of Figure \ref{fig:cpts_frequent_changepoint.proj1}. 
\noindent The wave signal also has 29 change points with varying slopes which is defined as follows:
\begin{itemize}
    \item for $1 \leq t \leq 50$, $f_t=-1+0.4\, t$ if $(t~ \trm{mod}~ 10) \in \{1, \ldots, 5\}$; $f_t=1-0.4\,t$, otherwise,
    \item for $51 \leq t \leq 150$, $f_t=-1+0.2\,t$ if $(t~ \trm{mod}~ 20) \in \{1, \ldots, 10\}$; $f_t=1-0.2\,t$, otherwise,
    \item for $151 \leq t \leq 250$, $f_t=-1+0.1\,t$ if $(t~ \trm{mod}~ 40) \in \{1, \ldots, 20\}$; $f_t=1-0.1\,t$, otherwise,
    \item for $251 \leq t \leq 500$, $f_t=-1+0.04\,t$ if $(t~ \trm{mod}~ 100) \in \{1, \ldots, 50\}$; $f_t=1-0.04\,t$, otherwise.
\end{itemize}
The top-right panel of Figure \ref{fig:cpts_frequent_changepoint.proj1} shows this signal.

We generated $1000$ independent samples of $y_{_t}$ in model \eqref{fmodel.proj1} with $\varepsilon_t \stackrel{\trm{i.i.d}} {\sim} N(0\, ,\, \sigma^2)$ for both signals. The mPRUTF algorithm was then applied to these samples to estimate their change point locations. Figure \ref{fig:cpts_frequent_changepoint.proj1} shows the histograms of the locations of these change points for the signals. The figure provides evidence that mPRUTF is unable to effectively detect change points in signals with frequent change points and short segments. It also shows that the results deteriorate when the noise variance $\sigma^2$ or the polynomial order $r$ increase.

It turns out that the success of the mPRUTF algorithm critically relies on its stopping rule. Equation \eqref{stop.rule.proj1} verifies that estimating the noise variance $\sigma^2$ and specifying  the threshold $x_{\alpha}$ from a Gaussian bridge process play crucial roles in the stopping rule. As discussed in \cite{fryzlewicz2018detecting}, the two widely used robust estimators of $\sigma$, Mean Absolute Deviation (MAD) (used here) and Inter-Quartile Range (IQR), overestimate $\sigma$ in frequent change point scenarios. In addition, determining the accurate value of the threshold $x_{\alpha}$ using \eqref{thresh.stop.rule} is affected in such scenarios. These two factors prevent the stopping rule from being effective in the mPRUTF algorithm and lead to the underestimation of change points for these scenarios. We must note that such poor performance in frequent change point scenarios is not specific to mPRUTF. As investigated in \cite{fryzlewicz2018detecting}, state-of-the-art methods such as PELT, WBS, MOSUM, SMUCE and FDRSeg are among the  approaches that fail in such scenarios.

\subsection{mPRUTF in Models With Dependent Error Terms}
\label{discussion:numeric.dependent.proj1}
How can mPRUTF's performance be affected by various types of random errors, such as non-Gaussian or dependent errors? This is of course an important question and will be the topic of future works.  Notice that the dual solution path of trend filtering is not impacted by the type of random errors. However, the type of random errors plays a key role in the stopping rule of mPRUTF because the stopping rule is built based on Gaussian bridge processes established by Donsker's Theorem.

\begin{figure}[!ht]
\begin{center}
\begin{subfigure}{.32\textwidth}
  \centering
  \includegraphics[width=\linewidth]{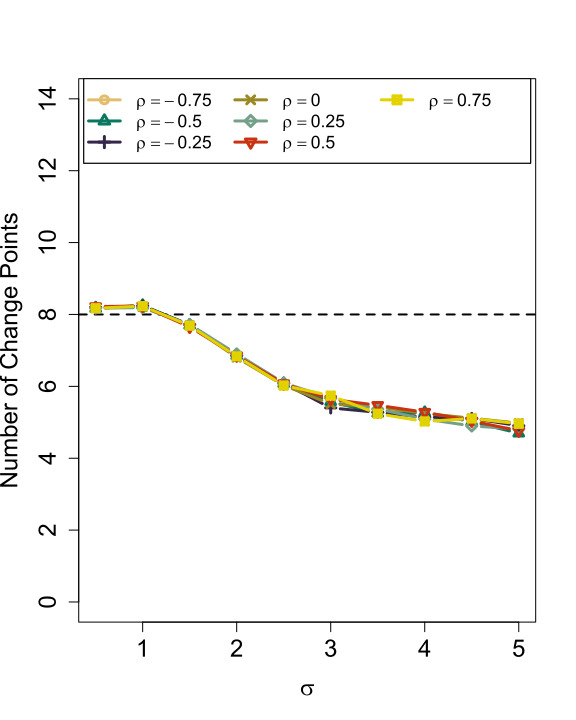}
  \label{fig:pwc.ncpts.supp.proj1}
\end{subfigure}
\begin{subfigure}{.32\textwidth}
  \centering
  \includegraphics[width=\linewidth]{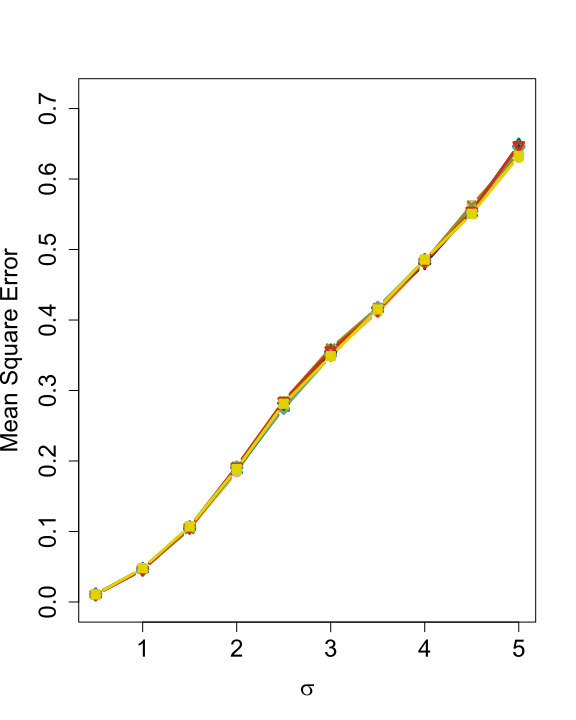}
  \caption{PWC signal }
  \label{fig:pwc.mse.supp.proj1}
\end{subfigure}
\begin{subfigure}{.32\textwidth}
  \centering
  \includegraphics[width=\linewidth]{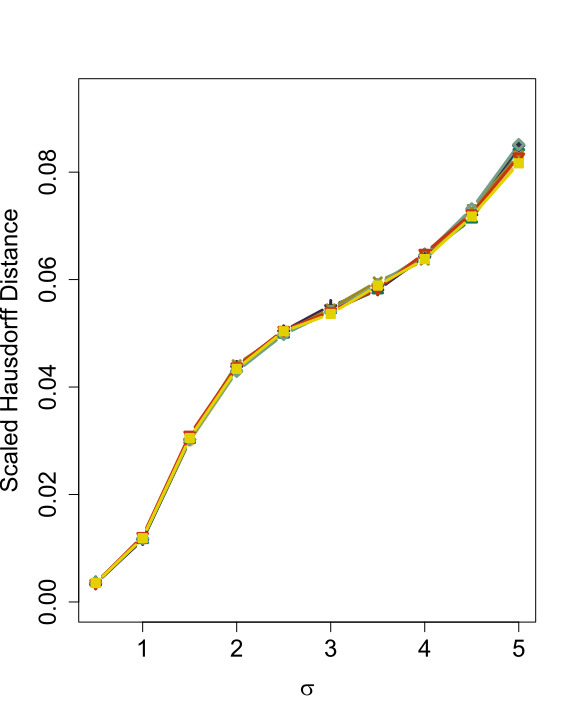}
  \label{fig:pwc.dh.supp.proj1}
\end{subfigure}
\newline
\begin{subfigure}{.32\textwidth}
  \centering
  \includegraphics[width=\linewidth]{img/ncpts_pc_dep.jpg}
  \label{fig:pwl.ncpts.supp.proj1}
\end{subfigure}
\begin{subfigure}{.32\textwidth}
  \centering
  \includegraphics[width=\linewidth]{img/mse_pc_dep.jpg}
  \caption{PWL signal }
  \label{fig:pwl.mse.supp.proj1}
\end{subfigure}
\begin{subfigure}{.32\textwidth}
  \centering
  \includegraphics[width=\linewidth]{img/dh_pc_dep.jpg}
  \label{fig:pwl.dh.supp.proj1}
\end{subfigure}
\caption[mPRUTF Results for PWC and PWL in Models With Dependent Random Errors]{The estimated average number of change points, MSEs and Hausdorff distances of various methods for both PWC and PWL signals. The results are based on weakly dependent observations and provided for various values of the error variability $\sigma$.}
\label{fig:simul-pc-pl_dep.proj1}
\end{center}
\end{figure}

To empirically investigate the performance of mPRUTF for weakly dependent random errors, a simulation study is carried out here. To this end, we generate $N=5000$ samples from model \eqref{fmodel.proj1} with the PWC and PWL signals. We consider errors $\varepsilon_i$ from an $AR(1)$ model with $\varepsilon_i = \rho \, \varepsilon_{i-1} + e_i$, for $i=1,\ldots,n$. Here, $e_i$'s are independent and identical random errors drawn from $N\big(0\, ,\, (1-\rho^2)\, \sigma^2\big)$ with $\rho \in \{-0.75,\, -0.5,\, -0.25,\, 0,\, 0.25,\, 0.5,\, 0.75 \}$ and $\sigma \in \big\{ 0.5,\, 1,\, 1.5,\, \ldots,\, 4.5,\, 5 \big\}$. The results of mPRUTF for both PWC and PWL signals are provided in Figure \ref{fig:simul-pc-pl_dep.proj1}. As can be seen, the results are very similar, in terms of the average number of change points, MSEs and the scaled Hausdorff distances, for various values of $\rho$. Therefore, it appears that the mPRUTF algorithm is quite robust against dependent error terms. Extensive studies of mPRUTF for non-Gaussian and dependent random errors will be carried out in future research.

\section{Discussion}
\label{sec:discussion.proj1}

This paper proposed an algorithm, PRUTF, to detect change points in piecewise polynomial signals using trend filtering. We demonstrated that the dual solution path produced by the PRUTF algorithm forms a Gaussian bridge process for any given value of the regularization parameter $\lambda$. This conclusion allowed us to derive an efficient stopping rule for terminating the search algorithm, which is vital in change point analysis. We then proved that when there is no staircase block in the signal, the method guarantees consistent pattern recovery. However, it fails in doing so when there is a staircase in the underlying signal. To address this shortcoming in such a case, we suggested a modification in the procedure of constructing the solution path, which effectively prevents false discovery of change points. Evidence from both simulation studies and real data analysis reveals the accuracy and the high detection power of the proposed method.


\appendix

\section*{Appendix A}
\renewcommand{\theequation}{A.\arabic{equation}}
\renewcommand{\thesubsection}{A.\arabic{subsection}}
\setcounter{equation}{0}

\subsection{Proof of Theorem \ref{thm:gaussian.bridge.proj1}}
\label{prf:gaussian.bridge.proj1}
For $\varepsilon_{_1},\,\dots,\,\varepsilon_{_n},\,\dots\overset{\rm i.i.d.}{\sim}N(0,\,\sigma^2)$, and a sequence $q_{_1},\,\dots,\,q_{_n},\,\dots$ of real numbers, let
\begin{align*}
    \nu_{_k}={\rm Var}\left(\sum\limits_{i=1}^{k}q_{_i}\,\varepsilon_{_i} \right)=\sigma^2\sum\limits_{i=1}^kq_{_i}^2\qquad \text{ for } \quad k \geq 1\,.
\end{align*}
Define the partial weighted sum process $\big\{ S_{_n}(t)\,:~ 0\leq\,  t\, \le 1 \big\}$ by
\begin{equation*}
S_{_n}(t)=\frac{1}{\sqrt{\nu_{_n}}}\sum\limits_{i=1}^{\lfloor nt\rfloor}q_{_i}\,\varepsilon_{_i}\,,\qquad \trm{for} \quad 0\leq t\le 1\,.
\end{equation*}
Obviously, for any $k\geq 1$, and any $0< t_{_1}<t_{_2}<\cdots<t_{_k}\leq 1$, the vector $\big(S_{_n}(t_{_1}),\ldots,S_{_n}(t_{_k})\big)$ has a multivariate normal distribution, and therefore $\big\{ S_{_n}(t)\,:\, 0\leq\, t\, \le 1\big\}$ is a Gaussian process for any given $n$.

\begin{enumerate}[label=\alph*)]
\item In our case, first note that
\begin{align}\label{w0j.process.epsilon.append}
   \left[\big(\mathbf{D}_{_{-\mathcal{A}}}\mathbf{D}_{_{-\mathcal{A}}}^T\big)^{-1}\mathbf{D}_{_{-\mathcal{A}}}\right]_{\lfloor mt\rfloor} \mathbf{y}
   & ~=~ \left[\big(\mathbf{D}_{_{-\mathcal{A}}}\mathbf{D}_{_{-\mathcal{A}}}^T\big)^{-1}\mathbf{D}_{_{-\mathcal{A}}}\right]_{\lfloor mt\rfloor} \big(\,\mathbf{y}- \mbf f\, \big)
   \\[8pt]
   & ~= ~\left[\big(\mathbf{D}_{_{-\mathcal{A}}}\mathbf{D}_{_{-\mathcal{A}}}^T\big)^{-1}\mathbf{D}_{_{-\mathcal{A}}}\right]_{\lfloor mt\rfloor} \bsy\varepsilon,
\end{align}
which is a partial weighted sum process of independent and identical Gaussian random variables $\varepsilon_i, ~ i=1\, ,\, \ldots\, ,\, n$. The first equality in \eqref{w0j.process.epsilon.append} is derived from the fact that the structure of the true signal $\mbf f$  remains unchanged within the $j$-th block, meaning that $\left[\mathbf{D}_{_{-\mathcal{A}}}\right]_{\lfloor mt\rfloor}\mathbf{f}=0$, for $\big(\tau_{_j}+r_{_a}\big)/m ~\leq~ t ~\leq~ \big(\tau_{_{j+1}}-r_{_b}\big)/m$, which in turn implies
\begin{align*}
    \left[\big(\mathbf{D}_{_{-\mathcal{A}}}\mathbf{D}_{_{-\mathcal{A}}}^T\big)^{-1}\mathbf{D}_{_{-\mathcal{A}}}\right]_{\lfloor mt\rfloor}\mathbf{f}=0.
\end{align*}

Thus, from the aforementioned argument for $\big\{ S_n(t) \big\}$, the process $\mbf W_j=\big\{ W_j(t):~(\tau_{_j}$ $+\, r_{_a})/m\leq t\leq (\tau_{_{j+1}}-r_{_b})/m\big\}$ is a Gaussian process, where
\begin{align*}
    W_j(t)= \big( \tau_{j+1}-\tau_j-r \big)^{-(2\,r+1)/2} \left[ \big(\mathbf{D}_{_{-\mathcal{A}}} \mathbf{D}_{_{-\mathcal{A}}}^T \big)^{-1} \mathbf{D}_{_{-\mathcal{A}}} \right]_{\lfloor mt\rfloor}\mathbf{y}.
\end{align*}
Additionally, with the conditions given in \eqref{wbridge.tails.proj1},  $\mbf W_{_j}$ is a Gaussian bridge process over the interval $(\tau_{_j}+r_{_a})/m ~\leq~ t ~\leq~ (\tau_{_{j+1}}-r_{_b})/m$. Furthermore, from \eqref{w0j.process.epsilon.append}, the mean vector and covariance matrix of $\mbf W_{_{j}}$ can be computed  as $\mbf 0$ and $\sigma^2 \, \big(\mathbf{D}_{_{-\mathcal{A}}}\mathbf{D}_{_{-\mathcal{A}}}^T\big)^{-1}$.

\item  Recall that the covariance matrix $\big( \mathbf{D}_{_{-\mathcal{A}}}\mathbf{D}_{_{-\mathcal{A}}}^T \big)^{-1}$ is a block diagonal matrix which states that the covariance matrix between two distinct blocks is zero. This completes the proof of the theorem.
\end{enumerate}
 
\subsection{Proof of Theorem \ref{thm:consistency.constraints.proj1}}
\label{prf:consistency.constraints.proj1}
 
\begin{enumerate}[label=\alph*)]
    \item For $t=1,\,\ldots,\tau_{_1}-r_{_b}$, and both signs $\pm 1$, according to the KKT conditions, the dual variables $\widehat{u}(t)$ must lie between $-\lambda$ and $\lambda$, that is, 
    \begin{align}
        -\lambda\leq \widehat{u}_{_0}^{\,\mathrm{st}}(t)-\lambda\left[\big(\mathbf{D}_{_{-\mca A}}\mathbf{D}_{_{-\mca A}}^T\big)^{-1}\mathbf{D}_{_{-\mca A}}\right]_t\,\mathbf{D}_{_{A_1}}^T \mbf 1\leq \lambda\,
    \end{align}
    which yields the constraint for the first block in \eqref{block.const.first}.
   
    \item Similar to the first block, for $t=\tau_{_{J_0}}+r_{_a},\,\ldots,\,m$, the constraint becomes
    \begin{align}
       -\lambda \leq \widehat{u}_{_{J_0}}^{\,\mathrm{st}}(t)+\lambda\left[\big(\mathbf{D}_{_{-\mca A}}\mathbf{D}_{_{-\mca A}}^T \big)^{-1} \mathbf{D}_{_{-\mca A}}\right]_t\,\mathbf{D}_{_{A_{_{J_0}}}}^T \mbf 1\leq \lambda\,,
    \end{align}
    which leads to the result of \eqref{block.const.last}.
   
    \item For $t=\tau_{_j}+r_{_a},\,\ldots,\,\tau_{_{j+1}}-r_{_b}$, and $j=1,\,\ldots,\,J_{_0}-1$ the constraint for the exact pattern recovery becomes
    \begin{align}
        \lambda\,s_{_j}\leq\widehat{u}_{_j}^{\,\mathrm{st}}(t)-\lambda \left[ \big(\mathbf{D}_{_{-\mca A}} \mathbf{D}_{_{-\mca A}}^T \big)^{-1}\mathbf{D}_{_{-\mca A}}\right]_t\, \left( \mathbf{D}_{_{A_{j+1}}}^T \mathbf{s}_{_{j+1}}+ \mathbf{D}_{_{A_j}}^T \mathbf{s}_{_j}\right) 
        \leq \lambda\,s_{_{j+1}}\,.
    \end{align}
Since the stochastic process $\widehat{u}_{_j}^{\,\mathrm{st}}(t)$ is symmetric around zero, when $s_{_{j+1}}$ and $s_{_j}$ have the opposite signs, this constraint reduces to \eqref{block.const.inter}. Otherwise, when $s_{_{j+1}}=s_{_j}$, which accounts for the staircase in block $j$, from \eqref{drift.term.staircase.proj1} the constraint becomes $\,\widehat{u}_{_j}^{\,\mathrm{st}}(t) \leq 0\,$ or $\,\widehat{u}_{_j}^{\,\mathrm{st}}(t) \geq 0\,$.
\end{enumerate}

\subsection{Proof of Theorem \ref{thm:consistency.proj1}}\label{prf:consistency.proj1}
\begin{enumerate}[label=(\alph*)]
  \item The PRUTF algorithm is consistent in pattern recovery if the event
  \begin{align}\label{pattern.recovery.event.proj1}
      \Big\{ \widehat{\bsy\tau}=\bsy\tau \Big\} \medcap \Big\{\, \trm{sign} \big(\mbf D_t \widehat{\mbf f}_n\, \big) \,=\, \trm{sign} \big(\mbf D_t \mbf f\, \big), ~\forall\, t \in \bsy\tau  \Big\},
  \end{align}
  occurs with probability approaching one. For ease of exposition, we first compute the probability of the statement in \eqref{pattern.recovery.event.proj1} for the piecewise constant case, $r=0$. We then extend this probability computation to an arbitrary piecewise polynomial $r \in \mbb N$. 
  
  {\bf Case  $r=0\,$:}
  In this case, the event in
  \eqref{pattern.recovery.event.proj1} is equivalent to $\big\{A_n\, \cap B_n\, \big\}$ where
  \begin{align}\label{An.event.proj1}
      A_n= \Big\{ \min_{t \in \bsy\tau}\, \big| \mbf D_t \widehat{\mbf f}_n \big| ~>~ 0 \Big\},
  \end{align}
  and 
  \begin{align}\label{Bn.event.proj1}
      B_n= \Big\{ \max_{t \in \bsy\tau^c}\, \big| \mbf D_t\, \overline{\bsy\varepsilon}_n \big| ~\leq~ 4\, \lambda_n \Big\}.
  \end{align}
  For $t \in \bsy\tau^c= \{1, \, \ldots\, ,\, m\}\backslash \bsy\tau$, observe that  $\mbf D_t \big(\, \widehat{\mbf f}_n - \mbf f \big) =0$; therefore, 
  \begin{align*}
      \big| \mbf D_t\,  \overline{\bsy\varepsilon}_n \big| & ~=~ \big| \mbf D_t\, \big( \overline{\mbf y}_n - \widehat{\mbf f}_n\, \big) +  \mbf D_t\, \big(\, \widehat{\mbf f}_n - \mbf f \big) \big| = \big| \mbf D_t\, \big( \overline{\mbf y}_n - \widehat{\mbf f}_n\, \big) \big|
      \\[9pt]
      & ~=~ \big| \mbf D_t\, \mbf D^T \widehat{\mbf u} \big| ~\leq~  4 \,\lambda_n,
  \end{align*}
  which is captured in event $B_n$.  The last inequality in the above equation occurs because, from Theorem \ref{thm:consistency.constraints.proj1}, we have $|\widehat{\mbf u}| \leq \lambda_n$ as well as the fact that $\sum_{i=1}^m \big| \big[ \mbf D_t\, \mbf D^T \big]_i \big|$ $= 2^{\,r+2}$, for an arbitrary $r$. In the following, we derive the conditions under which the probabilities of both events $A_n$ and $B_n$ converge to 1.
  \begin{itemize}
      \item To compute the probability of $A_n$, we first note that, for every $t\in \bsy\tau$,
      \begin{align*}
        \big| \mbf D_t\, \widehat{\mbf f}_n \big| & ~=~ \big| \overline{\mbf y}_{n,\,t} - \overline{\mbf y}_{n,\, t-1} - \lambda_n \big( s_t -s_{t-1} \big) \big| 
        \\[9pt]
        & ~=~ \big| \mbf D_t\, \overline{\bsy\varepsilon}_n + \mbf D_t\, \mbf f - \lambda_n \big( s_t -s_{t-1} \big) \big|
        \\[9pt]
        & ~\leq~  \big| \mbf D_t\, \overline{\bsy\varepsilon}_n  - \lambda_n \big( s_t -s_{t-1} \big) \big| + \big|  \mbf D_t\, \mbf f \big|,
      \end{align*}
      where $\overline{\mbf y}_{n,\,t}$ is the average of observations in the segment created by block $t$. The last inequality in the above statement is derived from the triangular inequality. Therefore, in order to verify $A_n$, it is enough to show that, with the probability approaching one,
      \begin{align}\label{bias.delta.proj1}
          \max\limits_{t \in \bsy\tau}\,  \big| \mbf D_t\, \overline{\bsy\varepsilon}_n  - \lambda_n \big( s_t -s_{t-1} \big) \big|  \leq \delta_n,
      \end{align}
      where $\delta_n= \min\limits_{t\, \in\, \bsy\tau}\, \big| \mbf D_t \mbf f \big|$ is the minimum jump between change points. Equivalently, it suffices to show that
      \begin{align}\label{max.bias.delta.proj1}
          \max_{t \in \bsy\tau}~ \lambda_n\, \big|  s_t - s_{t-1}  \big|  \leq \delta_n/2,
      \end{align}
      and 
      \begin{align}\label{max.errorbar.proj1}
       \max_{t \in \bsy\tau}~ \big| \mbf D_t\, \overline{\bsy\varepsilon}_n \big| \leq \delta_n/2.
    \end{align}
    The inequality in \eqref{max.bias.delta.proj1} holds if $\lambda_n \leq \dfrac{\delta_n\, \underline{L}_n}{4}$, where $\underline{L}_n= \min\limits_{j=0,\,\ldots,\,J_{_0}}~  \big|\tau_{_{j+1}}-\tau_{_j}\big|$. Applying the union and Gaussian tail bounds, the probability of the complement of the event in \eqref{max.errorbar.proj1} can be computed as
    \begin{align}\label{pmax.epsilon.pwc.proj1}
      \Pr \big( \max_{t \in \bsy\tau} \big| \mbf D_t\, \overline{\bsy\varepsilon}_n \big| & \,\geq\, \delta_n/2 \big) 
       ~\leq~ \sum_{t\in \bsy\tau} \Pr \big( \big| \mbf D_t\, \overline{\bsy\varepsilon}_n \big| \geq \delta_n/2 \big)
      \\[9pt]
      & ~=~ \sum_{t\in \bsy\tau} \Pr \left( \big| Z_n \big| \geq \frac{\delta_n\, \sqrt{\underline{L}_n}}{ 2\, \sqrt{2}\, \sigma_n} \right)
      ~\leq~ 2\, J_0\, \exp \left( -\frac{\delta^2_{n}\, \underline{L}_n}{ 16\, \sigma_n^2} \right)
      \\[9pt]
      & ~=~ 2\, \exp \left( -\frac{\delta^2_{n}\, \underline{L}_{n}}{ 16\, \sigma_n^2} + \log(J_0) \right).
  \end{align}
  The probability in \eqref{pmax.epsilon.pwc.proj1} converges to zero if, for some $\xi>0$,
  \begin{align}
      \frac{\delta_{n}\, \sqrt{\underline{L}_{n}}}{ \sigma_n} \longrightarrow \infty
      \qquad\quad \trm{and}  \qquad\quad
      \frac{ \delta_{n}\, \sqrt{\underline{L}_{n}}}{\sigma_n\, \sqrt{\log (J_0)}} ~>~ \sqrt{16}\, (1+\xi).
  \end{align}

  \item Next, we verify conditions under which $\Pr (B_n) \longrightarrow 1$. Equivalently, it is enough to determine the conditions under which the following probability converges to zero.
  \begin{align}
      \Pr \big(B_n^{\,c} \big) & ~=~ \Pr \Big( \max_{t \in \bsy\tau^c}\, \big| \mbf D_t \overline{\bsy\varepsilon}_n \big| \geq 4\, \lambda_n \Big) 
      \,\leq\, \sum_{t\in \bsy\tau^c}~ \Pr \big( \big| \mbf D_t\, \overline{\bsy\varepsilon}_n \big| \geq 4\, \lambda_n \big)
      \\[9pt]
      & ~=~ \sum_{t\in \bsy\tau^c}~ \Pr \left( \big| Z_n \big| \geq \frac{4\, \lambda_n\, \sqrt{\underline{L}_n}}{ \sqrt{2}\,\sigma_n} \right)
      \\[9pt]
      & ~\leq~ 2\, \big( n-J_0 \big)\, \exp \left( -\frac{4\, \lambda_n^2\,\, \underline{L}_{n}}{ \sigma_n^2} \right)
      \\[9pt]
      & ~=~ 2\, \exp \left( -\frac{4\, \lambda_n^2\,\, \underline{L}_{n}}{ \sigma_n^2} + \log \big( n-J_0 \big) \right).
  \end{align}
  The above probability converges to zero if, for some $\xi>0$,
  the following conditions hold,
  \begin{align}
      \frac{\lambda_n\, \sqrt{\underline{L}_{n}}}{ \sigma_n} \longrightarrow \infty
      \qquad\quad \trm{and}  \qquad\quad
      \frac{2\, \lambda_n\, \sqrt{ \underline{L}_{n}}}{\sigma_n\, \sqrt{\log (n-J_0)}} ~>~  (1+\xi).
  \end{align}
  \end{itemize}
  
  {\bf Case arbitrary $r \,$:} For the piecewise polynomial of order $r$ $(r \in \mbb N)$, we note that, for any $t \in [\tau_j\, ,\, \tau_{j+1})$,
\begin{align*}
    \big|  \mbf D_t\, \widehat{\mbf f}_n \Big|  & ~=~ \big|  \mbf D_t\, \big( \mbf y_n - \mbf D_{_{-\mca A}}^T\, \widehat{\mbf u}\big)  \Big|
    \\[9pt]
    & ~=~ \bigg|  \mbf D_t\, \Big[ \big(  \mbf I - \mbf P_{D} \big) \mbf y_n ~-~ \lambda_n\,\mbf P_{D}\left( \mbf{D}_{_{A_{j+1}}}^T \mbf{s}_{_{j+1}}+ \mbf{D}_{_{A_j}}^T \mbf{s}_{_j}\right)  \Big] \bigg|
    \\[9pt]
    & ~=~ \bigg|  \mbf D_t\, \Big[ \big(  \mbf I - \mbf P_{D} \big) \big(\mbf f +\bsy\varepsilon_n \big) ~-~ \lambda_n\,\mbf P_{D}\left( \mbf{D}_{_{A_{j+1}}}^T \mbf{s}_{_{j+1}}+ \mbf{D}_{_{A_j}}^T \mbf{s}_{_j}\right)  \Big] \bigg|,
\end{align*}
where $\mbf P_D=\mbf D_{_{-\mca A}}^T \big( \mbf D_{_{-\mca A}} \mbf D_{_{-\mca A}}^T \big)^{-1}\mbf D_{_{-\mca A}}$ is the projection map onto the row space of $\mbf D_{_{-\mca A}}$. In the preceding statement, the second equality is derived by plugging in the statement in \eqref{usegment} in place of $\widehat{\mbf u}$. From \eqref{Dprojection.map.proj1}, recall that $\mbf I - \mbf P_{D} $ is equivalent to the prediction matrix in the $r$-th polynomial regression of  $\mbf y$ onto indices $ \tau_j+1\, ,\, \tau_j+1\, ,\, \ldots\, ,\, \tau_{j+1}$. This fact allows us to derive an upper bound for the variance of $\mbf D_t\, \big(  \mbf I - \mbf P_{D} \big)\, \bsy\varepsilon_n$ \citep{yu2020localising},
\begin{align*}
    \max_{t\in \bsy\tau}~ \trm{Var} \Big(  \mbf D_t\, \big(  \mbf I - \mbf P_{D} \big)\, \bsy\varepsilon_n \Big) ~\leq~ 2^{\,r+1}\, \frac{ n^{2r}\, \sigma_n^2}{\underline{L}_n^{2r+1}}.
\end{align*}
Following a procedure similar to that used in the case $r=0$, 
\begin{align}
     \Pr \left( \max_{t \in \bsy\tau}\, \big| \mbf D_t\, \big(  \mbf I - \mbf P_{D} \big)\, \bsy\varepsilon_n \big|  \,\geq\, \frac{\delta_n}{2} \right) 
       ~\leq~  2\, \exp \left( -\frac{\delta_{n}^2\,\, \underline{L}_{n}^{2r+1}}{ 2^{\,r+4}\, n^{2r}\, \sigma_n^2} + \log \big(J_0 \big) \right).
 \end{align}
   For the case of an arbitrary $r$, there is a slight modification in the definition of event $B_n$: 
  \begin{align}\label{Bn.event.genera.r.proj1}
      B_n= \Big\{ \max_{t \in \bsy\tau^c}~ \big| \mbf D_t\, \big(  \mbf I - \mbf P_{D} \big)\, \bsy\varepsilon_n \big| ~\leq~ 2^{\,r+2}\,\lambda_n \Big\}.
  \end{align}
  Again, in the same manner
  \begin{align}
      \Pr \big(B_n^{\,c} \big) ~\leq~ 2\, \exp \left( -\frac{2^{\, r+2}\, \lambda_n^2\,\, \underline{L}_{n}^{2r+1}}{ n^{2r}\, \sigma_n^2} + \log \big( n-J_0 \big) \right).
  \end{align}
    Therefore, for an arbitrary $r$, the PRUTF algorithm is consistent in pattern recovery if, in addition to
    \begin{align*}
        \lambda_n < \dfrac{\delta_n\, \underline{L}_n^{2r+1}}{n^{2r}\, 2^{\,r+2}},
    \end{align*}
    the conditions in \eqref{consistent.conditions1.proj1} and \eqref{consistent.conditions2.proj1} hold.

    \item As shown in part (c) of Theorem \ref{thm:consistency.constraints.proj1}, in staircase blocks, the violation of the KKT conditions boils down to crossing the zero line for a Gaussian bridge process. Suppose $j$-th block is a staircase block; therefore, PRUTF can attain the exact discovery if $\widehat{u}_{_j}^{\,\mathrm{st}}(t) \leq 0$ or  $\widehat{u}_{_j}^{\,\mathrm{st}}(t) \geq 0$, for all $(\tau_{_j}+r_{_a})/m\, \leq\, t\, \leq\, (\tau_{_{j+1}}-r_{_b})/m$. Hence
    the probability of this event occurring is equal to $\Pr\Big(\max\limits_{0\, \leq\, t\, \leq\, L_{_j}} \widehat{u}_{_j}^{\,\mathrm{st}}(t) \leq 0\Big)$. According to \cite{beghin1999maximum},
    \begin{align}
        \Pr\Big(\max\limits_{0\, \leq\,  t\, \leq\, L_{_j}} \widehat{u}_{_j}^{\,\mathrm{st}}(t) \leq a\Big)=1-\exp\Big(-\frac{2\,a^2}{S_{_r}^2(L_{_j})}\Big),
    \end{align}
    where $S_{_r}^2(L_{_j})$ is the $L_{_j}$-th diagonal element of the matrix $\sigma^2 \big(\mathbf{D}_{_{A_j}}\mathbf{D}_{_{A_j}}^T\big)^{-1}$. As a result, the probability converges to zero as $a$ vanishes. This result implies that the PRUTF algorithm fails to consistently recover the true pattern in the presence of staircase patterns.
 
\end{enumerate}

\newpage
\renewcommand\refname{Bibliography}
\bibliographystyle{apalike}
\bibliography{Project1}

\end{document}